\documentclass[11pt,showpacs]{article}

\newfont{\tenmsb}{msbm10 scaled\magstep1}

\let\ssection=\subsection
\renewcommand{\subsection}{\setcounter{equation}{0}\ssection}

\usepackage{graphics,amssymb,amsmath,amsthm,amscd,array}
\usepackage{graphicx,subfigure}
\usepackage{makeidx}
\usepackage{bm}
\usepackage{amsmath,dsfont}

\font\BBBig=cmr10 scaled\magstep4
\font\small=cmr9

\def\parag{\hfil\break} 
\def\kikezd{\parag\underbar}
\def\IR{{\mathds{R}}} 
\def\cL{{\cal{L}}}
\def\cS{{\cal{S}}}
\def\cA{{\cal{A}}}
\def\cB{{\cal{B}}}
\def\cE{{\cal{E}}}

\def\cG{{\cal{G}}}

\def\vbeta{{\vec{\beta}}}
\def\vgamma{{\vec{\gamma}}}
\def\smallover#1/#2{\hbox{$\textstyle\frac{#1}{#2}$}} %
\def\smallcirc{{\,\raise 0.5pt \hbox{$\scriptstyle\circ$}\,}}
\def\2{{\smallover1/2}}
\def\={{\!=\!}}
\def\O{{\rm O}}
\def\o{{\rm o}}
\def\const{{\rm const}}
\def\p{{\partial }}
\def\dAlembert{\vcenter {
    \hbox {\vrule height8pt width0.4pt depth0.0pt
           \vrule height8pt width7.2pt depth-7.6pt
           \vrule height8pt width0.4pt depth0.0pt
           \kern-8pt
           \vrule height0.4pt width8pt depth0.0pt
          \,}}}

\def\vD{{\vD}}

\def\and{\qquad\hbox{\small and}\qquad}
\def\where{\qquad\hbox{\small where}\qquad}

\def\rot{{\rm curl\ }}
\def\grad{{\rm grad\ }}

\def\vnabla{{\overrightarrow{\nabla}}}
\def\vJ{{\vec{J}}}
\def\vD{{\overrightarrow{D}}}
\def\vA{{\vec{A}}}
\def\vcA{{\overrightarrow{\cal A}}}

\def\vx{{\vec{x}}}
\def\vX{{\overrightarrow{X}}}

\def\vB{{\overrightarrow{B}}}
\def\vE{{\overrightarrow{E}}}
\def\vR{{\overrightarrow{R}}}
\def\ve{{\vec{e}}}
\def\vr{{\vec{r}}}
\def\vj{{\vec{\jmath}}}
\def\vv{{\vec{v}}}
\def\vp{{\vec{p}}}

\def\va{{\vec{a}}}
\def\vb{{\vec{b}}}
\def\vc{{\vec{c}}}
\def\cM{{\cal{M}}}
\def\cH{{\cal{E}}}
\def\cE{{\cal{E}}}
\def\cP{{\cal{P}}}
\def\cG{{\cal{G}}}
\def\cK{{\cal{K}}}
\def\cJ{{\cal{J}}}
\def\cN{{\cal{N}}}
\def\cD{{\cal{D}}}

\def\D{
{D\mkern-2mu\llap{{\raise+0.5pt\hbox{\big/}}}\mkern+2mu}
}  
\def\vsigma{{\vec{\sigma}}}

\begin{document}
\setlength{\baselineskip}{15pt}

\title{\BBBig Vortices in (abelian)
\\[18pt]
 Chern-Simons gauge theory
\\[18pt]}

\author{
Peter A. HORVATHY
\\[12pt]
Laboratoire de Math\'ematiques et de Physique Th\'eorique
\\[12pt]
Universit\'e de TOURS (France)
\\[20pt]
Pengming ZHANG
\\[12pt]
Institute for Modern Physics, Chinese Academy of Sciences
\\[12pt]
LANZHOU (China)
}

\maketitle

\begin{abstract}
The vortex solutions of various classical planar field theories with
(Abelian) Chern-Simons term are reviewed. Relativistic vortices, put
forward by Paul and Khare, arise when the Abelian Higgs model is
augmented with the Chern-Simons term. Adding a suitable sixth-order
potential and turning off the Maxwell term provides us with pure
Chern-Simons theory with both topological and non-topological
self-dual vortices, as found by Hong-Kim-Pac, and by
Jackiw-Lee-Weinberg. The non-relativistic limit of the latter leads
to non-topological Jackiw-Pi vortices with a pure fourth-order
potential. Explicit  solutions are found by solving the Liouville
equation.

The scalar matter field can be replaced by spinors, leading
to fermionic vortices.

Alternatively, topological vortices in external field
are constructed in
the phenomenological model proposed by Zhang-Hansson-Kivelson. Non-relativistic Maxwell-Chern-Simons vortices
are also studied.

The Schr\"odinger symmetry of Jackiw-Pi vortices, as
well as the construction of some
time-dependent vortices, can be explained
by the conformal properties of non-relativistic space-time,
derived in a Kaluza-Klein-type framework.

\bigskip

\end{abstract}


 \vskip15mm\noindent 
 {\sl Physics Reports} {\bf 481} 83 (2009)\\
e-print~: \texttt{arXiv:0811.2094 [hep-th]}

\newpage
\null\newpage
\tableofcontents
\newpage\null\newpage

\section{INTRODUCTION}

\subsection{Chern-Simons Form and the Hall Law}\label{CSHall}

The mathematical machinery developed by Maxwell to formulate the principles of Faraday's electromagnetism has been,
subsequently, applied to pure mathematics, namely to describe bundles
over \textit{even dimensional} manifolds, using characteristic classes \cite{KoNo}.
 One hundred years later, in the early 1970s,
S. S. Chern, and J. Simons \cite{Chern}, proposed to study
bundles over \textit{odd-dimensional} manifolds using
secondary characteristic classes.
In $3$ space-time dimensions (the only case we study here),
the (Abelian) Chern-Simons three-form is \footnote{
Three-dimensional space-time indices are denoted by
$\alpha, \beta,\dots=0,i$.
In the relativistic case, we work with a metric of signature
$(-1,1,1,)$, with $\epsilon_{012}=1$. Hence
\begin{equation}
\epsilon^{\alpha\beta\gamma}
A_\alpha F_{\beta\gamma}=
-A_0\epsilon_{ij}F_{ij}-2\epsilon_{ij}A_iF_{j0}
=
-2A_0B-2\vA\times\vE,
\label{CSconvention}
\end{equation}
where
$B=\epsilon_{ij}\p_iA_j$ and
$E_i=F_{i0}=\p_iA_0-\p_0A_i$.
This same convention is kept in the non-relativistic case.

The non-Abelian generalization (not considered here)
of the Chern-Simons form  is
$$
\frac{1}{4}
{\rm Tr}\left(
A_\alpha F_{\beta\gamma}
-\frac{2}{3}A_\alpha A_\beta A_\gamma\right)
dx^\alpha\wedge dx^\beta\wedge dx^\gamma.
$$}
\begin{equation}
\hbox{(CS form)}=
\frac{1}{4}
A_\alpha F_{\beta\gamma}\,
dx^\alpha\wedge dx^\beta\wedge dx^\gamma,
\label{CSform}
\end{equation}
where $A=A_\alpha dx^\alpha$ is some real vector potential
and $F_{\alpha\beta}$ is its curvature.

The first applications of the Chern-Simons form to physics
 came in the early 1980s, namely in
\textit{topologically massive gauge theory} \cite{CSem,DJT}. It was
realized that, in $(2+1)$ space-time dimensions,  (\ref{CSform}) can  be added to
 the usual Maxwell term in the electromagnetic action,
\begin{eqnarray}
S=S_{em}+S_{CS}
=\int \frac{1}{4}F_{\alpha\beta}F^{\alpha\beta}
\,d^3x
-\kappa \int
\frac{1}{4}\epsilon^{\alpha\beta\gamma}
A_\alpha F_{\beta\gamma}\,d^3x,
\label{CSemAction}
\end{eqnarray}
where $\kappa$ is some coupling constant, which determines
the relative strength of the Maxwell and Chern-Simons
dynamics.

The unusual feature of (\ref{CSemAction})
is that, unlike the conventional
Maxwell term, the Chern-Simons form
(\ref{CSform}) is \textit{not} invariant under a gauge transformation
$A_\alpha\to A_\alpha+\p_\alpha\lambda$. Explicitly,
\begin{equation}
\hbox{(CS form)}\to\hbox{(CS form)}
+\frac{1}{4}
(\p_\alpha\lambda)F_{\beta\gamma}\,
dx^\alpha\wedge dx^\beta\wedge dx^\gamma.
\label{CSchange}
\end{equation}
It yields, nevertheless, consistent physics~:
 the field equations,
\begin{equation}
\p_\alpha F^{\alpha\gamma}+\frac{\kappa}{2}
\epsilon^{\gamma\alpha\beta}F_{\alpha\beta}=0,
\label{CSFE}
\end{equation}
\textit{are} gauge invariant. This can be understood by noting that,
using the sourceless Maxwell equations
$\epsilon^{\alpha\beta\gamma}\p_\alpha F_{\beta\gamma}=0$
[which merely follow from our using potentials,
$F_{\mu\nu}=\p_\mu A_\nu-\p_\nu A_\mu$],
the integrand in  (\ref{CSemAction})
changes by a surface term,
\begin{equation}
\Delta \cL=-\p_\alpha\left(\lambda\,\frac{\kappa}{4}
\epsilon^{\alpha\beta\gamma}F_{\beta\gamma}\right),
\label{DeltaCS}
\end{equation}
and defines, therefore,  a satisfactory gauge theory.  Moreover,
it can be inferred from (\ref{CSFE}) that the Chern-Simons
dynamics endows the gauge field $A_\mu$ with a gauge invariant ``topological
mass'' \footnote{The
Chern-Simons form exhibits a behavior similar to that of  a
Dirac monopole, for which no global vector potential
exists, yet the classical action is satisfactorily defined.
In the non-Abelian context, and over a compact space-time
manifold, this leads to the quantization of the
Chern-Simons coefficient, interpreted as the topological mass
 \cite{DJT,JCS,HoNash}, in an analogous way to the Dirac quantization of the monopole charge \cite{HAction}.}.

The Chern-Simons term can thus be used
to sup\-ple\-ment, or even replace, the usual
Maxwellian dynamics.
The resulting dynamics is ``poorer'', since it allows
no propagating modes.  It has, on the other hand, larger symmetries~:
while the Maxwell term $(1/4)F_{\alpha\beta}F^{\alpha\beta}$ is only meaningful if
a metric $g_{\alpha\beta}$ is given, the Chern-Simons
term is \textit{topological}, i.e.,  the integral
$$
\int\epsilon^{\alpha\beta\gamma}A_\alpha F_{\beta\gamma}\,d^3x
$$
is \textit{independent} of the metric we choose.
Thus, while the Maxwell theory has
 historically been at the
origin of (special) relativity, the Chern-Simons term can accommodate
both relativistic and non-relativistic (or even mixed)
theories.

Note that Galilean field theory has been investigated, starting in
the mid-eighties  \cite{Hagen1, Hagen2}.

\kikezd{The  Hall Effect}

The Hall Effect, predicted and first observed around 1879
 \cite{Hall} says that, in a thin conducting layer subjected
to a planar electric field and to a magnetic field
 perpendicular to the sample,
the electric current and the field are related by
the {\it off-diagonal} relation
\begin{equation}\fbox{$
\vec{\jmath}=\sigma{\vE},
\qquad
\sigma=\left(\begin{array}{cc}
0&\sigma_H
\\
-\sigma_H&0
\end{array}\right)\ ,
$}
\label{HallLaw}
\end{equation}
where $\sigma_H$ is the Hall conductivity.

One hundred years after Hall's original discovery, a similar experiment
was performed under rather different conditions
 \cite{QHE,GirvQHE,StoneQHE,EzawaQHE}. It was found that, in a thin semiconductor placed into a
 strong perpendicular magnetic field, the longitudinal resistance drops to zero
 if the magnetic field takes some specific
 values, called ``plateaus''. Furthermore, the
Hall conductivity, $\sigma_H$, is {\it quantized}.
 In the integer Quantum Hall Effect
(IQHE) $\sigma_H$ is an integer multiple of some basic unit while in
the Fractional Quantum Hall Effect (FQHE), it is a rational
multiple.

The  explanation of this surprising, and unexpected,
quantization, provided by Laughlin's  ``microscopic'' theory
\cite{Laughlin},
 involves quasiparticles and quasiholes, which are
composite objects that carry both  (fractional) electric charge and
a magnetic flux~: they are \textit{charged vortices}. For further
details, the reader is referred to the literature
\cite{QHE,GirvQHE,StoneQHE,EzawaQHE}.

Vortices (uncharged, however) have been encountered before in ``ordinary'' superconductivity and superfluidity. Their phenomenological description
is  provided by Lan\-dau-\-Ginzburg theory \cite{LiPi,NiOl} -- the Cooper
pairs formed by the electrons are represented by a scalar matter field,
whose charge is twice that of the electron.
The pairs interact through their electromagnetic fields,
governed by the Maxwell equations.

The similarities between the Fractional Quantum Hall Effect
 and superconductivity led condensed matter physicists to call for a phenomenological, effective theory  of the FQHE \cite{Girv1}.
 In the first instance, such a
 ``Landau-Ginzburg'' theory  of the Hall effect must reproduce
 Hall's law, (\ref{HallLaw}).
Now, adding the usual current term  $-\textit{e}j^\alpha A_\alpha$
to (\ref{CSemAction}) and suppressing the conventional Maxwell term,
the variational equations read, in $(2+1)d$ form,
\begin{eqnarray}\fbox{$
\kappa E_i=-\textit{e}\epsilon_{ik}j^k, \qquad \kappa
B=\textit{e}\rho, $} \label{2+1CSequationbis}
\end{eqnarray}
The spatial components here  are
precisely the \textit{Hall law},
 (\ref{HallLaw})  with $\kappa=\textit{e}\sigma_H$.

 The main physical application of Chern-Simons gauge theory is,
in fact, to  the \textit{Quantum Hall Effect}
\cite{GirvQHE, StoneQHE,EzawaQHE,Girv1, Girv2,
JackHallCS,Friedman,ZHK,Read,LeeZhang,ZhangRev,EHI91/2,
EHI91/1,Tafel}.

Conversely, the Hall law, combined with current conservation and Faraday's law, together imply the Chern-Simons
equations and
 \textit{require}, therefore,  the Chern-Simons form
 (\ref{CSform}) \cite{Frohlich}.

Using the Chern-Simons term in the Hall context has been advocated
by high-energy physicists \cite{JackHallCS,Friedman}. However, their
efforts had seemingly little effect on condensed matter physicists,
who went their own way to arrive, independently, at similar
conclusions. The evolution has been parallel and (almost) unrelated
in high-energy/mathematical physics and in condensed matter physics,
for at least a decade.

It is interesting to compare the early progress in the two fields.
Similar ideas arose, independently and almost simultaneously, see Table \ref{tableau}.
The main difference has been
that while condensed matter physicists were more interested in
the physical derivation, and in its application to the Hall effect,
high-energy/mathematical physicists explored the existence and
the construction of solutions.
\begin{table}[thp]
\begin{tabular}{|l|l|}
\hline
FIELD THEORY (hep-th)
&CONDENSED MATTER (cond-mat)\\
\hline 1981 Schonfeld; Deser-Jackiw-Templeton: & 1980-1982 v.
Klitzing et al.; Tsui et al.:
 \\
topologically massive gauge theory&Integer/Fractional Quantum
 Hall effect\\
\hline
1984-85 Hagen: Galilei-invariant field&1983 Laughlin:\\
 theory in (2+1)d; Jackiw; Friedman et al.:
&microscopic theory of FQHE\\
relation to Hall effect
&ground-state wave functions\\
\hline 1986 Paul-Khare; De Vega-Schaposnik: &1986-87
Girvin-MacDonald:
\\
vortices in Maxwell/YM + CS
&
effective ``Landau-Ginzburg'' theory\\
\hline 1990-91 Hong et al, Jackiw et al: &1989 Zhang, Hansson,
Kivelson (ZHK):
\\
relativistic topological/non-topological
&time-dependent LG theory  with\\
self-dual vortices
& vortex solutions\\
\hline 1990 Jackiw-Pi: exact non-relativistic
&\\
non-topological vortices&\\
\hline 1991 Ezawa et al., Jackiw-Pi: time-dependent
&1991-93 Ezawa et al., Tafelmayer: topological\\
 breather vortices in external field
&vortices in the Zhang et al. model\\
\hline
1997 Manton: NR Maxwell-CS vortices in&\\
external transport current
&\\
\hline
\end{tabular}
\caption{\it The Chern-Simons form in
field theory and in condensed matter physics.}
\label{tableau}
\end{table}

The first, static, ``Landau-Ginzburg'' theory for the QHE was
put forward by Girvin and  MacDonald \cite{Girv1,Girv2} on phenomenological grounds.
An improved theory, allowing for time-dependence, was derived
from Laughlin's microscopic theory by Zhang,
Hansson and Kivelson \cite{ZHK,Read,LeeZhang,ZhangRev} (see Section \ref{LGQHE}).

Condensed matter physics is intrinsically  non-relativistic. This
was pointed out, for example,
 by Feynman in his Lectures on Statistical Mechanics
\cite{Feynman}. Ordinary Landau-Ginzburg theory does not admit any
interesting non-relativistic extension, owing to the intrinsically
relativistic character of the Maxwell dynamics. The topological
nature of the Chern-Simons term makes it possible, however, to
accommodate both relativistic as well as non-relativistic field
theories. The theory of Zhang et al. in Ref. \cite{ZHK} is
non-relativistic.

Relativistic Chern-Simons theories with topological vortex solutions
(Section \ref{RVort}) have been considered starting from the mid
eighties, by high-energy physicists \cite{PaKh, NAbCS,KuKh}.

We first review some aspects of the Abelian Higgs model \cite{NiOl}
with a Chern-Simons term \cite{PaKh,Boya}, and then proceed to the
relativistic, pure Chern-Simons model studied in Refs.
\cite{HoKiPa,JaWe,JaLeWe}. These simplified models have the
advantage of accommodating self-duality (`Bogomol'ny') equations,
which yield, in turn, ``Liouville-type'' equations reminiscent of
those in the Abelian Higgs model.

Then (Section \ref{NRVort}) the non-relativistic limit of the
relativistic theory can be considered, providing us with an
explicitly solvable model~: for a particular choice of the potential
and for a specific value of the coupling constant, solutions to the
second-order field equations can be found by solving, instead, the
first order ``self-dual'' equations \cite{JaPi1,JaPi2,Fund5,Fund6}.
The problem can in fact be reduced to solving the \textit{Liouville
equation}. Not all solutions are physically admissible, however -
only those which correspond to \textit{rational functions}. This
provides as with a \textit{quantization theorem of the magnetic
charge}, as well as with a \textit{parameter counting}.

The subtle problem of symmetries and the construction of a conserved
energy-momentum tensor is revisited in Section \ref{JPmodel}.

Similar ideas work for vortices in a constant  background field, see Sec. \ref{background}. These models are important, since
they correspond precisely to those proposed
in the Landau-Ginzburg theory of
the Fractional Quantum Hall Effect \cite{ZHK,ZhangRev}.

Finally, we consider spinorial models \cite{ChoRelSpin,LiBha,DHPSpinor1,DHPSpinor2}.
Again, explicit solutions are found and their symmetries are
studied using the same techniques as above.

For completeness, we would like to list a number of related issues not discussed here.

First of all, many of the properties studied here
 can be generalized to non-Abelian
interactions \cite{NAbCS,KuKh} which have, of course, many
further interesting aspects. Jackiw-Pi vortices,
for example, can be generalized to $SU(N)$ gauge theory
leading to generalizations of the Liouville equation.
See, e.g., Refs. \cite{JPDT,Fund6}.

Experimentally, superconducting
vortices arise often as lattices in a
finite domain. Within the
Jackiw-Pi model, this amounts to selecting doubly-periodic
solutions of the Liouville equations \cite{Olesen}.

The relation to similar models which arise in
condensed matter physics could also
be developed \cite{MaPaSo,double-layer}.
Other interesting aspects concern anomalous coupling
\cite{Torres}, as well as various self-duality properties
\cite{Proca, SDTopMass1,SDTopMass2,SDTopMass3}.

Returning to the Abelian context, we should mention the study of the
\textit{dynamics of vortices}
\cite{SoliDynM1,SoliDynM2,ShellRub,Ruback, KMLVd,
KimMin,HuaChou,YKimLee,Dziarmaga,Liu,LiuStern,BakLee}.

\subsection
{Feynman's ``Another point of view'' }\label{Feynman}

Let us start our investigations with a historical excursion.
Feynman, in his {\sl Lectures on Statistical Mechanics}
\cite{Feynman}, proposed his ``another point of view'' to describe
superconductivity. He argued  that the ground states of
superconductor should be described by a non-relativistic wave
function $\psi$, which would
 solve a \textit{gauged, non-linear
Schr\"odinger equation},
\begin{equation}
i\p_t\psi=\left[-\frac{\vD^2}{2m}+qV -\alpha(1-\varrho)\right]\psi.
\label{FNLSch}
\end{equation}
The covariant derivative,
$
\vD=\vnabla-iq\vA,
$
 involves  $q=2e$, the charge of the Cooper pairs.
$\varrho=\psi^*\psi$ is the 
\textit{particle or ``number'' density}.

Feynman notes that the system admits a
hydrodynamical transcription \cite{Madelung}. Setting
$
\psi=\sqrt{\varrho}\,e^{i\phi},
$
the imaginary and real parts of equation (\ref{FNLSch})
become,
\begin{eqnarray}
\p_t\varrho+\vnabla\cdot(\varrho\,\vv)&=&0,
\label{hydroconteq}
\\[6pt]
\p_t\phi+\frac{m}{2}{\vv}^2&=&-qV-U, \qquad U=
-\frac{1}{2m}\frac{\bigtriangleup\sqrt{\varrho}}
{\sqrt{\varrho}}-\alpha(1-\varrho), \label{hydroEuler}
\end{eqnarray}
where the velocity field is,
\begin{eqnarray}
\vv=\frac{1}{m}\left(\vnabla\phi-q\vA\right).
\label{velocityfield}
\end{eqnarray}
Since the current is,
\begin{equation}
\vj=\frac{1}{2mi}
\big(\psi^*\vD\psi-\psi(\vD\psi)^*\big)
=\varrho\,\vv,
\end{equation}
equation (\ref{hydroconteq}) can also be presented as,
\begin{equation}
\p_t\varrho+\vnabla\cdot\vj=0.
\label{hydroconteqbis}
\end{equation}
Thus, the number density, $\varrho$,  satisfies,
 with the current $\vj$,  the
continuity equation.
The gradient of (\ref{hydroEuler}) yields, furthermore,
\begin{equation}
m\left[\p_t\vv+(\vv\cdot\vnabla)\vv\right]
=q\big(\vE+\vv\times\vB\big)-\vnabla U,
\label{hydroEulerbis}
\end{equation}
where $\vE=-\vnabla V-\p_t\vA$ and
$\vB=\vnabla\times\vA$ are the electric and
magnetic fields, respectively.

Here we recognize the
hydrodynamical equations of a charged fluid in
an ({external}) electromagnetic field.
$U$ is interpreted as the pressure.

The system proposed by Feynman can be viewed as classical field theory~:
(\ref{FNLSch}) can be derived from the Schr\"odinger Lagrangian
\begin{eqnarray}
\cL=
-i\psi^*\p_t\psi+\frac{1}{2m}|\vD^2\psi|^2+
qV\psi^*\psi+\frac{\alpha}{2}\,\big(1-\psi\psi^*\big)^2.
\label{FNLSchLag}
\end{eqnarray}

Equivalently, the hydrodynamical
equations (\ref{hydroconteq})-(\ref{hydroEuler}) derive
from the ``hydrodynamical Lagrangian'' \cite{JHydro}
\begin{eqnarray}
\cL=
\frac{1}{2i}\p_t{\varrho}+\varrho\,\p_t{\phi}
+\displaystyle\frac{m\varrho}{2}\,\vv^2
+\frac{1}{2m}\big(\vnabla\sqrt{\varrho}\big)^2
+qV\varrho+\frac{\alpha}{2}\,\big(1-\varrho\big)^2.
\end{eqnarray}

In Feynman's approach the gauge field is external, however~; it has
no {proper dynamics} as yet. Can the coupled matter plus gauge field
system be promoted to a self-consistent one~?

The first, natural, idea would be to posit that the  electromagnetic
field satisfies the Maxwell equations. But  the Galilean invariance
of the Schr\"odinger equation (\ref{FNLSch})  is inconsistent with
the {Lorentz} invariance of the Maxwell equations~: the coupled
system would have no clear symmetry \footnote{The Galilean invariant
electromagnetism proposed by Le Bellac, L\'evy-Leblond \cite{LBLL}
 could, in principle, yield a consistent
matter+gauge field system,
but it would provide a too poor dynamics for interesting
physical applications.}.

The problem could not be resolved in Feynman's time,
as the Chern-Simons form had not yet been
discovered. The topological character of the latter
allows it to couple ``Chern-Simons'' electromagnetism
to either relativistic or non-relativistic matter.

\subsection{Landau-Ginzburg Theory of the Quantum
Hall Effect}\label{LGQHE}

Feynman's ideas did not attract too much attention at that time~: superconductivity was thought to be
 satisfactorily understood by the
London-, Landau-Ginzburg-, and BCS theories.
The 1980s, with the unexpected discoveries of the
\textit{Quantum Hall Effect} \cite{QHE,GirvQHE,StoneQHE,EzawaQHE} and of
\textit{high-temperature superconductivity} \cite{highTc},
witnessed, however, a sudden rise in interest in alternative theories.

Motivated by the analogies between the Hall
Effect and superfluidity,
Girvin, and MacDonald \cite{Girv1,Girv2}  called, in particular, for a ``Landau--Ginzburg''
theory for the Quantum Hall Effect.
On phenomenological grounds, they  suggested representing the
``off-diagonal long range order'' (ODLRO) by a scalar field
$\psi({\vx})$ on the plane,
and the ``frustration due to deviations away from the quantized Laughlin density'' by
an effective gauge potential, ${\va}({\vx})$.
They proposed to describe this
static, planar system by the Lagrange density \footnote{Girvin and MacDonald \cite{Girv2} use $\kappa=\theta/8\pi^2$.},
\begin{equation}
\cL_{GM}=-i\phi\big(|\psi|^2-1\big)
+\2|(\vnabla-i\,{\va})\psi|^2
+i\frac{\kappa}{2}\Bigl(\phi\,\vnabla\times{\va}
+{\va}\times\vnabla\phi
\Bigr),
\label{GMlag}
\end{equation}
where $\phi$ is a scalar field.
The equations of motion read
\begin{eqnarray}
\2\vD^2\psi&=&-i\phi\,\psi,
\label{GirNLS}
\\[6pt]
\kappa\, b&=&-1+|\psi|^2,
\label{GirGauss}
\\[6pt]
\kappa\,\vnabla\times (i\phi)&=&\vj,
\qquad
\vj=\frac{1}{2i}\big(\psi^*\vD\psi-\psi(\vD\psi)^*\big)
\label{GirAmpHall},
\end{eqnarray}
where
$\vD=\!\vnabla -i\,{\va}$ is the gauge-covariant derivative and
 $b=\!\vnabla\times{\va}$ is an effective magnetic field.

\vskip2mm
 We make the following observations.
\begin{enumerate}
\item
The first of these equations is a static, Schr\"odinger-type,
 gauged wave equation for the matter field.

\item
The second equation relates the effective magnetic
field to the particle density. The ``physical gauge field'' is
\begin{equation}
\vA=\va-\vcA
\where
\rot\vcA=-\frac{1}{\kappa}\ .
\label{Girpot}
\end{equation}
Then equation (\ref{GirGauss}) can be presented as
\begin{equation}
\kappa\, B=|\psi|^2,
\label{physGauss}
\end{equation}
where $B=\vnabla\times\vA=
b-\cB$. Here
\begin{equation}
\cB=\vnabla\times\vcA=-\frac{1}{\kappa}
\end{equation}
can be viewed
as a ``background magnetic field'',
 and the physical magnetic field, $B$,
is  determined by the
density $\varrho=|\psi|^2$ and by $\cB$. It corresponds to the value
determined by the Laughlin ground states of the FQHE.

Integrating (\ref{physGauss}) over the whole plane shows that the
total flux and the electric charge,
$$
\Phi=\displaystyle\int\! d^2\vx\, B
\qquad\hbox{\small and}\qquad
Q=\displaystyle\int\!d^2\vx\,\varrho,
$$
 are proportional,
\begin{equation}
\kappa\,\Phi=Q.
\end{equation}
Anticipating the existence of vortex solutions [Section \ref{ZHKvort}], the latter can be identified with Laughlin's quasiparticles. Taking
$\kappa=1/m$ where $m$ is an integer shows that our ``anyon''
will carry  magnetic flux $\Phi$ and $1/m$ units of fractional charge.

\item
Introducing the ``effective electric field''
\begin{equation}
\ve=i\vnabla\phi,
\end{equation}
we recognize, in the last equation, (\ref{GirAmpHall}),
the \textit{Hall law} (\ref{HallLaw}),
$$
j_i=\kappa\epsilon_{ij}e_j,
$$
 with
$\kappa$ identified as  the {\it Hall conductance} $\sigma_H$.
\end{enumerate}

The remarkable fact (apparently unknown to Girvin and MacDonald)
is that setting
\begin{equation}
a_0=i\phi,
\qquad
f_{\alpha\beta}=\p_\alpha a_\beta-\p_\beta a_\alpha,
\end{equation}
the last term in their
Lagrangian (\ref{GMlag}) is {precisely} the static, Chern-Simons form,
\begin{equation}
i\frac{\kappa}{2}\,\Bigl(\phi\,\vnabla\times{\va}
+{\va}\times\vnabla\phi\Bigr)
=
-\frac{\kappa}{4}\,\epsilon^{\alpha\beta\gamma}a_\alpha f_{\beta\gamma}.
\end{equation}

\kikezd{The Zhang-Hansson-Kivelson (ZHK) model}

Soon after the proposal of Girvin and MacDonald (GM)
\cite{Girv2},
 Zhang {\it et al.} (ZHK) \cite{ZHK},
argued that the Girvin-MacDonald model is merely
 a first step in the right direction, and put forward
an ``improved'' Landau-Ginzburg model for the QHE.
They derived their theory directly from the microscopic theory \cite{ZHK, ZhangRev}.
 Their starting point is the Hamiltonian
of a planar system of polarized electrons,
\begin{equation}
H_{pe}=\frac{1}{2m}\sum_a\left[\vp_a-e\vcA(\vx_a)\right]^2+
\sum_ae\cA_0(\vx_a)+\sum_{a<b}V(\vx_a-\vx_b),
\label{polelHam}
\end{equation}
where $\cA_\alpha$ is the [symmetric gauge] vector potential
for the constant external magnetic field $\cB$, and
$\cA_0$ is the scalar potential for the external
electric field,
$$
\cA_i=-\2\cB\,\epsilon_{ij}x^j,
\qquad
\cE_i=-\p_i\cA_0,
$$
respectively.
$V$ is the two-body
interaction potential between the electrons; the common assumption is that $V$ is Coulombian.

The many-body wave function satisfies the
Schr\"odinger equation,
\begin{equation}
H_{pe}\Psi(\vx_1,\dots,\vx_N)=E\,\Psi(\vx_1,\dots,\vx_N),
\label{peSch}
\end{equation}
and is assumed, consistently with the Pauli principle, to be
\textit{totally antisymmetric} w.r.t. the interchange
of any two electrons.

The clue of Zhang et al. \cite{ZHK} is to map
the problem onto  a \textit{bosonic} one. In fact, let us
 consider the bosonic system with Hamiltonian
\begin{equation}\begin{array}{ll}
H_{bos}=&\displaystyle\frac{1}{2m}\sum_a\left[\vp_a-e(\vcA(\vx_a)-\va(\vx_a))\right]^2
\\[20pt]
&+ \displaystyle\sum_ae\big(\cA_0(\vx_a)+a_0(\vx_a)\big)
+ \displaystyle\sum_{a<b}V(\vx_a-\vx_b),
\end{array}
\label{bosHam}
\end{equation}
where the new vector potential, $a_\alpha$, describes a
gauge interaction of a specific form among the particles,
\begin{equation}
\va(\vx_a)=\frac{\Phi_0}{2\pi}\frac{\theta}{\pi}
\sum_{b\neq a}\vnabla\gamma_{ab},
\label{statvecpot}
\end{equation}
where $\theta$ is a
(for the moment unspecified) real parameter,
and $\gamma_{ab}=\gamma_a-\gamma_b$ is the difference of the
 polar angles of electrons \underbar{a} and \underbar{b}
 w.r.t. some origin and polar axis. $\Phi_0=h/ec$ is the
flux quantum. The $N$-body bosonic wave function $\phi$
is required to be \textit{symmetric} and satisfies
\begin{equation}
H_{bos}\phi(\vx_1,\dots,\vx_N)=E\,\phi(\vx_1,\dots,\vx_N).
\label{bosSch}
\end{equation}

Let us now  consider the singular gauge transformation
\begin{eqnarray}
\Psi(\vx_1,\dots,\vx_N)=U\,\phi(\vx_1,\dots,\vx_N),
\qquad
U=\exp\left[-i\sum_{a<b}\frac{\theta}{\pi}\gamma_{ab}\right].
\label{singgauge2}
\end{eqnarray}
Then it is easy to check that
\begin{eqnarray}
U\left[\vp_a-e(\vcA-\va)\right]U^{-1}=
\vp_a-e\vcA,
\end{eqnarray}
so that
\begin{equation}
U\,H_{bos}\,U^{-1}=H_{pe}.
\end{equation}
It follows that $\phi$ satisfies the bosonic Schr\"odinger equation (\ref{bosSch}) precisely
when $\Psi$ satisfies the polarized-electron
eigenvalue problem (\ref{peSch}) with the same eigenvalue.

To conclude our proof, let us observe that $\Psi$
is antisymmetric precisely when the parameter $\theta$
is an \textit{odd multiple} of $\pi$,
\begin{equation}
\theta=(2k+1)\pi.
\end{equation}

Having replaced the fermionic problem by a
bosonic one with the strange interaction
(\ref{statvecpot}), Zhang et al. proceeded to
derive a mean-field theory.
 Their model also involves  a scalar field $\psi$ coupled to both
an external electromagnetic field $\cA_{\mu}$ and to a ``statistical'' gauge field, $a_{\mu}$. It also includes a
potential term, and is time-dependent. Their Lagrangian reads \footnote{In the notation of \cite{ZHK}, $\kappa={e^2}/{2\theta}$. }
\begin{equation}\fbox{$
\begin{array}{ll}
\cL_{ZHK}\;=
-i\psi^*D_{t}\psi
+\frac{1}{2}|\vD\psi|^{2}+U(\psi)-
\displaystyle\frac{\kappa}{2}\epsilon^{\mu\nu\sigma}a_{\mu}\p_{\nu}a_{\sigma},
\\[6pt]
\end{array}
$}
\label{ZHKLag}
\end{equation}
where
\begin{equation}
D_\alpha\psi=(\p_\alpha-ieA_\alpha)\psi,
\qquad
A_\alpha=a_\alpha-\cA_\alpha
\label{bcovder}
\end{equation}
is the covariant derivative.
Note that while $D_\alpha$ involves
both the statistical and the background vector potentials combined into the
\textit{total} vector potential $A_\alpha$,
the Chern-Simons potential only involves the statistical field
$a_\alpha$.

The field equations of the ZHK model read
\begin{eqnarray}
iD_t\psi&=&-\frac{1}{2}\vD^2\!\psi+\frac{\delta U}{\delta\psi^*},
\label{ZHKNLS}
\\[6pt]
\kappa\,\epsilon_{ik}\,e_k&=&ej_i,
\qquad
\vj=\frac{1}{2i}\big(\psi^*\vD\psi-\psi(\vD\psi)^*\big),
\label{ZHKFCI}
\\[6pt]
\kappa b&=&e\varrho,\qquad\varrho=|\psi|^2,
\label{ZHKGauss}
\end{eqnarray}
where  $e_k = \partial_k a_0 - \partial_t a_k$ and $ b = 1/2
\epsilon_{ij} (\partial_i a_j-\partial_j a_i)$. I.e., the matter
field satisfies a non-linear Schr\"odinger equation, which is
supplemented by the field-current identities of pure Chern-Simons
theory \footnote{Remember that
 that the covariant derivative
involves the total gauge field, cf. (\ref{bcovder}).}.

These equations describe a scalar field in a coupled statistical and background field and a non-linear potential, appropriate
for the FQHE. Its
vortex-type solutions, predicted by Zhang et al.
\cite{ZHK,LeeZhang,ZhangRev}, will be studied in Section \ref{background}.

Note that the ZHK Lagrangian is \textit{first-order
in the time derivative} of the scalar field. It is
indeed non-relativistic see Section
\ref{background}.

The scalar potential of Zhang et al. is the
self-interaction potential
\begin{equation}
U(\psi)=-\mu\vert\psi\vert^2+\frac{\lambda}{4}\vert\psi\vert^{4}+\const.,
\label{ZHKpot}
\end{equation}
where the constant is chosen so that the minimum of $U$ is $0$. The
term $\mu\vert\psi\vert^2$ ($\mu\geq0$) here is a chemical
potential, while the quartic term is an effective interaction,
coming from the non-local expression in the second-quantized
Hamiltonian,
$$
\2\int \psi^\star({\vx})\psi^\star({\vx}')
V({\vx}-{\vx}')
\psi({\vx})\psi({\vx}')d^2{\vx}d^2{\vx}'
$$
when the two-body potential is approximated by a delta function,
$$
V({\vx}-{\vx}')=\frac{\lambda}{2}\,\delta({\vx}-{\vx}').
$$

It is worth mentioning that,
for a static system in a purely magnetic
background and for $U(\psi)\equiv0$, the ZHK and
GM models are
 mathematically (but not physically) equivalent \cite{HHY2}.
Let us indeed assume that all fields are static,
$\p_t(\;\cdot\;)=0$. Setting
\begin{eqnarray}
\cA_0=0,\qquad
\cB=-\frac{1}{\kappa},\qquad
A_0=a_0=i\phi,\qquad
\va=\vA+\vcA,\qquad
\label{substi}
\end{eqnarray}
the GM Lagrangian (\ref{GMlag}) becomes,
\begin{eqnarray*}
\cL_{GM}=-A_0|\psi|^2+A_0+
\2\big|\big(\vnabla-i(\vA+\vcA)\big)\psi\big|^2
\\[8pt]
+\frac{\kappa}{2}\Big(A_0\vnabla\times\vA+A_0\vnabla\times\vcA
+\vA\times\vnabla A_0+\vcA\times\vnabla A_0\Big).
\end{eqnarray*}
However,  integrating, by parts the third term and using
$\vnabla\times\vcA=\cB$,
$$
A_0+\frac{\kappa}{2}\big(
A_0\vnabla\times\vcA+\vcA\times\vnabla A_0
\big)=
A_0\underbrace{\left(1+\frac{\kappa}{2}(\vnabla\times\vcA+\vnabla\times\vcA)\right)}_0+\;\hbox{\small surface terms}.
$$
Dropping surface terms leaves us with
\begin{equation}
{\cal L}_{GM}=-i\psi^*(-iA_0)\psi+
\2\big|\big(\vnabla-i(\vA+\vcA)\big)\psi\big|^2
+
\frac{\kappa}{2}\Big(A_0\vnabla\times\vA
+\vA\times\vnabla A_0\Big).
\nonumber
\end{equation}
For $e=1$ and setting $\kappa=e^2/4\theta$, this is precisely the static ZHK Lagrangian (\ref{ZHKLag}),
\textit{up to the substitution}
\begin{equation}
A_\alpha\leftrightarrow a_\alpha.
\label{inter}
\end{equation}

An even simpler proof is obtained using  the field equations. The substitution (\ref{substi}), followed by the
interchange (\ref{inter}), carries  the GM field equations
(\ref{GirNLS})-(\ref{GirGauss})-(\ref{GirAmpHall}) into the ZHK equations
(\ref{ZHKNLS})-(\ref{ZHKGauss})-(\ref{ZHKFCI}).

From now on we only consider the ZHK theory.

\section{RELATIVISTIC CHERN-SIMONS VORTICES}\label{RVort}

After this short excursion into condensed matter physics, we turn now to a study of field-theoretical aspects.

\subsection{Abelian Higgs Model with  Chern-Simons Term}\label{PaulKhare}

The first (Abelian)\footnote{Non-Abelian
vortices have also  been studied, around the same time, cf.
\cite{NAbCS, Fund6}.}  Chern-Simons model
is obtained  \cite{PaKh} by simply adding the Chern-Simons term to the usual Abelian Higgs model, defined on
$(2+1)$-dimensional Minkowski space, with metric
$(g_{\mu\nu})={\rm diag}\,(-1,1,1)$,
\begin{eqnarray}
\cL_{PK}=&\cL_{H}+\cL_{CS}=&
\frac{1}{4}F_{\alpha\beta}F^{\alpha\beta}
+\frac{1}{2}D_\alpha\Psi(D^\alpha\Psi)^*
+U(\Psi)
-\displaystyle
\frac{\kappa}{4}\epsilon^{\alpha\beta\gamma}A_\alpha F_{\beta\gamma},
\label{CSMLag}
\\[8pt]
&&U(\Psi)=\displaystyle\frac{\lambda}{4}\big(1-|\Psi|^2)^2,
\label{quarticpot}
\end{eqnarray}
where
$
D_\alpha\Psi=\partial_\alpha\Psi-ieA_\alpha\Psi
$
is the covariant derivative, with
$e$  the electric charge of the field $\Psi$.
 The Chern-Simons term is coupled through the ``Chern-Simons'' coupling constant $\kappa$.

Below, we shall refer to this as the Paul-Khare (PK)
model.

The equations of the Abelian Higgs model and
of the pure Chern-Simons-Maxwell system (\ref{CSFE}), respectively, are merged into
\begin{eqnarray}
\frac{1}{\sqrt{g}}D_\alpha\left(\sqrt{g}D^\alpha\Psi\right)&=&
-\lambda(1-|\Psi|^2)\Psi,
\\[8pt]
\frac{1}{\sqrt{g}}\p_\nu (\sqrt{g}F^{\nu\alpha})+\frac{\kappa}{2}\,
\sqrt{g}
\epsilon^{\alpha\mu\nu}F_{\mu\nu}
&=&-\frac{e}{2i}\left(\Psi^*D^\alpha\Psi-\Psi(D^\alpha\Psi)^*\right).
\label{PKeqmot}
\end{eqnarray}

Solutions can be sought along the same lines
as in the Abelian Higgs model. Consider the static, radial Ansatz
\begin{equation}
A_0=nA_0(r),
\qquad
A_r=0,
\qquad
A_\theta=nr\,A(r),
\qquad
\Psi(r)=f(r)e^{in\theta}.
\label{PKradAns}
\end{equation}
Upon using
\begin{eqnarray}
\begin{array}{lll}
(g_{\mu\nu})={\rm diag}(-1,1,r^2),\qquad
&(g^{\mu\nu})={\rm diag}(-1,1,r^{-2}),
&\sqrt{g}=r,
\\[12pt]
D_0\Psi=-ienA_0\Psi,
&D_{r}\Psi=D^r\Psi=f'\,e^{in\theta},&
\\[12pt]
D_{\theta}\Psi=in(1-erA)\Psi,\qquad
&D^\theta=r^{-2}D_{\theta},&
\\[12pt]
F_{0r}=-F^{0r}=-nA_0',
&F_{0\theta}=0,&
\\[12pt]
F_{r\theta}=n(A+rA'),
&F^{r\theta}=r^{-2}F_{r\theta},&
\\[12pt]
j^0=ne^2f^2A_0,
&j^r=0,
&j^\theta=-\displaystyle\frac{n}{r}e^2\left(A-
\displaystyle\frac{1}{er}\right)f^2,
\end{array}
\label{PKmetricquant}
\end{eqnarray}
the static, $2^{nd}$-order equations become \footnote{The corresponding equations, (5a-c),
of Ref. \cite{PaKh} are obtained by the substitution
$A^{(PK)}=-nr\,A$, \hfill\break ${A_0}^{(PK)}=-nA_0,\
\mu=-\kappa$.},
\begin{eqnarray}
f''+\displaystyle\frac{f'}{r}-e^2n^2(A-\displaystyle\frac{1}{er})^2f+n^2e^2A_0^2f+\lambda(1-f^2)f=0,
\label{PKradf}
\\[9pt]
A''+\displaystyle\frac{A'}{r}-\displaystyle\frac{A}{r^2}-
e^2f^2\big(A-\displaystyle\frac{1}{er}\big)+\kappa A_0'=0,
\label{PKradA}
\\[8pt]
A_0''+ \displaystyle\frac{A_0'}{r}-e^2f^2A_0
+\kappa(A'+\displaystyle\frac{A}{r})=0.
\label{PKradA0}
\end{eqnarray}

These equations, like those in the
 usual Abelian Higgs model [to which they reduce for $\kappa=0$
 and $A_0=0$],  have not yet been solved analytically,
but Paul and Khare argue that a finite-energy
solution does exist for each integer $n$.
The solutions,  produced numerically (see below),
 carry both \textit{quantized magnetic flux and electric charge},
\begin{equation}
\Phi=\frac{2\pi}{e}\,n,\quad
\qquad
Q=\kappa\,\frac{2\pi}{e}\,n=\kappa\,\Phi,
\label{PKFluxCharge}
\end{equation}
The first statement follows from
\begin{equation}
\Phi=\oint_S A_\theta d\theta,
\label{flux}
\end{equation}
where the integration is along the circle at infinity,
when the conditions at infinity (\ref{PKfinEn}) below, are taken into account.

Expressing the current through the gauge potentials
by equation (\ref{PKradA0}), we have,
using the boundary and small-$r$ conditions (\ref{PKfinEn}) and  (\ref{PKsmallrf})-(\ref{PKsmallrA})
-(\ref{PKsmallrA0}), respectively,
\begin{eqnarray*}
Q &=&\int j^0d^2x
=n\int d^2\vx\, e^2A_0|\Psi|^2
=
n\int d^2x\left(
A_0^{\prime\prime}+\frac{A_0^{\prime}}{r}+
\kappa\big(A^{\prime}+\frac{A}{r}\big)
\right)
\\[8pt]
&=&2\pi n\underbrace{\int_0^\infty\!(rA_0^{\prime })^{\prime}dr}_0+
2\pi n\kappa\int_0^\infty (rA)^{\prime}dr
=\frac{2\pi\kappa}{e} n,
\end{eqnarray*}
proving
the flux-charge relation (\ref{PKFluxCharge}).
Note that this proof is, in fact, the radial
version of the general statement in Section \ref{CSHall}.

Hence, our objects
represent \textit{charged topological vortices}, sitting at the origin.

\kikezd{Energy of a static configuration}.

Let us consider a static field configuration $(A_\mu,\Psi)$.
The general formula gives, for the energy density,
$
\Theta^{00}=
\cL.
$
\begin{itemize}
\item{} The conventional Maxwell + matter terms in (\ref{CSMLag})
contribute, after ``improvement'',
$$
\frac{1}{2}F_{0i}^2+\frac{1}{4}F_{ij}^2
-\frac{1}{2}e^2A_0^2\,\vert\Psi|^2+
\frac{1}{2}\vert\vD\Psi\vert^2+U(\Psi).
$$

\item The contribution of
the Chern-Simons term is, in turn,
$$
-\frac{\kappa}{4}\left(
\epsilon^{0ij}A_0F_{ij}+2\epsilon^{0ij}A_iF_{j0}\right).
$$
But $F_{j0}=\p_jA_0$ since the configuration is static, so that
the second term in the bracket is
$$
2\left(\partial_j\left(\epsilon^{j0i}A_iA_0\right)+
\epsilon^{0ij}(\p_iA_j)A_0\right)
=\partial_j\big(2\epsilon^{j0i}A_iA_0\big)
+\epsilon^{0ij}F_{ij}A_0.
$$
\end{itemize}
Dropping the surface term, the Chern-Simons contribution is, therefore,
\begin{equation}
-\frac{\kappa}{2}\epsilon^{0ij}A_0F_{ij}=
\kappa\,A_0B,
\end{equation}
where $B=\2\epsilon^{0ij}F_{ij}=-F_{12}$
is the magnetic field. Hence, the total energy is,
\begin{equation}
\cE=\int\!\Big\{
\2{\vE}^2+\2B^2+
\2|\vD\Psi|^2
-\2e^2A_0^2|\Psi|^2+U(\Psi)+\kappa A_0B\Big\}
d^2{\vx}.
\label{CSMEnDens}
\end{equation}

The energy of the static, radial configuration (\ref{PKradAns}) is, in particular,
\begin{equation}
\begin{array}{lll}
\cE&=\pi\displaystyle\int_0^\infty\!rdr\!\!
&\left\{
\left
(n^2\big(A'+ \displaystyle\frac{A}{r}\big)^2
+\big(f'\big)^2+
e^2n^2\big(A-\displaystyle\frac{1}{er}\big)^2f^2
+
\displaystyle\frac{\lambda}{2}\big(f^2-1\big)^2\right)
\right.
\\[20pt]
&&\left. +\;n^2\left((A_0')^2-e^2A_0^2f^2
+
2\kappa A_0\big(A'+\displaystyle\frac{A}{r}\big)\right)
\right\}.
\end{array}
\label{PKradEnergy}
\end{equation}
The system (\ref{PKradf})-(\ref{PKradA})-(\ref{PKradA0}) should be
supplemented with the
 finite-energy asymptotic conditions
\begin{equation}
A_0(r)\to0,
\qquad
A(r)-\frac{1}{er}\to0,
\qquad
f(r)\to1
\qquad\hbox{as}\quad r\to\infty.
\label{PKfinEn}
\end{equation}
Inserting the asymptotic values (\ref{PKfinEn}) into the
radial system yields approximate solutions valid for large $r$.

Firstly, taking $A_0\approx0,\,A-{1}/{er}\approx0$
and dropping higher-order terms in (\ref{PKradf})
yields, for $\varphi=1-f$,
the same equation  as in the
usual Abelian model with no Chern-Simons term,
namely Bessel's modified equation of order $0$,
$$
\varphi''+\frac{\varphi'}{r}-2\lambda\,\varphi\approx0.
$$
Hence, once again,
the scalar field approaches its
asymptotic value exponentially with characteristic length
the Higgs mass,
\begin{equation}
\varphi\approx
 \frac{c}{\sqrt{r}}\,e^{-m_\psi\,r},
\qquad
m_\psi=\sqrt{2\lambda}.
\label{largerPKf}
\end{equation}
Next, setting $f\approx1$ in
(\ref{PKradA})-(\ref{PKradA0})  yields, for
\begin{equation}
a=-\big(A-\frac{1}{er}\big)\,r^{1/2}
\qquad\hbox{\small and}\qquad
a_0=A_0\,r^{1/2},
\end{equation}
the coupled system
\begin{equation}
\left\{\begin{array}{lll}
a''-\big(e^2+\displaystyle\frac{3}{4r^2}\big)a
&=&\kappa\big(a_0'-\displaystyle\frac{1}{2r}a_0\big),
\\[10pt]
a_0''-\big(e^2-\displaystyle\frac{1}{4r^2}\big)a_0
&=&\kappa \big(a'+ \displaystyle\frac{1}{2r}a\big).
\end{array}\right.
\label{aa0}
\end{equation}
For very large $r$,  all terms involving inverse powers of $r$
can be dropped,
$$
\left\{\begin{array}{lll}
a''-e^2a &=&
\kappa a_0',
\qquad
\\[8pt]
a_0''-e^2a_0
&=&\kappa a'.
\end{array}\right.
$$
Adding and subtracting, we obtain,
$$
\left\{\begin{array}{lll}
(a+a_0)''-\kappa(a+a_0)'-e^2(a+a_0)&=&0,
\\[8pt]
(a-a_0)''+\kappa(a-a_0)'-e^2(a-a_0)&=&0.
\end{array}
\right.
$$
Both $a$ and $a_0$ are small for large $r$.
Searching the solution in the exponential form, we find that  those  which
remain bounded are,
\begin{eqnarray}
A(r)&\sim& \frac{1}{er}
-\frac{\textrm{e}^{-\2\sqrt{\kappa^2+4e^2}\,r}}{\sqrt{r}}
\Big(c_1\textrm{e}^{\2\kappa r}+c_2\textrm{e}^{-\2\kappa r}\Big),
\\[10pt]
A_0(r)&\sim&
\frac{\textrm{e}^{-\2\sqrt{\kappa^2+4e^2}\,r}}{\sqrt{r}}
\Big(c_1\textrm{e}^{\2\kappa r}-c_2\textrm{e}^{-\2\kappa r}\Big).
\end{eqnarray}
Thus, the decay is two-fold exponential, with
masses,
\begin{eqnarray}
{\big(m_A\big)}_\pm=e\big(\sqrt{1+\frac{\kappa^2}{4e^2}}\mp\frac{\kappa}{2e}\big).
\label{PKmassA}
\end{eqnarray}
The long-range tail of the Chern-Simons field splits, therefore,
the Abelian Higgs value  $m_A=e$ of the mass;
the latter is recovered when $\kappa\to0$.
Note that both values of the mass are positive,
so that the fields decay exponentially for both choices.

Mass splitting is characteristic
for Chern-Simons theories \cite{PisRao}.

One can be puzzled if both asymptotic behaviors are physical. It has
been shown by a subtle analysis \cite{Inoz,LoMaSch,JaKhKuPa} that
the only one allowed  for a regular, finite-energy solution is
\begin{eqnarray}
m_A={(m_A)}_-=\frac{|\kappa|}{2}
\big(\sqrt{1+\frac{4e^2}{\kappa^2}}-1\big).
\label{PKgoodmassA}
\end{eqnarray}

The small-r behavior is obtained along the same lines as for the
Abelian Higgs model. Inserting the power-series expansion into the
field equations, (\ref{PKradf})-(\ref{PKradA})-(\ref{PKradA0}), we
get,
\begin{eqnarray}
f(r)&=&f_nr^{|n|}-\frac{f_n}{4(|n|+1)}(\lambda
+2e|n|^2a_1+e^2|n|^2b_0^2)r^{|n|+2}+O(r)^{|n|+4},
\label{PKsmallrf}
\\[8pt]
A(r)&=&a_1r+\frac 18 (\kappa ^2a_1-\delta_1^{|n|}e f_1^2)r^3+O(r)^5,
\label{PKsmallrA}
\\[8pt]
A_0(r)&=&b_0-\frac 12\kappa a_1r^2-\frac 1{32}
(\kappa ^3a_1-\delta _1 ^{|n|}
 e f_1^2 (\kappa-2 e b_0))r^4+O(r)^6.
 \label{PKsmallrA0}
\end{eqnarray}
It is important to underline that finite-energy
regular solutions require $b_0=A_0(0)\neq0$
\cite{LoMaSch,JaKhKuPa}.

Compared to  the Abelian Higgs model,
the changes arise due to the mixing of the Maxwell and
Chern-Simons terms and, in particular, to the presence of a nonvanishing $A_0$.

\begin{figure}
\begin{center}
\includegraphics[scale=.77]{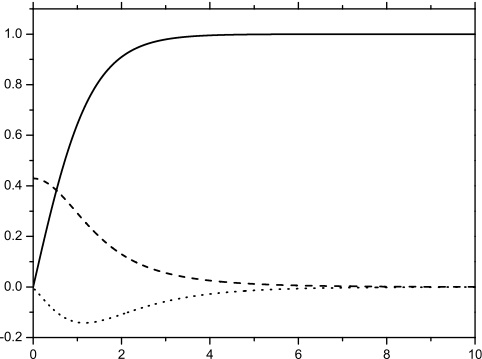}
\end{center}\vskip-4mm
\caption{\it Radial vortex in the PK model ($n=1$).
The heavy line represents the
the scalar, the dashed the magnetic,
and the dotted the radial electric field.}
\label{PKplot}
\end{figure}

The numerical solution for the scalar, the magnetic and the
electric field,
\begin{equation}
|\Psi(r)|=f(r), \qquad B(r)=n\big(A'(r)+\frac{A(r)}{r}\big), \qquad
E(r)=nA_0'(r), \label{fBE}
\end{equation}
is plotted on Figure \ref{PKplot}.

For further discussion of vortices in the Chern-Simons-modified
Abelian Higgs model, the reader should consult Refs.
\cite{Inoz,LoMaSch,JaKhKuPa,Boya}.

\subsection{Pure Chern-Simons Vortices
}\label{J(L)W}

While the Abelian Higgs model with Chern-Simons term has interesting
properties and supports, in particular, vortex solutions, the model
suffers from the mathematical difficulty of having to solve
second-order field equations. Apart from this technical point, one
may also argue, together with Hong-Kim-Pac \cite{HoKiPa}, Jackiw and
Weinberg \cite{JaWe}, and Jackiw, Lee and Weinberg \cite{JaLeWe},
that, for large distances the Chern-Simons term, which is of
lower-order in derivatives, will dominate the traditional Maxwell
one. It can, therefore,  make sense to
 turn off the Maxwell term altogether.

 For further convenience, the standard, fourth-order, self-interaction scalar potential (\ref{quarticpot}) will be
replaced by the planar analog of the $6^{th}$ order potential
considered before by Christ and Lee in their one-dimensional bag
model \cite{Christ}, and which, together with the quartic one, is
the only renormalizable choice \cite{JaWe,JaLeWe}. It is also
consistent with supersymmetry \cite{LeLeWe,BumSUSY}. Hence, the
Lagrangian proposed by Hong et al. \cite{HoKiPa}, and by Jackiw et
al. \cite{JaWe,JaLeWe} is, \footnote{Our $\kappa$ here is one half
of that of Jackiw and Weinberg in \cite{JaWe}. },
\begin{eqnarray}
\cL&=&
\frac{1}{2}(D_\mu\psi)^*D^\mu\psi+U(\psi)
-
\frac{\kappa}{4}\epsilon^{\alpha\beta\gamma}A_\alpha F_{\beta\gamma},
\label{RCSLag}
\\[8pt]
&&U(\psi)=\frac{\lambda}{4}|\psi|^2\big(|\psi|^2-1\big)^2.
\label{6pot}
\end{eqnarray}
Below we shall refer to this  as the
 HKP-JLW model.

The Euler-Lagrange equations read
\begin{eqnarray}
D_\mu D^\mu\psi&=&\displaystyle\frac{\lambda}{2}(|\psi|^2-1)\big(3|\psi|^2-1\big)\psi,
\label{relCSEL1}
\\[8pt]
\2\kappa\,\epsilon^{\mu\alpha\beta}F_{\alpha\beta}&=& - j^\mu,
\label{relCSEL2}
\end{eqnarray}
where
$j^\mu\equiv(\varrho,\vec{\jmath})$ is the
conserved current,
\begin{equation}
j^\mu=\frac{e}{2i}\big[\psi^*D^\mu\psi-\psi(D^\mu\psi)^*\big],
\qquad
\partial_\mu j^\mu=0.
\label{relcurrent}
\end{equation}

The first  equation in  (\ref{relCSEL1}) is the (by now
familiar) nonlinear Klein-Gordon equation with non-linearity
derived from the sixth-order potential (\ref{6pot});
the {\it Field-Current Identities} (FCI), in (\ref{relCSEL2}),
replace the Maxwell equations.
The space components of the FCI reproduce precisely the Hall law
(\ref{HallLaw}), of Section \ref{CSHall}. Such a theory should, therefore,  describe long-range, Hall-type physics.
Note that, unlike in the Maxwell case, our
equations here are of the first order
in the vector potential.

These equations are consistent with
those in Section \ref{PaulKhare} when the Maxwell term is switched off.

\goodbreak
\kikezd{Finite-energy  configurations}

The energy  of a static configuration is now
\begin{equation}
\cE=\int\!d^2{\vx}\Big\{\2D_i\psi(D^i\psi)^*
-\2e^2A_0^2|\psi|^2+\kappa A_0B+U(\psi)\Big\}.
\label{CSrelEn}
\end{equation}
Although this expression is {\it not} positive definite,
static solutions of the equations of motion (\ref{relCSEL1})-(\ref{relCSEL2})
are, nevertheless, stationary points of the energy.
Variation of (\ref{CSrelEn}) w.r.t. $A_0$ yields one of the equations of motion, namely
\begin{equation}
-e^2A_0|\psi|^2+\kappa B=0.
\label{CSrelconstr}
\end{equation}
Eliminating $A_0$ from (\ref{CSrelEn}) using this constraint,
a {\it positive definite} energy functional,
\begin{equation}
\cE=\int\!d^2\vec{r}\,\Big\{\2D_i\psi(D^i\psi)^*
+\frac{\kappa^2}{2e^2}\,\frac{B^2}{|\psi|^2}
+U(\psi)\Big\},
\label{CSrelEnbis}
\end{equation}
is obtained.

We are interested in finding static, finite-energy,
configurations.
The usual symmetry-breaking requirement
$
U(\psi)\to0
$
as $r\to\infty$
can be met in two different ways,
\begin{equation}\left\{
\begin{array}{ll}
\hbox{(a)} \qquad
&|\psi|\to1
\\[6pt]
\hbox{(b)}
&\psi\to0
\end{array}\right.
\qquad\hbox{as}\qquad
r\to\infty.
\label{twocases}
\end{equation}
In the first case, the scalar field takes its vacuum value, $1$, in
a non-trivial way and we get {\it topological vortices}. They will
be our main concern in this Section. {\it Non-topological vortices},
case (b), will be shortly hinted at the end of the Section.

\kikezd{Topological Vortices}

Finite energy at infinity is guaranteed by the conditions,
\begin{equation}
\left\{\begin{array}{llll}
i.)&\qquad\vert\psi\vert^2-1
&\quad=&\quad{\rm o}\big({1/r}\big),
\\[8pt]
ii.)
&\qquad B&\quad=&\quad{\rm o}\big({1/r}\big),
\\[8pt]
iii.)&\qquad\vD\psi
&\quad=&\quad{\rm o}\big({1/r}\big).
\end{array}\right.\qquad
{r\to\infty}.
\label{relfenercond}
\end{equation}
Therefore,  the $U(1)$ gauge symmetry is broken  for large $r$.
In particular, the scalar field $\psi$ is
covariantly constant, $\vD\psi\approx0$.
This equation is solved  by parallel transport,
\begin{equation}
\psi({\vx})=\exp\Big[ie\int_{{\vx}_0}^{{\vx}}A_idx^i\Big]\,\psi_0,
\label{partrans}
\end{equation}
which is well-defined whenever,
\begin{equation}
e\oint A_idx^i=
e\int_{\IR^2}\!d^2{\vx}\,B
\equiv e\,\Phi
=2\pi n,
\qquad
n=0,\pm1,\ldots.
\label{Windfluxquant}
\end{equation}
Thus, {\it the magnetic flux is quantized}.
The integer $n$ here is the
{\it winding number}, also called
{\it topological charge}, or {\it vortex number}.

The asymptotic values of the Higgs field
provide us with a mapping from the circle at infinity
into the vacuum manifold, which is again a circle,
$|\psi|^2=1$.
The property
$B\approx 0$ for large $r$ implies that the vector potential is asymptotically
a pure gauge, so that we have,
\begin{equation}
A_j\approx-\frac{i}{e}\,\partial_j\log\psi.
\label{freegauge}
\end{equation}

Spontaneous symmetry breaking generates mass \cite{DeYa}. Expanding
$j^\mu$ around the vacuum expectation value of $\psi$ we find
$j^\mu=-e^2A^\mu$, so that (\ref{relCSEL2}) is approximately
$$
\2\kappa\,\epsilon^{\mu\alpha\beta}F_{\alpha\beta}\approx e^2A^\mu
\qquad\Rightarrow\qquad
F_{\alpha\mu}\approx
({e^2/\kappa})\,\epsilon_{\alpha\mu\beta}A^\beta.
$$
Inserting $F_{\alpha\beta}$ and applying
 $\partial^\alpha$,
we find that the gauge field $A^\mu$
satisfies the Klein-Gordon equation
$$
\dAlembert A^\mu
\approx
\Big(\frac{e^2}{\kappa}\Big)^2A^\mu,
$$
showing that the mass of the gauge field is
\footnote{Re-inserting the speed of light and a non-trivial
vacuum value, the masses are
\begin{equation}
m_A=\frac{e^2}{|\kappa|c}\mu^2
\qquad
m_\psi=\frac{\sqrt{2\lambda}}{c}\mu^2.
\label{cgfmass}
\end{equation}
}
\begin{equation}
m_A=\frac{e^2}{|\kappa|}.
\label{gfmass}
\end{equation}

The Higgs mass is found, in turn, by expanding $\psi$ around its
expectation value, chosen
as $\psi_0=(1,0)$,
 $(\psi_r,\psi_ \theta)=(1+\varphi,\vartheta)$,
 \begin{equation}
U=\underbrace{U(1)}_{=0}
\quad+\quad
\underbrace{
\frac{\delta U}{\delta|\psi|}\Big\vert_{|\psi|=1}}_{=0}\,\varphi
\quad+\quad
\frac{1}{2}\underbrace{
\left(\frac{\delta^2U}{\delta|\psi|^2}\right)
\Big\vert_{|\psi|=1}}_{m_\Psi^2}
\varphi^2,
\label{potexp}
\end{equation}
since $|\psi|=1$ is a critical point of $U$. We conclude that
the mass of the Higgs particle is \footnote{This can also be seen by considering the radial equation (\ref{CSrelradeqf})-(\ref{CSrelradeqA})-(\ref{CSrelradeqA0})
below.},
\begin{equation}
m_\psi^2=\frac{\delta^2U}{\delta|\psi|^2}\Big\vert_{|\psi|=1}
={2\lambda}.
\label{Higgsmass}
\end{equation}
\goodbreak

\kikezd{Radially symmetric  vortices}

For the static, radially symmetric Ansatz
(\ref{PKradAns}), i.e. for
$$
A_0=nA_0(r),
\qquad
A_r=0,
\qquad
A_\theta=nrA(r),
\qquad
\psi(r)=f(r)e^{in\theta},
$$
the equations of motion (\ref{relCSEL1})-(\ref{relCSEL2}) read, using (\ref{PKmetricquant}) \footnote{Note the absence of second-order derivatives in the fields $A$ and $A_0$. The formulae in \cite{JaLeWe}
 are obtained setting $A_0^{JLW}=-nA_0$ and
$A^{JLW}=-nerA$.},
\begin{eqnarray}
f''+\displaystyle\frac{f'}{r}
-e^2n^2\left(A-\displaystyle\frac{1}{er}\right)^2\!f+e^2n^2A_0^2f+\displaystyle\frac{\lambda}{2}f(1-f^2)(3f^2-1)=0,
\label{CSrelradeqf}
\\[8pt]
-e^2f^2\left(A-\displaystyle\frac{1}{er}\right)
+\kappa A_0'=0,
\label{CSrelradeqA}
\\[8pt]
-e^2f^2A_0+\kappa\left(A'+\displaystyle\frac{A}{r}\right)
=0,\label{CSrelradeqA0}
\end{eqnarray}
supplemented with the large-r boundary conditions
\begin{equation}
f(r)\to1,
\quad
A(r)-\displaystyle\frac{1}{er}\to0,
\quad
A_0(r)\to0,
\qquad
\hbox{\small as}\qquad
r\to\infty.
\label{relpCSlargar}
\end{equation}

 The static, radial equations (\ref{CSrelradeqf})-(\ref{CSrelradeqA})-(\ref{CSrelradeqA0})
can also be derived using the radial form of the energy.
In  the radial case, we have,
\begin{equation}
\pi\int_0^\infty\!\!rdr
\left\{2U(f)+
(f')^2+e^2n^2\big(A-\frac{1}{er}\big)^2f^2+
n^2\big(2\kappa A_0(A'+\frac{A}{r})-e^2A_0^2f^2\big)
\right\}.
\label{radtoten}
\end{equation}
Then
(\ref{radtoten}) is obtained from (\ref{CSrelEn})
using the handy formulae (\ref{PKmetricquant}). Then
variation w.r.t. $f,\ A,$ and $A_0$ yields the
radial equations (\ref{CSrelradeqf})-(\ref{CSrelradeqA})-(\ref{CSrelradeqA0}).

Eliminating, as above, $A_0$ from the energy using
 the  FCI (\ref{CSrelconstr}), i.e.,
 $$\kappa B=e^2f^2A_0,
 $$
yields the radial form of the positive expression (\ref{CSrelEnbis}),
\begin{equation}
\cE=\pi\int_0^\infty\!\! rdr\left\{2U(f)+
(f')^2+n^2e^2\left(A-\frac{1}{er}\right)^2f^2+
\frac{\kappa^2n^2}{e^2f^2}
\left(A'+\frac{A}{r}\right)^2\right\}.
\label{relradsymen}
\end{equation}

Then variation of (\ref{relradsymen}) with respect to $A$ and $f$ provides us with two more equations.

No analytic solutions have been found so far. Numerical solutions
are shown on Fig. 2. Approximate solutions for large-$r$ can be
obtained by inserting the asymptotic forms of the field.

\vskip2mm
$\bullet$ The deviation of $f$ from its asymptotic value,
$\varphi=1-f$, is
found by inserting $\varphi$ into
(\ref{CSrelradeqf}) and
expanding to first order in $\varphi$,
\begin{equation}
\varphi''+\frac{1}{r}\varphi'-2\lambda\varphi\approx0
\qquad\Rightarrow\qquad
\varphi\approx GK_0(\sqrt{2\lambda}\,r).
\label{phieq}
\end{equation}
Hence, the asymptotic behavior is exponential drop-off
 with characteristic length
$(m_\psi)^{-1}$,
\begin{equation}
\varphi\approx\frac{G}{\sqrt{r}}\,e^{-m_\psi r}.
\label{phival}
\end{equation}

$\bullet$ Setting
$f\approx1$ into the last two equations yields
\begin{equation}
A'+\frac{A}{r}-\frac{e^2}{\kappa}A_0\approx0,
\qquad
A_0'-\frac{e^2}{\kappa}\left(A-\frac{1}{er}\right)\approx0,
\label{relappeq1}
\end{equation}
from which we infer, firstly, that
\begin{equation}
\frac{d^2A_0}{d\rho^2}
+
\frac{1}{\rho}\frac{dA_0}{d\rho}-A_0\approx0,
\qquad
\rho\equiv(e^2/|\kappa|)r.
\label{relappeq2}
\end{equation}
where we recognize the modified Bessel equation of order zero. Hence,
\begin{equation}
A_0\approx CK_0(\frac{e^2}{|\kappa|}r).
\label{A_0Bess}
\end{equation}

Similarly, setting
\begin{equation}
a=A-\displaystyle\frac{1}{er}\ ,
\label{adef}
\end{equation}
we find,
\begin{equation}
a''+\frac{a'}{\rho}
-\Big(1+\frac{1}{\rho^2}\Big)a\approx0,
\label{alphaeq}
\end{equation}
which is the modified Bessel equation of order $1$.
 Thus $a\approx CK_1(\rho)$ so that
 \footnote{Another way of deriving this result is to express $A$,
from the second equation in (\ref{relappeq2}), as
$$
A=\frac{1}{er}+\frac{\kappa}{e^2}\,A_0',
$$
and use the recursion relation
$
K_0'=-K_1
$
of the Bessel functions.}
\begin{equation}
A\approx\frac{1}{er}-C
K_1\big(\frac{e^2}{|\kappa|}\,r\big).
\label{AeqBess}
\end{equation}

Note that there is {\it no mass splitting} in the
present, purely-CS, system since, unlike in the
Maxwell -- Chern-Simons case \footnote{Switching off the Maxwell term
 can be achieved by letting $\kappa$ become
big so that $\frac{4e^2}{\kappa^2}\ll1$.
Then (\ref{PKgoodmassA}) becomes
approximately ${e^2}/{|\kappa|}$ as here above.
The  mass with the plus sign
behaves  approximately as $\approx|\kappa|\gg 1$
and diverges therefore.}, (\ref{aa0}),
the asymptotic $A$ and $A_0$ equations
(\ref{relappeq2}) and  (\ref{alphaeq}) are uncoupled.

An even coarser approximation is
obtained by eliminating the first derivative
terms in  (\ref{alphaeq}) [and in (\ref{relappeq2})],
 setting $a=u\rho^{-1/2}$ [and $A_0=u\rho^{-1/2}$] and
dropping terms with inverse powers of $r$.
Then {\it both} equations reduce to
$$
u''=\big(\frac{e^2}{\kappa}\big)^2u,
$$
so that
\begin{equation}
A_0\sim\frac{D}{\sqrt{r}}\,\textrm{e}^{-m_Ar}, \qquad
A\sim\frac{1}{er}+\frac{D}{\sqrt{r}}\,\textrm{e}^{-m_Ar} \qquad
r\to\infty, \label{ueq}
\end{equation}
for some constant $D$.
The fields $A_0$ and $A$ thus approach their asymptotic values also
exponentially, with characteristic length determined by the gauge field mass $m_A=e^2/|\kappa|$
\footnote{
Equation (\ref{ueq}) is consistent with the asymptotic
expansion of Bessel functions.}.

The penetration depths of the
gauge and scalar fields are, therefore,
\begin{equation}
\eta=\frac{1}{m_A}=\frac{|\kappa|}{e^2}
\qquad\hbox{resp.}\qquad
\xi=\frac{1}{m_\psi}=\frac{1}{\sqrt{2\lambda}},
\label{penet}
\end{equation}
respectively.

Similarly, the
usual analysis yields  the small-$r$ behavior,
\begin{equation}\left\{
\begin{array}{l}
f(r)=f_0r^{|n|}+\frac 1{8(|n|+1)}(\lambda
-2n^2e^2b_0^2)f_0r^{|n|+2}+O(r^{|n|+4}),
\\[12pt]
A(r)=\frac 1{2|n|+2}\frac{e^2}\kappa b_0f_0^2r^{2|n|+1}+\frac 1{8(|n|+1)(|n|+2)}%
\frac{e^2}\kappa (\lambda -2n^2e^2b_0^2-\delta _{|n|}^1\frac{4ef_n^2}{\kappa
b_0})b_0f_0^2r^{2|n|+3}+O(r^{2|n|+5}),
\\[14pt]
A_0(r)= b_0-\frac 1{2|n|}\frac e\kappa f_0^2r^{2|n|}-\frac
e{8(|n|+1)^2}(\lambda-2n^2e^2b_0^2)f_0^2r^{2|n|+2}
+O(r^{2|n|+4}),
\end{array}\right.
\label{relpCSsmallr}
\end{equation}
where $b_0$ and $f_0$ are free constants.
Note that $f(r)\to0$ and $A(r)\to 0$, but
$A_0(r)\to b_0=\const\neq0$ as ${r\to0}$.
In conclusion, the fields behave, as,
\begin{equation}\begin{array}{lllll}
|\psi(r)|&\equiv f(r)\qquad
&\propto\qquad
&\left\{\begin{array}{c}
r^{|n|}\qquad
\\[8pt]
1-\frac{G}{\sqrt{r}}\,\textrm{e}^{-m_\psi r}
\end{array}\right.\qquad
&\begin{array}{l}
r\sim0
\\[8pt]
r\to\infty
\end{array}
\\[28pt]
|\vE(r)|&=|nA_0'(r)|\qquad
&\propto\qquad
&\left\{\begin{array}{c}
r^{2|n|-1}
\\[8pt]
\frac{1}{\sqrt{r}}\,\textrm{e}^{-m_Ar}+\dots
\end{array}\right.\qquad
&\begin{array}{l}
r\sim0
\\[8pt]
r\to\infty
\end{array}
\\[28pt]
|B(r)|\qquad
&=n\left(A'+\displaystyle\frac{A}{r}\right)\qquad
&\propto\qquad
&\left\{\begin{array}{c}
r^{2|n|}
\\[8pt]
\frac{1}{\sqrt{r}}\,\textrm{e}^{-m_Ar}+\dots
\end{array}\right.
&\begin{array}{l}
r\sim0
\\[8pt]
r\to\infty
\end{array}
\end{array}
\label{approxfield}
\end{equation}
\begin{figure}
\begin{center}
\includegraphics[scale=.7]{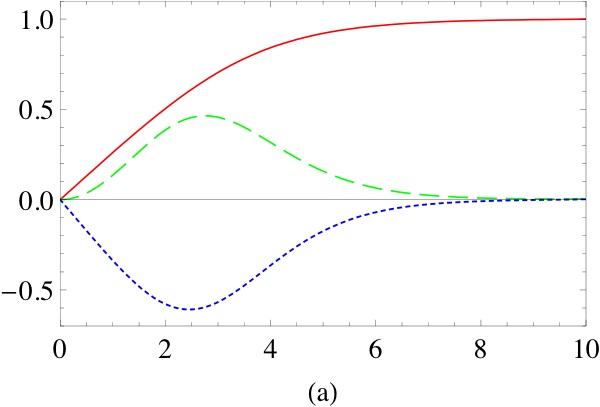}\quad
\includegraphics[scale=.7]{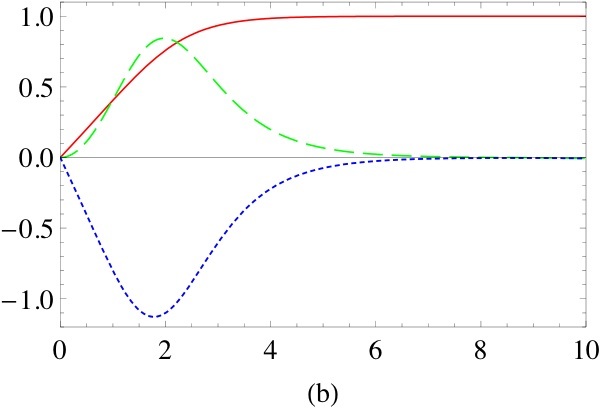}
\end{center}\vskip-3mm
\caption{\it Radial topological vortices in the
HKP-JLW model
(a) below, $\lambda=\2\lambda_{SD}$, and (b) above,
$\lambda=2\lambda_{SD}$, the critical coupling
$\lambda_{SD}$. The heavy/dotted/dashed
lines refer to the scalar/electric/magnetic fields. }
\label{JLWnonSD}
\end{figure}

\goodbreak
\kikezd{Self-dual vortices}

In the Abelian Higgs model
an important step has been to recognize that, for a specific value of
the coupling constant, the field equations could be reduced
to first-order ones \cite{Bogo,VeSch}.  Intuitively, this
happens when the gauge- and scalar field masses are equal,
so that the Higgs attraction is precisely canceled by the
electromagnetic repulsion, opening the way to static
multivortex solutions.

A similar equipoise of the forces can also be achieved
 by a suitable
modification of the model \cite{HoKiPa,JaWe,JaLeWe},
as discussed
below in some detail.

Let us suppose that the scalar and gauge fields have equal masses,
\begin{equation}
m_\psi=m_A=\frac{e^2}{c|\kappa|}\equiv m,
\label{SDCSmass}
\end{equation}
and, hence, equal penetration depths, $\xi=\eta$
[the velocity of light has been inserted for further use].
This amounts to choosing the
coefficient of the potential as
\begin{equation}
\frac{\lambda_{SD}}{4}=
\frac{e^4}{8\kappa^2c^4}\ .
\label{SDCSlambda}
\end{equation}
Returning to units where  $c=1$ we observe that
the Bogomol'ny trick applies, i.e.,
using the identity
\begin{equation}
\vert\vD\psi\vert^2=
\vert(D_1\pm iD_2)\psi\vert^2
\pm eB\,|\psi|^2 \pm \vnabla\times\vj\ ,
\label{SDidentity1}
\end{equation}
the energy (\ref{CSrelEnbis}) can be rewritten in the form
\begin{eqnarray}
\cE=&\displaystyle\int\! d^2\vx\left\{
\2\Big|(D_1\mp iD_2)\psi\Big|^2
+\2\Big\vert\frac{\kappa}{e}\,\frac{B}{\psi}
\mp\frac{e^2}{2\kappa}\psi^*(1-|\psi|^2)\Big\vert^2\right\}\nonumber
\\[8pt]
&+\displaystyle\int\! d^2\vx\,
\frac{1}{4}\left(\lambda-\frac{e^4}{2\kappa^2}\right)
|\psi|^2\big(1-|\psi|^2\big)
\pm\displaystyle\int\! d^2\vx\;\left(
\frac{eB}{2}+\2\vnabla\times\vj\,\right).
\label{relCSBogodec}
\end{eqnarray}
The current in the last term can be dropped,
yielding the magnetic flux,
\begin{equation}
\pm\int\! d^2\vx\,\frac{eB}{2}
=
\pm\,\frac{e}{2}\Phi\ .
\end{equation}

For the specific value [consistent
with (\ref{SDCSlambda})],
\begin{equation}
\lambda=\lambda_{SD}=\frac{e^4}{2\kappa^2}\ ,
\label{JLWSDlambda}
\end{equation}
of the coupling constant, the
indefinite third term drops out, leaving us
with the first, non-negative integral.
Thus,  we
have  established that the energy admits a Bogomol'ny bound,
\begin{equation}
\cE\;\geq\;\frac{e|\Phi|}{2}=\pi|n|.
\label{relBogobound}
\end{equation}
Equality here is only attained if the {\it self-duality equations},
\begin{eqnarray}
D_1\psi\pm iD_2\psi=0,
\label{relSD1}
\\[6pt]
eB\mp\frac{m^2}{2}\,|\psi|^2(1-|\psi|^2)=0,
\label{relSD2}
\end{eqnarray}
hold, with the upper/lower sign chosen for
$n$ positive/negative.

It is readily verified directly
that the solutions of the first-order equations,
(\ref{CSrelconstr}) and (\ref{relSD1})-(\ref{relSD2}), solve
the second-order field equations
(\ref{relCSEL1})-(\ref{relCSEL2}).

Equation (\ref{CSrelconstr}) i.e. $\kappa B=e^2A_0|\psi|^2$, is one
(namely the time) component of the field equations (FCI)
(\ref{relCSEL2}).
Turning to the space component of the field-current identity,
\begin{equation}
\kappa\,\epsilon^{ij0}F_{j0}=-\jmath^i,
\label{FCIi}
\end{equation}
one first shows  that, for SD/ASD fields,
\begin{equation}
{\vj}_{SD}=\mp\frac{e}{2}\,\vnabla\times|\psi|^2. \label{SDcurrent}
\end{equation}
On the other hand, using again (\ref{CSrelconstr}),
$$
\kappa F_{j0}=\kappa\p_jA_0=\kappa\p_j\left(
\frac{\kappa B}{e^2|\psi|^2}\right).
$$
Hence, by $\epsilon^{ij0}=-\epsilon_{ij}$ and by the SD equation
(\ref{relSD2}), the r.h.s. of the FCI
is precisely minus the $i$-component of the current, as required.

As to
the non-linear Klein-Gordon equation (\ref{relCSEL1}),
expressing $A_0$ from (\ref{CSrelconstr}) and
using  (\ref{relSD2}), we get,
$$
D_0D^0\psi=e^2A_0^2\,\psi=\frac{(eB)^2}{m^2|\psi|^4}\,\psi
=\frac{m^2}{4}\big(1-|\psi|^2)^2\psi.
$$
Then, using $ \vD^2\psi=\mp eB\,\psi, $
 the l.h.s. of (\ref{relCSEL1}) becomes, by
(\ref{relSD1}),
$$
D_\mu D^\mu\psi= \frac{m^2}{4}\left(1-|\psi|^2\right)
\left(1-3|\psi|^2\right)\psi,
$$
which is the r.h.s of (\ref{relCSEL1}) as it should be. \vskip2mm

Let us first study the radial case.
Introducing  cf. (\ref{adef})  and using (\ref{PKmetricquant}),
\begin{equation}
a=A-\frac{1}{er}
\quad\Rightarrow\quad
B=n\left(A'+\frac{A}{r}\right)=
n\left(a'+\frac{a}{r}\right),
\label{adefbis}
\end{equation}
the radial self-duality equations become,
\begin{equation}\left\{
\begin{array}{l}
f'\pm en\,af=0,
\\[8pt]
en\left(a'+\displaystyle\frac{a}{r}\right)\mp
\displaystyle\frac{m^2}{2}f^2(1-f^2)=0.
\end{array}\right.
\label{relradSD}
\end{equation}
By (\ref{relpCSlargar}) and (\ref{relpCSsmallr}), the boundary conditions read
\begin{equation}
\begin{array}{cccccc}
a(\infty)&\sim0,\qquad
&f(\infty)&\sim1,
\\[10pt]
a(0)&\sim-\displaystyle\frac{1}{er},
&f(0)&\sim0.
\end{array}
\label{bigsmallrbeh}
\end{equation}

It is worth noting that the first-order equations
(\ref{relradSD}) can also be obtained by using the radial ``Bogomol'ny''
decomposition of the energy density in (\ref{relradsymen}),
\begin{eqnarray}
&\pi r&\left\{
\big[f'\pm enaf\big]^2
+
\frac{1}{m^2f^2}
\left[en\big(a'+\frac{a}{r}\big)
\mp\frac{m^2}{2}f^2(1-f^2)\right]^2
\right\}\nonumber
\\[6pt]
&&\mp\,\pi en(arf^2)'
\pm\pi en \big(ar)',
\label{relCSBogdecomp}
\end{eqnarray}
valid for $
U(f)=\frac{m^2}{8}\,f^2\big(f^2-1\big)^2
$.

For $n=0$ the only solution is the (asymmetric)
vacuum, as it can be seen using a simple scaling argument \cite{JaLeWe}.

For $n\neq0$, which we assume henceforth (except in our discussion, below, of the non-topological vortices), no analytic solution has been found so far.
The radial first-order equations (\ref{relradSD}), are
readily solved numerically, however.
\begin{figure}
\begin{center}
\includegraphics[scale=.68]{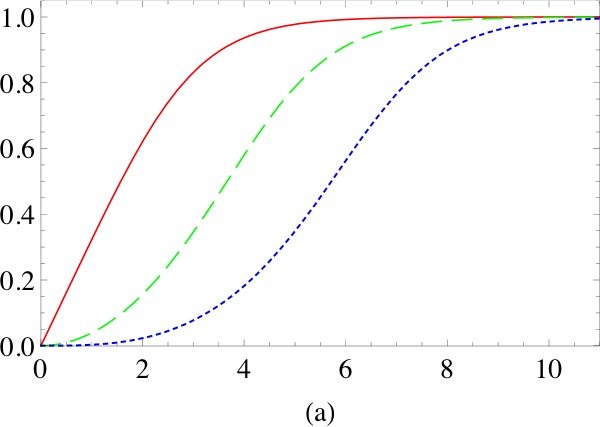}\qquad
\includegraphics[scale=.70]{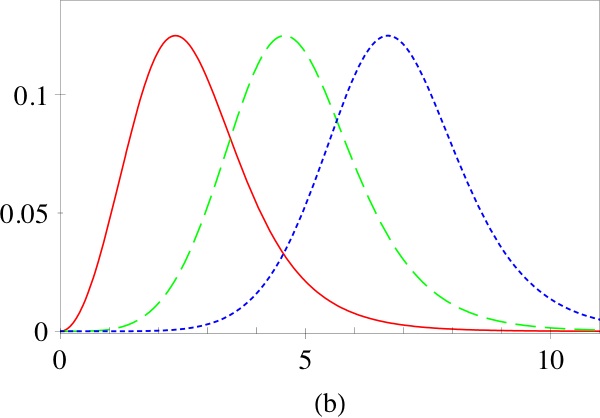}
\end{center}\vskip-5mm
\caption{\it (a) Scalar (b) magnetic fields of
radially symmetric, SD, topological, HKP-JLW  vortices
for winding numbers $n={\bf 1,2,3}$.
The magnetic field has a ``doughnut-like'' shape,
as $B$ vanishes at the origin.}
\label{SDRCSfield}
\end{figure}
According to Figures \ref{SDRCSfield} and \ref{SDRCSendens}, and also confirmed by calculation,
the magnetic field, and the
energy density, respectively, both vanish at the origin.
Consider the SD equation,
$$
eB=\pm\frac{m^2}{2}\,|\psi|^2(1-|\psi|^2).
$$
Hence, the magnetic field is zero where the scalar field is in a
ground state (i.e., either vanishes or equals 1). This is what
happens, in particular, for a radial vortex at the origin. The
magnetic field
 has, therefore,  a doughnut-like shape.

Similarly, the energy density of a radial SD configuration is,
 by  (\ref{relCSBogdecomp}),
\begin{equation}
w_{SD}\,d^2\vx=\mp\pi en(arf^2)'dr
\pm\pi eB rdr.
\end{equation}
Here the first term vanishes at the origin because
 $f(0)=0$ and the last term is
proportional to $B(0)=0$.

Some more insight is gained by studying the asymptotic
behavior.
\vskip2mm
$\bullet$ To study the large-$r$ behavior, set
$\varphi\equiv1-f$.
Inserting $f\approx1$, Eqs. (\ref{relradSD}) reduce to
\begin{equation}
\varphi'\approx\pm en{a},
\qquad
en\left(a'+\frac{a}{r}\right)\approx\pm m^2\varphi.
\label{largerradSD}
\end{equation}
Differentiating, we get,
$$
\begin{array}{lll}
\varphi''+\displaystyle\frac{1}{r}\varphi'-m^2\varphi\approx0
\qquad&\Rightarrow\qquad
&\varphi\approx CK_0(mr),
\\[10pt]
a''+\displaystyle\frac{a'}{r}-\left(m^2+\frac{1}{r^2}\right)a\approx0
\qquad&\Rightarrow\qquad
&a\approx CK_1(mr).
\end{array}
$$
Thus, for large $r$, using again (\ref{largerradSD})
and the known relation $K_0'=-K_1$,
\begin{equation}
f\approx1-CK_0(mr)
\qquad\hbox{\small and}\qquad
A\approx\frac{1}{er}
-\frac{m}{e|n|}CK_1(mr)
\label{AsymprRelSol}
\end{equation}
with some constant $C$.

\vskip2mm
$\bullet$ For small $r$, on the other hand, equation
(\ref{relradSD}) yields the expansion,
\begin{equation}\begin{array}{lll}
f(r)&=&f_0r^{|n|}-\displaystyle\frac{m^2}{8(|n|+1)^2}f_0^3r^{3|n|+2}+\displaystyle\frac{m^2}{8(2|n|+1)^2}%
f_0^5r^{5|n|+2}+\displaystyle\frac{m^4}{64(|n|+1)^4}f_0^5r^{5|n|+4}+O(r^{7|n|+4})
\\[16pt]
A&=&
\displaystyle\frac{m^2}{4|n|(|n|+1)e}f_0^2r^{2|n|+1}-\displaystyle\frac{m^2}{4|n|(2|n|+1)e}%
f_0^4r^{4|n|+1}-\displaystyle\frac{m^4}{32|n|(|n|+1)^3}f_0^4r^{4|n|+3}+O(r^{6|n|+3})
\\
\end{array}
\label{relCSsmallr}
\end{equation}

The results are consistent with (\ref{relpCSsmallr}).
\begin{figure}
\begin{center}
\includegraphics[scale=.8]{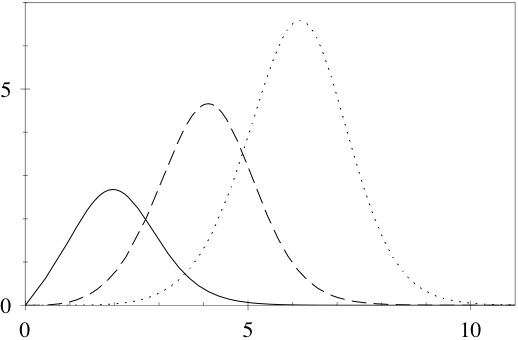}
\end{center}\vskip-5mm
\caption{\it Energy (= mass) densities of
radially symmetric, topological HKP-JLW vortices
with  winding numbers $n={\bf 1,2,3}$.
The ``doughnut-like'' shape is confirmed for all $n$.}
\label{SDRCSendens}
\end{figure}
The asymptotic behavior  (\ref{bigsmallrbeh}) is actually valid in full generality, without assuming radial symmetry. Putting $\psi=\sqrt{\varrho}\,e^{i\omega}$, the vector potential can be expressed
from the self-duality condition $(D_1\pm iD_2)\psi=0$ as,
\begin{equation}
e{\vA}=\vnabla\omega\pm \frac 1 2 \vnabla\times\log\varrho,
\end{equation}
and inserting it into (\ref{relSD2}), we  get the Liouville-type
equation \footnote{It is worth emphasizing that, unlike in the
non-relativistic case discussed later, the density, $\varrho$, is
{\it not}
 the time component of a conserved current so that
 the total mass, $\int\varrho d^2\vx$ may not be conserved. The time
component of the conserved current, $j_0$
in (\ref{relcurrent}), is different from $\varrho$,
and is {\it not} positive definite.},
\begin{equation}\fbox{$
\bigtriangleup\log\varrho=m^2\varrho(\varrho-1)\ .\
$}
\label{CSRLioutype}
\end{equation}

In the radial case, we can also proceed as follows.
Differentiating the first of the equations in
(\ref{relradSD}) and using the
second one, the gauge field can be eliminated to yield the
radial form of (\ref{CSRLioutype}),
\begin{equation}
\bigtriangleup\log f=\frac{m^2}{2}\,f^2(f^2-1).
\label{radCSLioutype}
\end{equation}
Where the scalar field
is small [as  near the origin for $n\neq0$, cf. (\ref{relCSsmallr})], the quartic term on
the r.h.s. can be neglected, leaving us with
the approximate equation,
\begin{equation}
\bigtriangleup\log f=-\frac{m^2}{2}\,f^2,
\label{JLWLiou}
\end{equation}
which is the classical Liouville equation.
Then, the short-$r$ behavior found above can
also be deduced by expanding in power series the
 known solutions, presented in Section \ref{LiouVort} below, of the latter.

The small-$r$ behavior of the SD
scalar field can also be obtained from the ``Liouville-type'' equation  \cite{JaLeWe}.

\begin{figure}
\begin{center}
\includegraphics[scale=.43]{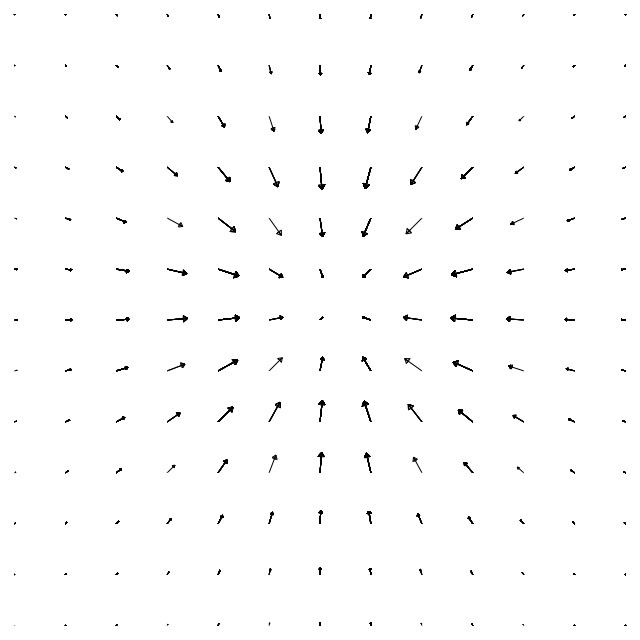}\qquad
\includegraphics[scale=.8]{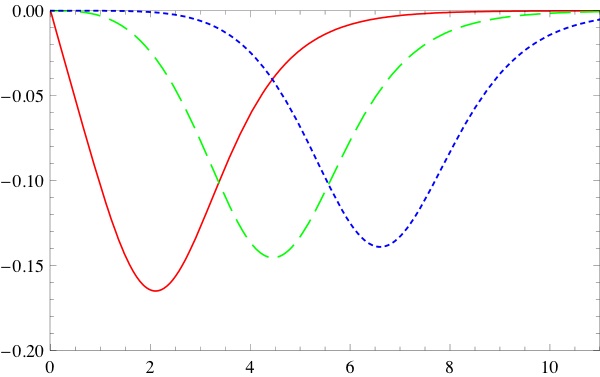}
\end{center}\vskip-5mm
\caption{\it (a) The electric field of the
radially symmetric HKP-JLW vortex is also radial.
Its magnitude is shown on (b) for $n=1,\ 2,\ 3$.}
\label{CSvecE1}
\end{figure}

Unlike classical Nielsen-Olesen vortices,
CS vortices also carry an electric field.
For our radial Ansatz, the latter, $\vE=-nA_0'\,\vr/r$,
shown on Figure \ref{CSvecE1}, is also radial.

\vskip2mm
Returning to the general (not necessarily radial)
vortices, one can ask if multivortices do exist.
The answer is yes.
Index-theoretical
calculations  \cite{HoKiPa,JaWe,JaLeWe}, similar to
those in the Nielsen-Olesen-Bogomol'ny case \cite{Weinberg}, allow one to deduce~:

\kikezd{Theorem}~: {\it For topological charge $n$,
the self-duality equations (\ref{relSD1}-\ref{relSD2})
admit a $2|n|$ parameter family of solutions}.

\vskip2mm
Pure Chern-Simons vortices in a uniform background field
are discussed in Ref. \cite{LeeYi}.
The existence of topological vortices
as well as a numerical construction has been
proved in Ref.  \cite{YiYa}.

\kikezd{Non-topological vortices}.

The boundary condition of case (b) in (\ref{twocases}),
$$
|\psi|\to0
\qquad\hbox{as}\qquad
r\to\infty,
$$
leads to {\it non-topological} vortices. They are quite subtle, and
therefore we limit ourselves  to the radial, and self-dual, case.
For more details, the reader is advised to consult Refs.
\cite{JaLeWe, YiYa,Spruck,KhaNontopJW}.

For $n\neq0$,
the radial Ansatz
(\ref{PKradAns}) can still be made, although
the integer $n$ is not a winding number: no topology
is involved. The
radial self-duality equations (\ref{relradSD}) still solve the equations of motion, and also imply the
Liouville-type equations (\ref{radCSLioutype}). The only
difference is the boundary condition $f(\infty)=0$.

By (\ref{adefbis}),
$$
\int\! B\,d^2\vx=
2\pi n\int\left(a'+\frac{a}{r}\right)rdr=
2\pi n\big[(ra)(\infty)-(ra)(0)\big].
$$
Using $(ra)(0)=-\frac{1}{e}$ cf. (\ref{bigsmallrbeh})
and setting \footnote{The definition is chosen
so that the formulae coincide with those in Ref. \cite{JaLeWe}. }
\begin{equation}
(ra)(\infty)=\frac{\alpha}{ne},
\label{excessflux}
\end{equation}
the flux is, hence,
\begin{equation}
\Phi=\Phi_n=\int B\,d^2\vx=
\frac{2\pi}{e}(n+\alpha).
\label{NTnflux}
\end{equation}
We stress that $\alpha$ here is an
unquantized parameter, so that the magnetic flux
of a non-topological vortex is {\it not} quantized.

Curiously, the problem also admits solutions
with $n=0$, when the radial Ansatz (\ref{PKradAns})
has to be modified according to,
\begin{equation}
\label{rad0Ansatz}
A_0=A_0(r),
\qquad
A=rA(r),
\qquad
\psi=f(r).
\end{equation}
The magnetic flux is  fractional,
\begin{equation}
\Phi_0=
\frac{2\pi}{e}\alpha,
\label{NT0flux}
\end{equation}
which is just (\ref{NTnflux}) with $n=0$.
Let us emphasize that $\alpha$ can, {\it a priori}
be positive or negative.

The radial self-duality equations read now
\begin{equation}\left\{
\begin{array}{l}
f'\pm e\,af=0,
\\[8pt]
e\left(a'+\displaystyle\frac{a}{r}\right)\mp
\displaystyle\frac{m^2}{2}f^2(1-f^2)=0.
\end{array}\right.
\label{rel0radSD}
\end{equation}

The radial Bogomol'ny decomposition
(\ref{relCSBogdecomp})
is still valid, and yields
$
\cE\geq\frac{e}{2}\Phi_0,
$
which is a physically admissible Bogomol'ny bound
as long as it is positive. Consequently, the
upper sign should be chosen for $\alpha>0$ and
the lower one for $\alpha<0$ so that the Bogomol'ny bound is, in fact,
\begin{equation}
\cE\;\geq\; \frac{e}{2}|\Phi_0|=\pi|\alpha|\geq 2\pi,
\label{Bogo0bound}
\end{equation}
since numerical study \cite{JaLeWe} indicates
that $|\alpha|\geq2$.

Let us assume that $\alpha>0$. For large $r$, $a(r)\sim\frac{\alpha}{er}$, so these equations imply
$
f(r) \sim r^{-\alpha}.
$
 A more detailed study \cite{JaLeWe}, consistent with the Liouville approximation (\ref{JLWLiou}),
  yields, in fact,
the large-$r$ expansion [consistent with the
solution of the Liouville equation],
\begin{eqnarray}
f(r)&=&\frac{f_0}{r^\alpha}-\frac{m^2f_0^3}{8(\alpha-1)^2}
\frac{1}{r^{3\alpha-2}}%
+O(r^{-5\alpha+4}),
\\[8pt]
a(r)&=&\frac{\alpha}{er}-\frac{m^2f_0^2}{4e(\alpha-1)}
\frac{1}{r^{2\alpha-1}}+O(r^{-4\alpha+3}).
\label{RCSNTlarger}
\end{eqnarray}

For small $r$ we find in turn
\begin{eqnarray}
f(r)&=&
f_0-\frac{m^2}8f_0^3(1-f_0^2)r^2+\frac{m^4}{128}%
f_0^5(1-f_0^2)(2-3f_0^2)r^4+O(r^6),
\label{RCSNTsmallrf}
\\[8pt]
a(r)&=&
\frac{m^2}{4e}f_0^2(1-f_0^2)r
-\frac{m^4}{32e}%
f_0^4(1-f_0^2)(1-2f_0^2)r^3
+0(r^6).
\label{RCSNTsmallra}
\end{eqnarray}

\begin{figure}
\begin{center}
\includegraphics[scale=.7]{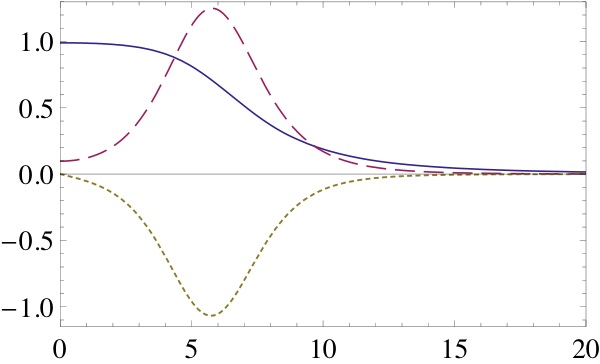}\quad
\includegraphics[scale=.73]{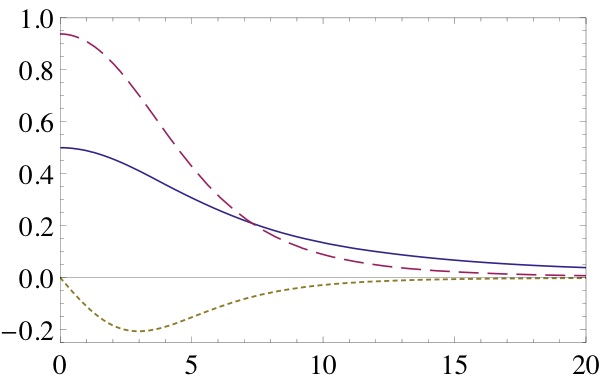}
\end{center}\vskip-5mm
\caption{\it Non-topological soliton with $n=0$
in the HKP-JLW model for (a) $f(0)=0.99$ (b)
$f(0)=0.5$.
The heavy/dashed/dotted line show
the scalar/magnetic/radial electric fields.}
\label{NTR0CSvort}
\end{figure}

\begin{figure}
\begin{center}
\includegraphics[scale=.72]{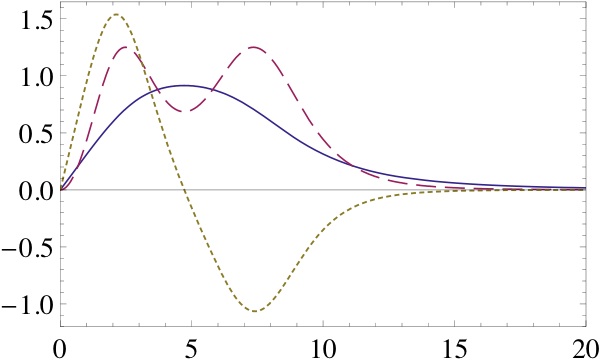}\quad
\includegraphics[scale=.72]{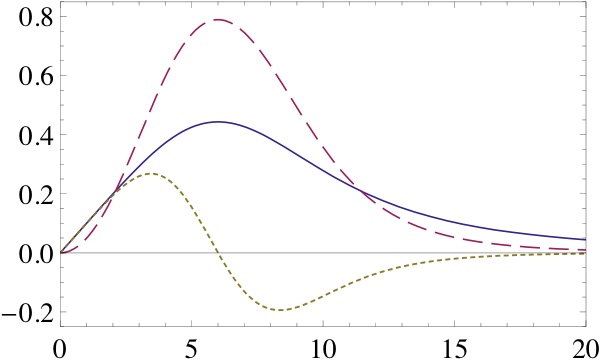}
\end{center}\vskip-5mm
\caption{\it Non-topological vortices with $n=1$
in the HKP-JLW model. (a) $f_0=0.31$ (b)
$f_0=0.1$. The heavy/dashed/dotted lines represent the scalar/magnetic/electric fields.}
\label{NTR1CSvort}
\end{figure}

$f_0\geq0$ can be assumed with no loss of generality.
The boundary conditions can only be
satisfied when $f_0\leq1$.
For $f_0=0$ and $f_0=1$, we get the symmetric and
the asymmetric vacuum solutions, respectively.
Interesting solutions, some of them plotted on Fig.
\ref{NTR0CSvort} arise, hence for $0<f_0<1$.
A similar analysis holds for $n\neq0$ \footnote{
Note that, when $n=1$, the self-duality equations
(\ref{relradSD}) is identical with
the one, (\ref{rel0radSD}), valid for $n=0$.
The solutions are distinguished by their
initial conditions~: the one with $n=0$ has
$f(0)\neq0$ and the one with $n=1$ has $f(0)=0$.}.
The self-duality equations are  those in
(\ref{relradSD}), and therefore the
small-$r$ expansion is still (\ref{relCSsmallr}).
Numerical solutions are shown on Fig.
\ref{NTR1CSvort}. They
are, roughly, mixtures of the $n=0$ non-topological soliton with
the topological vortex with $n\neq0$
\cite{JaLeWe,KhaNontopJW}.

Many theoretical questions, including
 the behavior of the fractional flux $\alpha$,
parameter-counting, the  relation to the non-relativistic
``Jackiw-Pi'' vortices discussed in the next Section,
etc., remain open, however.

\section{NON-RELATIVISTIC NON-TOPOLOGICAL VORTICES}\label{NRVort}

\subsection{The Jackiw-Pi Model}\label{JPmodel}

Now we proceed to a systematic study of non-relativistic
 vortices coupled to Chern-Simons electrodynamics.
 The one due to Jackiw and Pi  \cite{JaPi1,JaPi2}
can be derived from the relativistic model
 of Section \ref{J(L)W}.
Our starting point is the relativistic, matter Lagrangian
(\ref{RCSLag})-(\ref{6pot}), where the self-dual value
$\lambda_{SD}=e^4/2\kappa^2c^4$ has been chosen. Re-introducing the velocity of light, the (relativistic) matter Lagrangian reads
\begin{eqnarray}
\cL^{R}_{matter}=
-\frac{1}{2c^2}|D_t\psi|^2+\frac{1}{2}|\vD\psi|^2
+\frac{e^4}{8\kappa^2c^4}|\psi|^2\big(|\psi|^2-\mu^2\big)^2,
\label{RCSmattLag}
\end{eqnarray}
where  $t=x^0/c$ is non-relativistic time and, to keep control of the various terms, the symmetry breaking
vacuum value $\mu\neq0$ has been introduced\footnote{
$\mu=1$ can always be achieved by a suitable rescaling.}.
 $M$, the mass of oscillations around the {\it symmetric} vacuum state
$\psi=0$, can be read off the coefficient of the quadratic term \footnote{
 $M$ is {\it different} from the mass, $m$,
 of oscillations around the {\it symmetry breaking} ground state $|\psi|=\mu$.  In fact, $M=m/2$.},
\begin{equation}
M^2=\frac{e^4\mu^4}{4\kappa^2c^6}\ .
\label{Msquare}
\end{equation}
Now the non-relativistic limit of the system  is found \cite{Fund5,Fund6}
by decomposing the relativistic matter field into a particle and antiparticle, $\Psi$ and $\bar\Psi$ respectively,
\begin{equation}
\psi=\frac{1}{\sqrt{M}}\left(e^{-iMc^2t}\Psi+e^{+iMc^2t}\bar\Psi^*\right).
\label{NRlim}
\end{equation}
Inserting (\ref{NRlim}) into the matter Lagrangian and
dropping the oscillating terms, we find that the quadratic
terms cancel. To leading order in $1/c$, we get \footnote{Note that
\begin{equation}
D_\mu\Psi=(\p_\mu-ieA_\mu)\Psi,
\qquad
D_\mu\bar\Psi=(\p_\mu+ieA_\mu)\bar\Psi,
\nonumber
\end{equation}
i.e., the particle and the antiparticle have
opposite charges.},
\begin{eqnarray}
\left\{\displaystyle
\frac{1}{2i}\left(\Psi^*D_t\Psi-\Psi(D_t\Psi)^*\right)
+\frac{|\vD\Psi|^2}{2M}\right\}
&+&\left\{\displaystyle
\frac{1}{2i}\left({\bar\Psi}^*D_t\bar\Psi-\bar\Psi(D_t\bar\Psi)^*\right)
+ \displaystyle\frac{|\vD\bar\Psi|^2}{2M}\right\}
\nonumber
\\[10pt]
-\displaystyle\frac{e^4\mu^2}{4\kappa^2c^4M^2}(|\Psi|^2+|\bar\Psi|^2)^2
&+&
\frac{e^4}{8\kappa^2c^4M^3}
\big(|\Psi|^2+|\bar\Psi|^2\big)^3.
\label{papLag}
\end{eqnarray}
Particles and antiparticles are separately conserved, however, and
can, therefore,  be
safely eliminated by setting $\bar\Psi=0$.
Expressing $\mu^2$ using (\ref{Msquare}) yields
the coefficient of the quartic term.
The last term  can, in turn, be dropped, leaving us
 with the non-relativistic matter Lagrangian
 with an attractive quartic potential,
\begin{equation}
\cL^{NR}_{{\small matt}}=
\frac{1}{2i}\left(\Psi^*D_t\Psi-\Psi(D_t\Psi)^*\right)
+\frac{|\vD\Psi|^2}{2M}
-\frac{\lambda}{2}|\Psi|^4,
\qquad
\lambda=\frac{e^2}{Mc|\kappa|}.
\label{NRCSmatLag}
\end{equation}

It will be shown in Section \ref{JPmodel} below
that the theory is indeed Galilei invariant.
In what follows, we again take $c=1$.

We first let the constant $\lambda$ be arbitrary.
A self-consistent system is obtained by
adding the matter and Chern-Simons actions,
(\ref{NRCSmatLag}) and (\ref{CSform}), respectively,
\begin{equation}\fbox{$
\cL^{{\small JP}}=
-i\Psi^\star D_t\Psi+ \displaystyle\frac{|\vD\Psi|^2}{2M}
-\frac{\lambda}{2}|\Psi|^4
-\displaystyle
\frac{\kappa}{4}\epsilon^{\alpha\beta\gamma}A_\alpha F_{\beta\gamma}.
$}
\label{NRCSlag}
\end{equation}
Variation w.r.t. $\Psi^\star$ yields the
{\it gauged, non-linear Schr\"odinger equation,}
\begin{equation}\fbox{$
i\partial_t\Psi=
\left[-\displaystyle\frac{\vD^2}{2M}-eA_t
-\lambda\Psi^\star\Psi\right]\Psi\ ,\
$}
\label{gNLS}
\end{equation}
and variation w.r.t. the gauge potential yields
the Chern-Simons equations written in non-relativistic notations,
called the Gauss' law and the field-current identity (FCI), respectively,
\begin{eqnarray}
\kappa\,B
=&{e}\,\varrho\qquad
&\hbox{Gauss},
\label{Gauss}
\\[10pt]
\kappa\, \epsilon_{ij}E_j= &{e}\,J_i\qquad &\hbox{FCI}, \label{FCI}
\end{eqnarray}
where
$\varrho=\Psi^*\Psi$
and
$
\vJ=\frac{1}{2Mi}\big[
\Psi^*\vD\Psi-\Psi(\vD\Psi)^*\big]
$
are the particle density and current, respectively.
Together, they satisfy the continuity equation
$
\p_t\varrho+\vnabla\cdot\vJ=0,
$
as it is shown in the usual way. The mass,
\begin{equation}
{\cal M}=\int\varrho\, d^2\vx,
\label{nontopmass}
\end{equation}
is, therefore, conserved, whenever the integral
converges.

\vskip5mm\goodbreak
\kikezd{Symmetries of non-relativistic vortices}

Before constructing vortex solutions, let us briefly study their
symmetry properties. Let us first remind the reader of the
definition~: a {\it symmetry} is a transformation which interchanges
the solutions of the coupled equations of motion. When applying it
to a solution of the field equations we get, hence, another
solution.

For a Lagrangian system, an infinitesimal space-time symmetry
can be represented by a vector field $X^\mu$ on space-time,
is a symmetry when it changes the Lagrangian by a surface term,
\begin{equation}
\cL\to\cL+\partial_{\alpha}K^\alpha,
\end{equation}
for some function $K$.
To each such transformation, N{\oe}ther's theorem
associates a conserved quantity, namely
\begin{equation}
C=\int\left(\frac{\delta\cL}{\delta(\partial_{t}\chi)}
\delta\chi-K^t\right)d^2{\vx},
\end{equation}
where $\chi$ denotes, collectively, all fields \cite{JaMa}.

The symmetry problem enters the non-relativistic theory
on (at least) two occasions.

Firstly, can we derive self-dual equations using a Bogomol'ny-type
decomposition of the energy~? Do they represent  absolute minima~?
This question is meaningful only
if a conserved energy-momentum tensor has been constructed. The
procedure is canonical in relativistic field theory, but
is rather subtle in the non-relativistic context. The expression
obtained by the usual procedure may not be the most convenient one and may be subject to further ``improvements'' \cite{JaPi2}.

Another (related) application of symmetry
is  the following. Do we have, at the
specific, ``self-dual'' value of the coupling constant,
solutions which are {\it not} self-dual~?
The (negative) answer is obtained  in a single
line, if the conformal symmetry of the theory is exploited \cite{JPDT}.

The Galilean symmetry of our Chern-Simons-theory follows from the
general quantum-field theoretical framework
\cite{Hagen1,Hagen2,JaPi2,SchrG1,SchrG2, SchrG3,MaCha}. (Another
approach \cite{DHP1} will be presented in Section \ref{Bargmann}.)

Each generator of the centrally-extended Galilei group
is associated, through the Noether theorem, with
a conserved quantity and, conversely, each conserved quantity generates
a space-time symmetry.  Jackiw and Pi \cite{JaPi2} construct all these
conserved quantities one-by-one, resorting, from time to time, to ``improvements''. This provides them with the rather
complicated-looking energy-momentum tensor,
\begin{equation}
\begin{array}{lll}
T^{00}&=&\displaystyle\frac{|\vD\Psi|^2}{2M}
-\displaystyle\frac{\lambda}{2}|\Psi|^4,
\\[12pt]
T^{i0}&=&-\displaystyle\frac{1}{2}\left(
(\vD_t\Psi)^*(D_i\Psi)+(D_i\Psi)^*D_t\Psi\right),
\\[12pt]
T^{0i}&=&
-\displaystyle\frac{i}{2}\left(\Psi^*D_i\Psi-(D_i\Psi)^*\Psi\right),
\\[12pt]
T^{ij}&=&-\frac{1}{2}\left(
(D_i\Psi)^*D_j\Psi+(D_j\Psi)^*D_i\Psi
-\delta_{ij}\,|\vD\Psi|^2\right)
\\[12pt]\qquad
&&+ \displaystyle\frac{1}{4}\left(\delta_{ij}\bigtriangleup
-2\p_i\p_j\right)(|\Psi|^2)+\delta_{ij}\,T^{00},
\end{array}
\label{Tmunu}
\end{equation}
whose conservation,
\begin{equation}
\p_\alpha T^{\alpha\beta}=0,
\end{equation}
can be checked by direct calculation.
(See also  \cite{JaPi2,Fund5,DHP1,MaCha}).
The tensor $T^{\alpha\beta}$ is symmetric only in the spatial indices,
\begin{equation}
T^{0i}\neq T^{i0},
\qquad
T^{ij}=T^{ji}.
\label{NRsymEM}
\end{equation}
$T^{ij}$ has been ``improved''
and satisfies, instead of the
tracelessness-condition $T^{\alpha}_{\ \alpha}=0$
of relativistic field theory, the non-relativistic trace condition
\begin{equation}
T^{i}_{\ i}=2T^{00}.
\label{tracecond}
\end{equation}

Turning to our concrete problem, we have the conserved Galilean quantities,
\begin{equation}
\begin{array}{lll}
\cH=&\displaystyle\int T^{00}d^2\vx,&\hbox{\small energy},
\\[10pt]
\cP_i=&\displaystyle\int T^{0i}\,d^2\vx,
\qquad
&\hbox{\small momentum},
\\[10pt]
\cJ=&\displaystyle\int \epsilon_{ij}x^iT^{0j}\, d^2\vx&\hbox{\small angular momentum},
\\[10pt]
\cG_i=&t\cP_i-M\displaystyle\int x_i\varrho\, d^2\vx
\qquad\qquad&\hbox{\small center of mass},
\\[10pt]
{\cal M}=&M\displaystyle\int\!\varrho\, d^2\vx
&\hbox{\small mass},
\end{array}
\label{GalGen}
\end{equation}
What is less expected is that the model admits
 two more conserved generators, namely,
\begin{equation}
\begin{array}{ll}
\cD=t\cH-\frac{1}{2}x_i\cP_i
&\hbox{\small dilatation},
\\[10pt]
\cK=-t^2\cH+2t\cD+\displaystyle\frac{M}{2}\int r^2\varrho\,
d^2\vx
\qquad\quad
&\hbox{\small expansion}.
\end{array}
\label{NRconfGen}
\end{equation}

The NRCS model of Jackiw and Pi share, therefore,  the larger
symmetry found previously for the free Schr\"odinger equation \cite{SchrG1,SchrG2,SchrG3},
as well as for the nonlinear Schr\"odinger equation with
quartic self-interaction potential.

It is worth mentioning that a Poisson bracket of conserved quantities
can be defined \cite{JaPi1,JaPi2,Fund5,HHY}. The resulting Lie algebra structure is that of
the Schr\"odinger group, the non-relativistic ``conformal'' extension of the
Galilei group  \cite{SchrG1,SchrG2,SchrG3}.
The unusual properties of the energy-momentum tensor highlighted
above are,
precisely, hallmarks of Schr\"odinger (rather then
Lorentz) conformal invariance \cite{JaPi2,DHP1, MaCha}.

In the present, CS, context, the energy functional
is,
\begin{equation}
\cH=
\int
\Big(\frac{|\vD\Psi|^2}{2M}-\frac{\lambda}{2}(\Psi^\star\Psi)^2\Big)d^2{\vx}\,.
\label{NRenfunc}
\end{equation}
Applying the Bogomol'ny trick using
 the identity
\begin{equation}
\vert\vD\Psi\vert^2=
\vert(D_1\pm iD_2)\Psi\vert^2
\pm M\vnabla\times\vJ\pm eB\,\varrho,
\label{identity1}
\end{equation}
[seen before as (\ref{SDidentity1})]
the energy (\ref{NRenfunc}) is  written as
\begin{equation}
\cH=
\int\!\left\{\frac{\big|(D_1\pm iD_2)\Psi|^2}{2M}
-\frac{1}{2}\big(\lambda-\frac{e^2}{M|\kappa|}\big)
(\Psi^\star\Psi)^2\right\} d^2\vx\,.
\label{NRBog}
\end{equation}
Thus, the energy is positive definite, $\cE\geq0$, if
\begin{equation}
\lambda\leq\frac{e^2}{M|\kappa|}\ ,
\label{posdefLambda}
\end{equation}
which we assume henceforth (together with
$\lambda>0$).

As already mentioned, additional symmetries play an important role
\cite{JPDT}.
Differentiating the expansion generator ${\cal K}$ in
(\ref{NRconfGen}) twice w.r.t. time shows in fact that
$$
\left(\frac{M}{2}\int r^2|\Psi|^2 d^2\vx\right)''
$$
is twice the Hamiltonian, and is, therefore,  time-independent.
It follows
that for fields that make $|\Psi|^2$ time-independent,
in particular for static fields, the energy {\it must} vanish.
When $\lambda$ takes the specific ``self-dual''
value, $\lambda_{SD}$ in (\ref{specLambda}) below, the energy is, by equation (\ref{NRBog}),
\begin{equation}
\cH=\int\!\frac{\big|D_\pm\Psi|^2}{2M}\, d^2\vx\ ,
\end{equation}
whose vanishing thus implies that  the solution is necessarily self-dual or anti-self-dual.

Let us remind the reader that, in the Abelian Higgs model, the
analogous theorem is rather difficult to prove \cite{Taubes2,Fund3}.

\kikezd{Self-dual non-relativistic non-topological
vortex solutions}

We would like to find static soliton solutions.

 The vacuum is clearly
\begin{equation}
\vA=0,\qquad\Psi=0.
\label{NRvacuum}
\end{equation}
To get finite energy, the following large-$r$
behavior is required~:
\begin{eqnarray}
|\Psi|\to0,\qquad\vD\Psi\to0
\qquad\hbox{\small as}\quad r\to\infty.
\label{finEnCond}
\end{eqnarray}
The second condition here implies
that the finite-energy vortices constructed below are \textit{non-topological}~:
$
\Psi|_\infty:S_\infty\to0.
$

The absolute minimum of the energy, namely zero, is
attained when both terms  in (\ref{NRBog}) vanish.
This requires
\begin{equation}
\begin{array}{lll}
D_+\Psi=0,\qquad\qquad
&\lambda=\displaystyle\frac{e^2}{\kappa M}
\qquad\qquad
&\hbox{\small if}\quad\kappa>0
\\[12pt]
D_-\Psi=0,\qquad
&\lambda=-\displaystyle\frac{e^2}{\kappa M}
\qquad
&\hbox{\small if}\quad\kappa<0
\end{array}
\end{equation}
where $D_\pm=D_1\pm iD_2$. More concisely, setting
 \footnote{The same value
(\ref{specLambda}) has been obtained from the
non-relativistic limit of the self-dual relativistic
theory taking $c=1$ in (\ref{NRCSmatLag}).}
\begin{equation}
\lambda=\lambda_{SD}=\frac{e^2}{M|\kappa|},
\label{specLambda}
\end{equation}
for which the absolute minimum of the energy is attained for {\it
self-dual} or {\it anti-self-dual fields},
\begin{equation}
D_\pm\Psi=0,
\label{NRSD}
\end{equation}
with $+/- =$ sign ($\kappa$).
Note that the Gauss law (\ref{Gauss}) should always be
assumed.

Do we get a static solution of the problem by minimizing
the energy~? Note first that the energy functional
(\ref{NRenfunc}) does not include the time component,
$A_t$, which should be fixed by the field equations.
To answer the question, let us first observe that
for the self-dual Ansatz (\ref{NRSD}), the current is
\begin{equation}
\vJ=\mp\frac{1}{2M}\vnabla\times\varrho,
\label{SDrho}
\end{equation}
cf. (\ref{SDcurrent}).
Using another identity, namely
\begin{equation}
\vD^2\Psi=
(D_-D_+-eB)\Psi
=
(D_+D_-+eB)\Psi,
\label{identity2}
\end{equation}
the static NLS equations can be written,
with the help of the Gauss law, as
\begin{eqnarray}
\left[\displaystyle\frac{D_-D_+}{2M}+\left(\lambda-
\displaystyle\frac{e^2}{2M\kappa}\right)\varrho
+eA_t\right]\Psi=0,\label{NLS-}
\\[8pt]
\left[\displaystyle\frac{D_+D_-}{2M}+\left(\lambda+
\displaystyle\frac{e^2}{2M\kappa}\right)\varrho +eA_t\right]\Psi=0.
\label{NLS+}
\end{eqnarray}
Then for the SD  Ansatz (\ref{NRSD}),
the FCI (\ref{FCI}) becomes, for $\lambda=\lambda_{SD}$,
$$
\vnabla \left(A_t\pm\frac{e}{2M\kappa}\varrho\right)=0,
$$
which is satisfied if
$$
A_t=\mp\frac{e}{2M\kappa}\varrho.
$$
Re-inserting this into (\ref{NLS-}-\ref{NLS+})
shows that the equations of motion are satisfied if \footnote{The sign choice in (\ref{NRCSSD}) is consistent with the one
we found before when we studied the energy.},
in addition to the Gauss law, (\ref{Gauss}) i.e.,
\begin{equation}
\fbox{$\kappa B=e|\Psi|^2\ ,\
$}
\label{Gaussconst}
\end{equation}
we also have,
\begin{equation}\fbox{$
\begin{array}{ll}
D_+\Psi=0\qquad
&\kappa>0
\\[8pt]
D_-\Psi=0\qquad\qquad
&\kappa<0
\end{array}\quad .\
$}
\label{NRCSSD}
\end{equation}
In conclusion, a static solution is obtained for
(\ref{NRCSSD}), provided the Gauss constraint
(\ref{Gaussconst}) holds, and when
the time component of the potential is chosen as
\begin{equation}
A_t=-\frac{e}{2M|\kappa|}\varrho. \label{SDAt}
\end{equation}

The coupled system (\ref{Gaussconst})-(\ref{NRCSSD})
can be solved as follows.
Separating the phase as \hfill\break
$\Psi=\sqrt\varrho\, e^{ie\omega}$, the SD equation is solved by
\begin{equation}
\vA=\pm\frac{1}{2e}\vnabla\times\log\varrho+
\vnabla\omega.
\label{NRvpot}
\end{equation}
Inserting $B=\vnabla\times\vA$ into the Gauss law,
(\ref{Gauss}), tells us that
 $\varrho$ has to solve the \textit{Liouville equation},
\begin{equation}\fbox{$
\bigtriangleup\log\varrho=\mp\displaystyle\frac{2e^2}{\kappa}\varrho,
$}
\label{Liouville}
\end{equation}
with the $-/+$ sign corresponding to $+/-$ in the SD/ASD equations.
Its solutions will be studied in the next Section.
Having solved (\ref{Liouville}), the scalar and
vector potentials are given by (\ref{SDAt}) and
(\ref{NRvpot}), respectively. In the latter, the
phase $\phi$ has to be chosen so that it
cancels the singularities due to the zeros of $\varrho$.
This property is related to the quantization of
the vortex charge, see Section \ref{LiouVort} below.
The vector potential will be given later in (\ref{genvecpot}).

For a self-dual solution, the conserved quantities (\ref{GalGen})-(\ref{NRconfGen}) are readily evaluated \cite{JaPi2}.
All densities are, in fact, expressed using the mass density,
\begin{equation}\left\{
\begin{array}{llll}
m&=&M\varrho
&\hbox{\small mass density},
\\[8pt]
w&=&\frac{1}{4M}\bigtriangleup\varrho
&\hbox{\small energy density},
\\[8pt]
\vp&=&\pm\frac{1}{2}\vnabla\times\varrho
&\hbox{\small momentum density},
\\[8pt]
j&=&
\mp\varrho\pm\frac{1}{2}\vnabla\cdot(\vx\varrho)\qquad\quad
&\hbox{\small angular momentum density},
\\[8pt]
\vec{g}&=&M\vx\varrho
&\hbox{\small boost density},
\\[8pt]
{d}&=&\frac{1}{4}\vx\times\vnabla\varrho
&\hbox{\small dilatation density},
\\[8pt]
{k}&=&\frac{M}{2}r^2\,\varrho
&\hbox{\small expansion density}.
\end{array}\right.
\label{SDcdens}
\end{equation}
Most of the charges including the energy vanish therefore, owing to the rapid fall-off of the density. Notable exceptions are
\begin{equation}\left\{
\begin{array}{llll}
\cM&=&M\displaystyle\int d^2\vx\varrho\qquad\quad
&\hbox{\small mass},
\\[12pt]
\cJ&=&\mp\; \displaystyle\frac{\cM}{M}
&\hbox{\small angular momentum}
\end{array}\right.
\label{SDcQ}
\end{equation}

We note, in conclusion, that a boost and/or an expansion, applied
to the static solution $\Psi_0({\vx})$ of Jackiw and Pi,
produces time-dependent solutions,  namely
\begin{equation}
\Psi(T,{\vx})=\frac{1}{1-kT}\exp\left\{-\frac{i}{2}\Big[
2{\vx}\cdot{\vb}+T\strut {\vb}^2+k\frac{({\vx}+{\vb}T)^2}{1-kT}
\Big]\right\}\,\Psi_0(\frac{{\vx}+{\vb}T}{1-kT}).
\label{expansImp}
\end{equation}

\subsection{Vortex Solutions of the Liouville equation}\label{LiouVort}

Non-relativistic vortices  are constructed out of
the  solutions of the Liouville equation (\ref{Liouville}) \footnote{It is tempting to imagine that these solutions
are the non-relativistic limits, for $|\alpha|=2$,
 of those non-topological relativistic
 pure Chern-Simons vortices of the previous Section.
 As plausible as it sounds, this
has not been rigorously proved yet, however.}.
The general theory shows that
solutions defined over the whole plane arise when the r.h.s.
has a negative sign.
Hence,
the upper sign has to be chosen in (\ref{Liouville}) when $\kappa>0$, and
the lower sign when $\kappa<0$. Note that this is precisely
what comes from our choice of the SD/ASD equation.

The general solution of the Liouville equation
reads
\begin{equation}\fbox{$
\varrho=\displaystyle\frac{4|\kappa|}{e^2}\,
\displaystyle\frac{|f'|^2}{(1+|f|^2)^2},
$}
\label{genLiouSol}
\end{equation}
where $f$ is a meromorphic function of $z=x+iy$.

\vskip2mm
$\bullet$ For the radial Ansatz
\begin{equation}
f(z)=cz^{-N},
\label{CSNRradAns}
\end{equation}
where $c$ is an arbitrary complex parameter,
we obtain, in particular, the radially symmetric
solution
\begin{equation}
\varrho(r)=\frac{4N^2|c|^2|\kappa|}{e^2}
\left(\frac{r^{(N-1)}}{|c|^2+r^{2N}}\right)^2.
\label{radrho}
\end{equation}
The profiles of solution with different $N$ are shown on the Fig. 8

Regularity of $\varrho$ at the origin requires that $N\geq1$.
But then $\log\varrho$ is singular as $r\to0$.
From (\ref{NRvpot}), a regular vector potential, $\vA$, can nevertheless
be obtained if the singularity coming from the first term,
proportional to $\vnabla\times\log\varrho$, is
cancelled, by choosing the [so far undetermined]
phase of $\Psi$ as,
\begin{equation}
\omega=(N-1)\theta,
\label{radvecpot}
\end{equation}
where $\theta$ is the polar angle of the position vector
\cite{JaPi2}. But then the field $\Psi$ is well-defined if
$N$ is an integer at least $1$.

For $N=1$ (Figure  \ref{vortex1}), the origin is
a maximum of $\varrho$; for $N\geq2$, it is a zero~:
the vortex has a ``doughnut-like'' shape, see
Figure  \ref{vortex2-4}. Equation  (\ref{radrho}) shows, furthermore, that the density $\varrho$
[and hence the magnetic field $B$] vanishes at the origin,
$\varrho(0)=0=B(0)$, except for $N=1$. In fact,
due to the Gauss law (\ref{Gauss}),
the magnetic field behaves as,
\begin{equation}
B\sim\varrho\sim r^{2(N-1)}
\qquad r\sim 0.
\end{equation}
\begin{figure}
\begin{center}
\includegraphics[scale=.7]{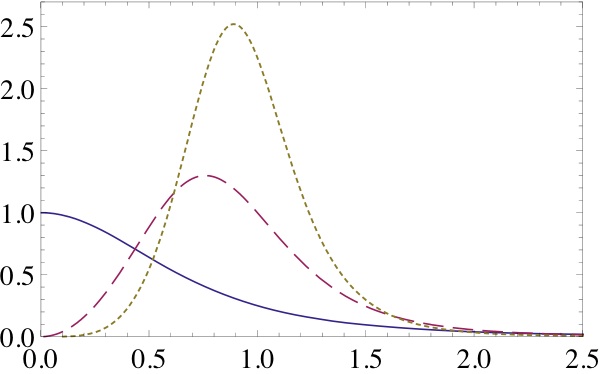}
\end{center}\vskip-5mm
\caption{\it Profiles of radially symmetric JP vortices with winding
numbers $N=1,2,3$ and scales $c=1$. For $n=1$ the vortex is
``hill-shaped'', but for $n\geq2$ it is ``doughnut-like''.}
\label{LrhoN123}
\end{figure}
\begin{figure}
\begin{center}
\includegraphics[scale=0.5]{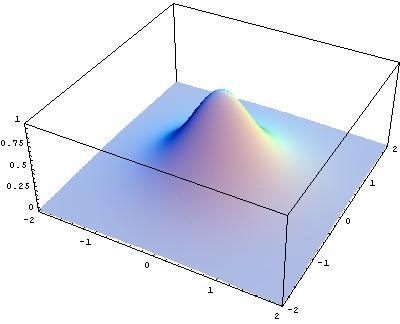}
\end{center}\vspace{-5mm}
\caption{\it The radially symmetric $N=1$
JP vortex is ``hill-like''~: it has a maximum at $r=0$.}
\label{vortex1}
\end{figure}
\begin{figure}\vskip-10mm
\begin{center}
\includegraphics[scale=0.5]{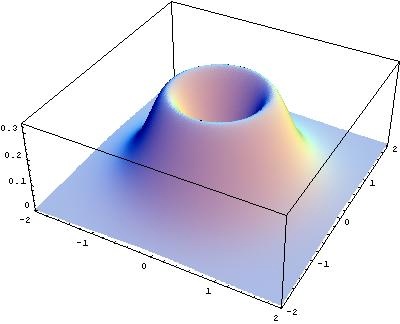}
\quad
\includegraphics[scale=0.5]{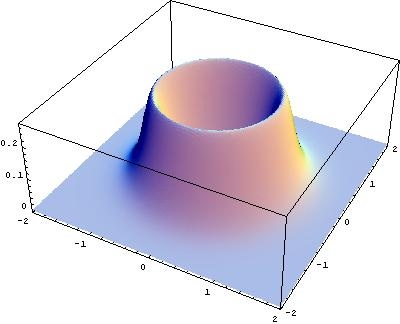}
\end{center}\vspace{-5mm}
\caption{\it For $N\geq2$, the radially symmetric
JP  vortices have a ``doughnut-like'' shape~: the particle density
vanishes at $r=0$. The figure shows the cases with $N=2$ and $N=4$.}
\label{vortex2-4}
\end{figure}
Returning to the general case, we observe that
{\it not} all meromorphic functions yield
physically interesting solutions, however.
The natural requirement is that
the magnetic and scalar fields, $B$ and $\Psi$, must be regular,
and that the magnetic flux,
$$
\Phi=\int\!B\,d^2\vx,
$$
must be finite. For the radial Ansatz (\ref{CSNRradAns}) we find, for example \footnote{In our units
$2\pi\hbar=h=1$.},
\begin{equation}
\Phi=\frac{4\pi\,({\rm sg}\,\kappa)}{e}\,N,
\label{NRCSflux}
\end{equation}
as seen by evaluating the flux integral
directly. Accordingly, the mass is
\begin{eqnarray}
\cM&=&\frac{4\pi M|\kappa|}{e^2}\,N\ .
\label{Nradmass}
\end{eqnarray}

More generally, we can ask, which functions $f$ yield regular,
finite-flux solutions~? How can we calculate the flux~? Is it
quantized~? How many independent solutions do we get for a fixed
value $\Phi$~? \footnote{This question has been studied in
\cite{CSflux,CSindex}. The proof given in \cite{CSflux} is based on
an asymptotic behavior, that is only valid in the radial case,
however. On the other hand, the parameter-counting, given in
\cite{CSindex}, uses an index theorem. Elementary proofs were
obtained in \cite{HY}.}

\kikezd{Theorem 1} \cite{HY}~: \textit{The meromorphic function $f(z)$ yields a regular vortex solution
with finite magnetic flux if, and only if, $f(z)$ is a rational function,
\begin{equation}
f(z)=\frac{P(z)}{Q(z)}
\qquad\hbox{\rm s.t.}\qquad {\rm deg}\, P< {\rm deg}\, Q,
\label{ratfunc}
\end{equation}
where the highest-order term in $Q$ can be normalized to $1$.}
\vspace{2mm}

\begin{figure}
\begin{center}
\includegraphics[scale=0.55]{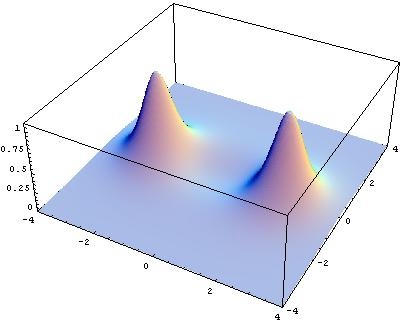}
\end{center}
\vspace{-7mm}
\caption{\it Two separated JP $1$-vortices with charge $N=2$.}
\label{JP1+1}
\end{figure}

\begin{figure}
\begin{center}
\includegraphics[scale=0.55]{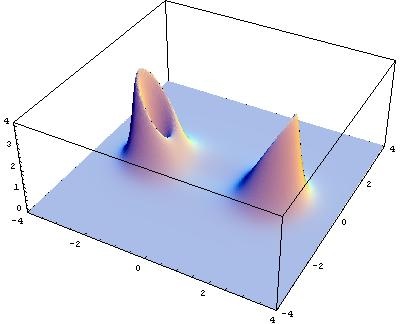}\quad
\includegraphics[scale=0.6]{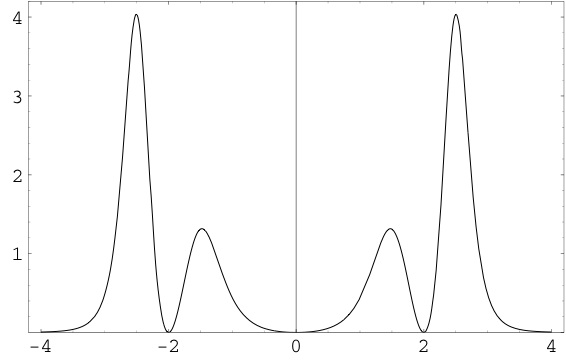}
\end{center}
\vspace{-7mm}
\caption{\it (a) Two separated charge-2 JP vortices
with total charge $N=4$. (b) The cross section of the mass density along the $x$ axis.}
\label{JP2+2}
\end{figure}

\begin{figure}
\begin{center}
\includegraphics[scale=0.55]{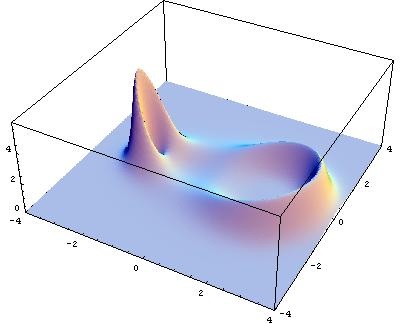}\quad
\includegraphics[scale=0.6]{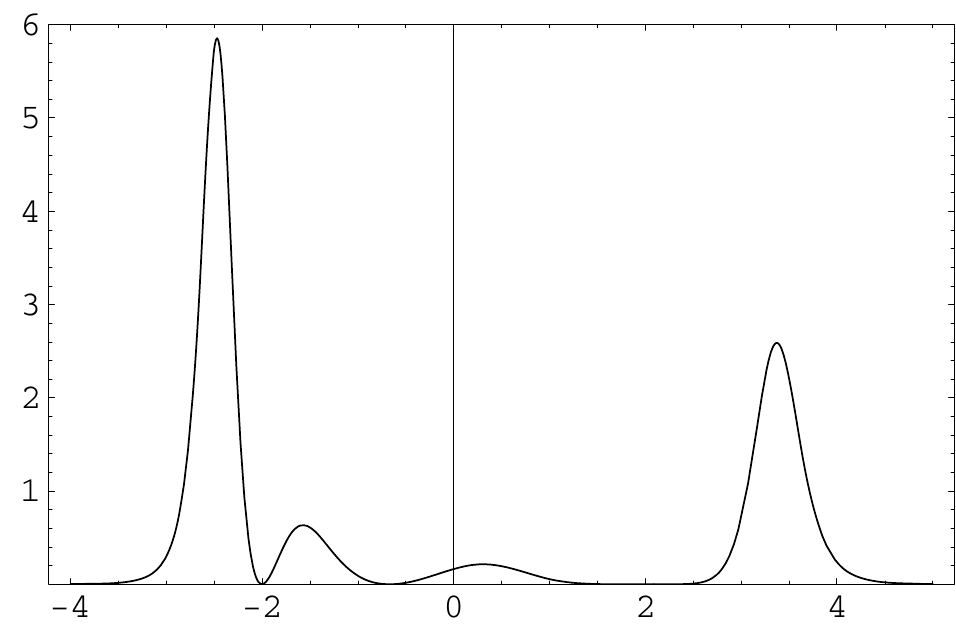}
\end{center}
\vspace{-7mm}
\caption{\it  Separated $2+4$ vortices and cross section along the $x$ axis.}
\label{JP2+4}
\end{figure}

\begin{figure}
\begin{center}
\includegraphics[scale=0.55]{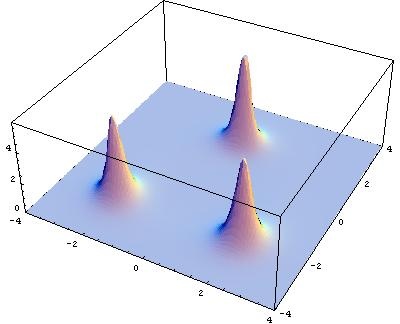}
\end{center}
\vspace{-7mm}
\caption{\it Symmetrically arranged separated $1$ vortices.}
\label{JP1+1+1}
\end{figure}

Various multivortex configurations are presented on Figures \ref{JP1+1}--\ref{JP2+2}--\ref{JP2+4}
and \ref{JP1+1+1}.

In particular, when all roots of $Q(z)$ are simple,
$f(z)$ can be expanded in partial fractions,
\begin{equation}
f(z)=\sum_{i=1}^N\frac{c_i}{z-z_i},
\label{JackAns}
\end{equation}
where the $c_i$ and the $z_i$ are $2n$ complex numbers, and
we get a $4N$-parameter family of
$N$ separated one-vortices \cite{JaPi2}.
Note that this formula
breaks down for superimposed vortices.

The proof \cite{HY} proceeds through a series of Lemmas, and
amounts to
showing that $f$ can only have a finite number of isolated singularities which cannot be essential
either at a finite point, or at infinity. Finally, a theorem
of complex analysis \cite{Ahlfors} says that $f$ is necessarily rational.

The density (\ref{genLiouSol}) is readily seen to be invariant
w.r.t. the transformation,
\begin{equation}
f\to\frac{f+c}{1-\bar{c}f}.
\end{equation}
In particular, taking $c$ to be imaginary and letting
it go to infinity, the density is invariant when
$f$ is changed into $1/f$. Hence, ${\rm deg}\, P\leq {\rm deg}\, Q$ can be assumed.
The case ${\rm deg}\, P={\rm deg}\, Q$ can be eliminated by a
suitable redefinition  \cite{JaPi2}.

\kikezd{Theorem 2} \cite{HY}~: \textit{
The magnetic flux of the solution generated by $P/Q$
is evenly quantized,}
\begin{eqnarray}
\Phi=2N ({\rm sign}\, \kappa)\,\Phi_0,
\qquad
N={\rm deg}\, Q,
\qquad\Phi_0=\frac{2\pi}{e}.
\label{NRfluxquant}
\end{eqnarray}

The proof amounts to showing that only the
roots of the denominator
\begin{equation}
Q(z)=(z-z_1)^{n_1}\dots(z-z_m)^{n_m},
\qquad(\sum_k n_k=N)
\end{equation}
contribute to the charge. The result (\ref{NRfluxquant}) is inferred
by transforming the flux  into a contour integral along the circle
at infinity $C$. The isolated zeros of $Q(z)$, $z_1,\dots,z_m$, are
identified with the ``positions'' of the vortices. Each  can be
encircled  by disjoint circles $C_k$, and the charge comes from
these zeros (see Fig. 15),
\begin{eqnarray}
\Phi=\oint_{C}=\sum_k\oint_{C_k}
=\sum_k n_k\left(({\rm sg}\,\kappa)\frac{4\pi}{e}\right)=
2N({\rm sg}\,\kappa)\Phi_0.
\end{eqnarray}
\begin{figure}
\begin{center}
\includegraphics[scale=0.45]{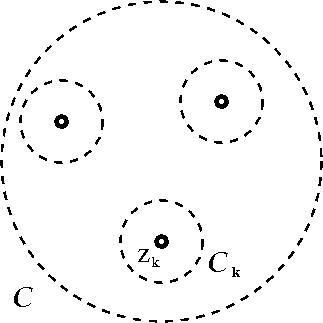}
\end{center}\vspace{-3mm}
\caption{\it The charge of a vortex is proportional to the sum of the
multiplicities of the zeros of the denominator $Q(z)$ in
(\ref{ratfunc}).}
\label{vcharge}
\end{figure}

Let us fix $N={\rm deg}\, Q> {\rm deg}\, P$.

\kikezd{Theorem 3.}~: \textit{
The solution generated by (\ref{ratfunc})  depends on $4N-1$ real parameters,
 where $N$ is the degree of the denominator $Q(z)$}.
\vspace{3mm}

The proof follows at once from Theorem 1~: $N$ is the degree
of the denominator, $Q(z)$, which, being normalized, has
$N$ complex coefficients. Since deg $P<$ deg $Q$, the numerator
also has $N$ complex coefficients (some of which can vanish).
The subtraction of $-1$ comes from noting that,
by (\ref{genLiouSol}), the general phase of $f$ is irrelevant, so that the highest coefficient of
$P$ can be chosen to be real.

\begin{figure}
\begin{center}
\includegraphics[scale=0.66]{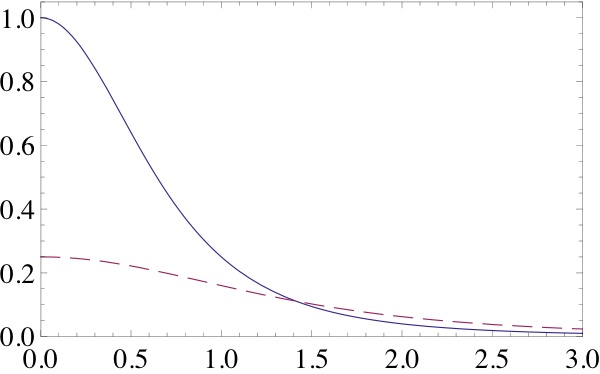}\quad
\includegraphics[scale=0.66]{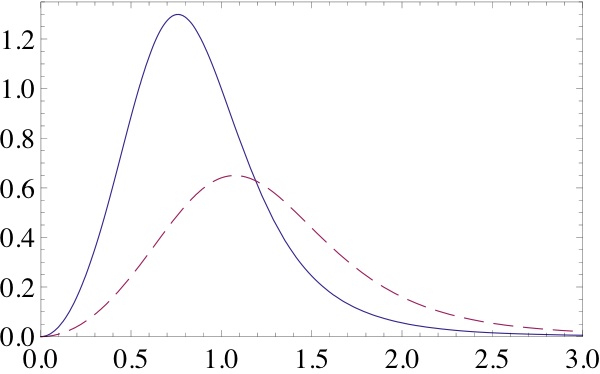}
\end{center}
\vspace{-4mm}
\caption{\it Changing the scale parameter of a
vortex ``smears it out'', whereas the
position and total charge remain unchanged.
This is illustrated on (a) charge-$1$
and (b) charge-$2$ vortices and scales
$c=1$ and $c=2$.}
\label{LrhoN1-2c1-2}
\end{figure}

The only degrees
of freedom of
 ``Nielsen-Olesen-Bogomol'ny'' vortices are their positions
 \cite{Weinberg,Taubes1,Fund3}~:
prescribing the zeros of the Higgs field
uniquely determines a self-dual vortex solution.
Chern-Simons vortices have
additional degrees of freedom. Prescribing the zeros
of the denominator $Q(z)$ in (\ref{ratfunc}) determines
the ``positions''. The coefficients of the polynomial $P(z)$ in the numerator are, however, still at our disposal.
In the widely separated case (\ref{JackAns}) we have,
for example, the complex numbers $c_i$. These can
be interpreted as {\it scales}. Changing them has no
influence on either on the location, or on the total charge of a vortex, but it does affect its {\it shape}.
Increasing $c$, for example, ``smears out'' the vortex, as
illustrated in Figure  \ref{LrhoN1-2c1-2}.

The Liouville equation (\ref{Liouville}) is singular
 where $\varrho$ vanishes, since
$\log\varrho\sim-\infty$ at such a point.
Equation  (\ref{Liouville}) is, therefore,
only  valid \textit{away of the zeros},
$z\neq z_j$, of $\varrho$.
If $z_j$ has multiplicity $N_j$, then,
in the neighborhood of a zero,
$\varrho(z_j)=0$, the  density behaves as
\begin{equation}
\varrho\sim|z-z_j|^{2(N-1)}.
\label{smallrJP}
\end{equation}
From this we infer that
equation (\ref{Liouville}) is \textit{singular}
when $N_j\geq2$. In fact,
\begin{equation}\fbox{$
\bigtriangleup\log\varrho\mp\displaystyle\frac{2e^2}{\kappa}\varrho=-4\pi\sum_j(N_j-1)\delta(z-z_j)\ ,
$}
\label{Liouvilledelta}
\end{equation}
since
$
\bigtriangleup\log r=-2\pi\,\delta(\vx)
$
in the plane.

\kikezd{The geometry of vortex solutions}.

These above results have a rather elegant
geometric interpretation \cite{topnontop}.
A rational function,
\begin{equation}
w=\frac{P(z)}{Q(z)}=\frac{a_mz^m+\dots+a_0}
{b_nz^n+\dots+b_0},
\label{ratfunc2}
\end{equation}
($a_m,b_n\neq0$) always has a limit
as $z\to\infty$, namely $\infty$ if
$m={\rm deg}\,P>{\rm deg}\,Q=n$,
$a_m/b_n$ if $m=n$, and
zero, if $m<n$. therefore,  such     function extends as a mapping,
still denoted by $f$, between the Riemann spheres,
\begin{equation}
f~:S_z\to S_w,
\end{equation}
 obtained by compactifying the complex
$z$ and $w$-planes by adding the point at infinity. Then
both $z$ and $w$ are stereographic coordinates.
 The $w$-sphere
carries, in particular, the canonical surface form
\begin{equation}
\Omega=2i\frac{dw\wedge d\bar{w}}{(1+w\bar{w})^2}.
\label{surf}
\end{equation}
Using the Gauss law $B=({e}/{\kappa})\varrho$,
the magnetic flux of the vortex  is, therefore,
\begin{eqnarray}
\Phi=\frac{{\rm sg}\,\kappa}{e}\displaystyle\int
\frac{4|f'|^2}{(1+|f|^2)^2}d^2z
=
\frac{2({\rm sg}\,\kappa)}{e}\displaystyle\int_{S_z}\!f^*\Omega,
\label{VortTopcharge}
\end{eqnarray}
where we recognize the \textit{topological charge}
of monopole theory \cite{monoptop1,monoptop2,monoptop3}. The integral in
(\ref{VortTopcharge}) is, in fact,
the same as the homotopy class of the mapping $f:~S_z\to S_w$.

Equivalently, the magnetic charge is the \textit{Brouwer degree} of $f$
[which is the number of times the image is covered].

Generalizing (\ref{radvecpot}),
 the regularity of the vector
potential requires choosing the phase $\phi$ so that \cite{HY}
\begin{equation}
(\p_x-i\p_y)\phi=\sum_{i=1}^{N_Q}\frac{n_i-1}{z-z_i}+
\sum_{k=1}^{N_P}\frac{n_k+1}{z-Z_k},
\label{genvecpot}
\end{equation}
where the $z_i, \,i=1,\dots n_Q$ are the distinct roots of the denominator
$Q(z)$ and  $n_i$ their respective multiplicities,  so that
$\sum_{i=1}^{N_Q}n_i=\hbox{deg}\,Q=N$ is the vortex number. The
$Z_k;\ k=1,\dots N_P$, are the roots of the numerator,
 $m_k$ their multiplicities , and
$\sum_{k=1}^{N_P}m_k=\hbox{deg}\,P<N$.

Let us stress, in conclusion, that the vortices constructed in this
section are rather different from those found before in the Abelian
Higgs model
 \cite{NiOl,Bogo,Taubes1,Fund3}.

First of all, they are \textit{non-topological}~;
the quartic potential is attractive and, therefore,
the scalar field has to vanish at large distances.
In other words, the field goes to the symmetric
``empty vacuum'' $\vA=0,\Psi=0$, as $r\to\infty$.
The gauge symmetry is, therefore,  {\it not} spontaneously broken at infinity. The magnetic flux
is, nevertheless, quantized, namely in even units,
cf. (\ref{NRfluxquant}).

They
also carry an electric charge. Hence, they are realizations of
 the quasiparticles and quasiholes, proposed
by Laughlin \cite{QHE} to explain the FQHE,
the main physical application of Chern-Simons gauge theory.

Remarkably, the self-dual solutions
of the $\O(3)$ non-linear sigma model \cite{BePo} are,
once again, precisely those described here.

\newpage

\section{NON-RELATIVISTIC VORTICES IN EXTERNAL FIELDS}\label{background}

\subsection{Topological Vortices in the ZHK Model}\label{ZHKvort}

Let us now return to the physically relevant
model of Zhang et. al. \cite{ZHK}, presented in Section \ref{LGQHE}. Roughly speaking, it is the non-relativistic counterpart of the (relativistic)
Abelian Higgs model with Chern-Simons term
and symmetry
breaking potential discussed in Section
\ref{PaulKhare}, -- but with  pure
CS interaction, as in Section \ref{J(L)W}.

Let us assume that all fields are time independent.
The static field equations read,
\begin{eqnarray}
ea_t\psi&=&-\frac{1}{2}\vD^2\!\psi-\frac{\lambda}{2}(1-|\psi|^2)\psi,
\label{ZHKNLSb}
\\[6pt]
\kappa\,\epsilon_{ik}\,\p_ka_t&=&
\frac{e}{2i}\big(\psi^*D_i\psi-\psi(D_i\psi)^*\big),
\label{ZHKFCIb}
\\[7pt]
\kappa (\p_1a_2-\p_2a_1)&=&e|\psi|^2,
\label{ZHKGaussb}
\end{eqnarray}
where the covariant derivative involves
the  total gauge field,
\begin{equation}
\vD=\vnabla-ie\vA,\qquad
\vA=\va-\vcA.
\label{Bcovder}
\end{equation}
Here $\va$ and $\vcA$ are the statistical and the
background gauge potentials, respectively.

Choose the gauge $\cA_0=0$.
In the absence of an external field, the system has no static and
uniform solution \cite{EHI91/1}.
 For
 $|\psi|^2=\varrho_0=\const$. and $\cB=0$, equation (\ref{ZHKGaussb}) implies that the statistical gauge
potential is (minus) the familiar expression in a constant magnetic field,
$$
a_i=-\frac{e\varrho_0}{2\kappa}\,\epsilon_{ij}x_j.
$$
The wave function can be chosen to be real
without any loss of generality, $\psi=\sqrt{\varrho_0}$,
since this can always be
achieved by a gauge transformation.
But then (\ref{ZHKFCIb}) allows us to infer  the temporal component,
$$
\epsilon_{ij}\,
\big(\p_ja_t-\frac{e^3\varrho_0^2}{2\kappa^2}x_j
\big)=0
\quad\Rightarrow\quad
a_t=\frac{e^3\varrho_0^2}{4\kappa^2}\,r^2+\const.
$$
Then (\ref{ZHKNLSb}) requires
$$
\psi\Big[\const+\frac{e^4\varrho_0^2}{8\kappa^2}r^2+
\frac{\lambda}{2}(1-\varrho_0)\Big]=0.
$$
But  the bracketed expression
is a function of the position,
so that necessarily $\psi\equiv0$.

The proof is plainly valid for
\textit{any} potential $U$.
Thus,
symmetry breaking is induced by the external field and
{\it not} by the nonlinear potential.

In a nonzero magnetic field,  $\cB\neq0$, we have two types of
uniform background solutions. One of them is
symmetry preserving, and the other is
symmetry breaking \cite{EHI91/1}.
\begin{equation}
\psi=0,
\qquad
a_0=\const,
\qquad
\va=0
\end{equation}
is a symmetric ground-state, and
\begin{equation}
\psi^{(0)}=\sqrt{\varrho_0},\quad
\varrho_0=\frac{\cB\kappa}{e},
\quad
a_t^{(0)}=\frac{\lambda}{2e}(\varrho_0-1),
\quad
a_i^{(0)}=-\frac{e\varrho_0}{2\kappa}\epsilon_{ij}x_j
=-\frac{\cB}{2}\epsilon_{ij}x_j
\label{ZHKvaca}
\end{equation}
is symmetry breaking one.
This is physically admissibly only if the density is positive,
 $\varrho_0>0$, so that the external magnetic field, $\cB$, the Chern-Simons coefficient,
$\kappa$, and the charge, $e$,
have to satisfy $\kappa\cB/e>0.$

The energy of a static configuration is \footnote{The Gauss constraint (\ref{ZHKGauss}),  is always assumed.}
\begin{equation}
\cE=\int\!\left\{
\frac{1}{2}|\vD\psi|^2+U(|\psi|^2)\right\}d^2\vx\ ,
\qquad
U(\varrho)=\frac{\lambda}{4}(1-\varrho)^2.
\label{ZHKen}
\end{equation}
It only differs from the energy in the
Abelian Higgs model in the absence of
the magnetic term $\2B^2$.

The energy  of the configuration
(\ref{ZHKvaca}), $\displaystyle\int U(\varrho_0)d^2\vr$, is finite only if the potential has its
absolute minimum, zero, at $\varrho=\varrho_0$. In the present case, this requires
\begin{equation}
\varrho_0=1\qquad\hbox{\small i.e.}\qquad\cB=\frac{e}{\kappa},
\label{bBkappa}
\end{equation}
which we assume henceforth.

It is convenient to re-express everything using the
total fields. Firstly, the ground state (\ref{ZHKvaca})
is,
\begin{equation}
\psi^{(0)}=1,
\qquad
A_0^{(0)}=0,
\qquad
\vA^{(0)}=0.
\label{ZHKvac}
\end{equation}
Next,  the static equations of motion
(\ref{ZHKNLSb})-(\ref{ZHKFCIb})-(\ref{ZHKGaussb}), read, in terms of the total fields,
\begin{eqnarray}
eA_t\psi&=&-\frac{1}{2}\vD^2\!\psi-\frac{\lambda}{2}(1-\varrho)\psi,
\label{ZHKstat1}
\\[6pt]
\kappa\,\epsilon_{ik}\,\p_kA_t&=&ej_i,
\label{ZHKstat2}
\\[7pt]
\kappa B&=&e(\varrho-1),
\label{ZHKstat3}
\end{eqnarray}
where the density and current are the by now familiar non-relativistic expressions,
$\varrho=\psi^*\psi$
and
$\vec{\jmath}=\frac{1}{2i}[\psi^*\vD\psi-\psi(\vD\psi)^*],
$
with the covariant derivative,
$\vD=\vnabla-i\vA$,
involving the total gauge field.
As a result of trading the Maxwell dynamics for a Chern-Simons one, the equations
(\ref{ZHKstat1})-(\ref{ZHKstat2})-(\ref{ZHKstat3})
differs substantially
from those
valid in the Abelian Higgs model. They also differ from their
relativistic counterparts, discussed in Section \ref{PaulKhare} and \ref{J(L)W}.

Let us  consider a general static configuration.
To get finite energy, we require that,
\begin{equation}
\vD\psi\to0,
\qquad |\psi|\to1,
\qquad\hbox{\small as}\quad r\to\infty.
\label{ZHKinfcond}
\end{equation}
The convergence should, moreover, be rapid enough to
make the energy integral (\ref{ZHKen}) converge.
The asymptotic conditions  (\ref{ZHKinfcond}) imply, just like in the Abelian-Higgs
case, that the total magnetic flux is quantized,
\begin{equation}
\int B\,d^2\vx=\Phi=\frac{2\pi}{e}\,n,
\label{ZHKflux}
\end{equation}
where $n$ is the winding number of the scalar field.
A related quantity, typical for the non-relativistic theory, is the {\it particle} or \textit{vortex number},
\begin{equation}
{\cal N}\equiv\int\big(\varrho-1\big)\,d^2{\vx}.
\label{vortexnumber}
\end{equation}
Assuming that the integral converges,
${\cal N}$ is conserved, since the current satisfies the continuity equation,
\begin{equation}
\partial_t(\varrho-1)+\vnabla\cdot\vj=0,
\end{equation}
as it can be shown using the equations of motion.
Using the Gauss constraint, (\ref{ZHKstat3}),
 the vortex number can be linked to the total
 magnetic flux,
\begin{equation}
{\cal N}=\frac{\kappa}{e}\,\Phi^{(total)}=
\frac{\kappa}{e}\int B\,d^2\vx.
\end{equation}
The assumptions of finite flux and finite
vortex number are, hence, equivalent.

Radially symmetric, topological solutions can be constructed along the lines
considered before. Let us start with the usual radial Ansatz
(\ref{PKradAns}) for the statistical field, i.e.,
\begin{equation}
A_t=nA_0(r),
\qquad
A_r=0,
\qquad
A_\theta=nr\,A(r),
\qquad
\psi(r)=f(r)e^{in\theta},
\label{ZHKradAns}
\end{equation}
whose winding number is $n$. The handy formulae
(\ref{PKmetricquant})
provide us with the radial equations
\begin{eqnarray}
f''+\displaystyle\frac{f'}{r}-e^2n^2f
\big(A-\displaystyle\frac{1}{er}\big)^2+\lambda(1-f^2)f+2enfA_0=0,
\label{ZHKradf}
\\[9pt]
-e^2f^2\big(A-\displaystyle\frac{1}{er}\big)+\kappa A_0'=0,
\label{ZHKradA}
\\[8pt]
e(1-f^2)+n\kappa\big(A'+\displaystyle\frac{A}{r}\big)=0.
\label{ZHKradA0}
\end{eqnarray}
Note that (\ref{ZHKradf}) only differs from
 (\ref{PKradf}), valid in the Paul-Khare case, in that
 the last term here is {\it linear} in $A_0$,
 while the corresponding one in  (\ref{PKradf})
is quadratic. This changes the large-$r$ behavior, see later.
(\ref{ZHKradA}) is identical to
(\ref{CSrelradeqA}), and (\ref{ZHKradA0})
only slightly differs from (\ref{CSrelradeqA0})
in the pure Chern-Simons model of Section
\ref{J(L)W}.

The energy of such a static, radial configuration is,
\begin{equation}
\cE=\pi\int_0^\infty\left\{(f')^2+
e^2n^2f^2\Big(\frac{1}{er}-A\Big)^2+2U(f^2)
\right\}rdr.
\label{ZHKraden}
\end{equation}

For small $r$ easy calculation yields the leading behavior
\begin{equation}\begin{array}{lll}
f&=&f_nr^{|n|}+\displaystyle\frac{(-\kappa\lambda+e|n|
(e-2\kappa b_0)f_n}{4(|n|+1)^2}\,r^{|n|+2}+
O(r^{|n|+4}),
\\[14pt]
A&=&\displaystyle\frac{e}{2|n|\kappa}\,r+
\displaystyle\frac{ef_n^2}{2|n|(|n|+1)}\,r^{2|n|+1}+
O(r^{2|n|+3}),
\\[14pt]
A_0&=&b_0-\displaystyle\frac{ef_n^2}{2\kappa|n|}
r^{2|n|}+O(r^{2|n|+2}),
\end{array}
\label{KZHsmallrExp}
\end{equation}
where $f_n$ and $b_0=A_0(0)$ are free parameters.

Due to the asymptotic conditions (\ref{ZHKinfcond}),
our solution tends, at infinity, to the symmetry-breaking vacuum (\ref{ZHKvac}),
\begin{equation}
f(r)\to1,
\qquad
A_0\to0,
\qquad
A-\frac{1}{er}\to0,
\qquad\hbox{\small as}\quad
r\to\infty.
\end{equation}
The large-$r$ behavior reveals some surprises, however \cite{EHI91/2,Tafel}. Setting,
\begin{eqnarray}
f=1-\varphi,
\qquad
a=A-\frac{1}{er},
\qquad
A_0=a_0,
\end{eqnarray}
yields indeed the system
\begin{eqnarray}
\varphi''+\frac{\varphi}{r}-2\lambda\varphi-2ena_0=0,
\label{ZHKvarphi}
\\[6pt]
a_0'-\frac{e^2}{\kappa}a=0,
\label{ZHKa0}
\\[8pt]
a'+\frac{a}{r}+\frac{2e}{n\kappa}\,\varphi=0.
\label{ZHKa}
\end{eqnarray}

The first surprise is that the scalar field
equation does {\it not} decouple.
The reason is that while, in (\ref{PKradf}), the term {\it quadratic} in $A_0$ could be dropped,
the one here is {\it linear} in  $a_0$, and
must be kept therefore.
The system (\ref{ZHKvarphi}-\ref{ZHKa0}-\ref{ZHKa})
can, nevertheless, be solved exactly \cite{Tafel}.
Our previous experience tells us that
the scalar field and the time component of the
vector potential behave similarly. Let us assume, therefore, that they are proportional,
\begin{equation}
2en a_0=-\,C\,\varphi,
\label{ZHKa0Ansatz}
\end{equation}
where $C$ is some constant [Sign and coefficients have been chosen for later convenience]. Inserting this Ansatz into (\ref{ZHKvarphi}) yields, once again, Bessel's
equation of order zero,
\begin{equation}
\varphi''+\frac{\varphi}{r}-(2\lambda-C)\varphi=0
\qquad\Rightarrow\qquad
\varphi\propto K_0(\sqrt{2\lambda-C}\,r).
\label{ZHKvarphi2}
\end{equation}
Next, deriving (\ref{ZHKa0Ansatz}) and inserting into
(\ref{ZHKa0}) allows us to express $a$ through $\varphi'$,
\begin{equation}
a=-\frac{\kappa c}{2e^3n}\,\varphi'.
\label{ZHKavarphi}
\end{equation}
Deriving once more and inserting into the last equation,
(\ref{ZHKa}), yields
$$
\varphi''+\frac{\varphi}{r}-\frac{4e^4}{\kappa^2C}\varphi=0,$$
which is consistent with (\ref{ZHKvarphi2}), provided
\begin{equation}
\frac{4e^4}{\kappa^2C}=2\lambda-C
\quad\Rightarrow\quad
C=\lambda\pm\sqrt{\lambda^2-\frac{4e^4}{\kappa^2}}.
\label{Cvalue}
\end{equation}
Hence
\begin{equation}
\varphi(r)\propto K_0(\sqrt{\mu}\,r),
\qquad
\mu=\lambda\mp\sqrt{\lambda^2-\frac{4e^4}{\kappa^2}}.
\label{varphivalue}
\end{equation}

Then $a_0$ is again a Bessel function of order zero,
due to (\ref{ZHKa0Ansatz}). Using ${K_0}^\prime=-K_1$,
(\ref{ZHKavarphi}) allows us to conclude that
\begin{equation}
a(r)\propto K_1(\sqrt{\mu}\,r).
\label{ZHKa2}
\end{equation}

Remarkably, the single parameter $\mu$
controls the large-$r$ behavior of {\it all} fields.
 This follows from the field equations. (\ref{ZHKstat3})
shows that the magnetic and the scalar fields
must decay in the same way; then (\ref{ZHKstat1})
extends this property to the time component
\footnote{
In contrast, the Gauss law in the relativistic
pure CS
model of Jackiw-(Lee)-Weinberg, (\ref{CSrelconstr}),
involves both $A_0$ and $\psi$,
allowing for different decay rates of the
fields.}. Remember that in  our previous
models, all fields drop off with the same characteristic length in the self-dual case only.

Physically, $\sqrt{\mu}$ is
the mass parameter. We have now two possibilities:
\begin{enumerate}
\item If
\begin{equation}
\lambda\geq\frac{2e^2}{|\kappa|}\ ,
\label{biglambda}
\end{equation}
the quantity below the inner square root in
(\ref{varphivalue}) is non-negative, so that
$\mu$ is real and is in fact positive, due
to our assumption $\lambda\geq0$. It is, hence, the mass of the fields. The latter
 takes, however, two values, corresponding
to the two signs,
\begin{equation}
m_\pm=\sqrt{\mu_\pm}=\sqrt{\lambda\mp\sqrt{\lambda^2-\frac{4e^4}{\kappa^2}}}\ .
\label{realZHKmass}
\end{equation}
The field decays as
\begin{equation}
f(r)\sim \frac{c}{\sqrt{r}}e^{-m_\pm r}.
\end{equation}
Thus we have,
just like in the (relativistic) Paul-Khare case, {\it mass splitting} -- except when the second square root is zero, so that $m_+=m_-$.
Remarkably, this happens precisely in the critical case
$\lambda=\lambda_{SD}$, see below.

\item If, however,
\begin{equation}
0\leq\lambda<\frac{2e^2}{|\kappa|}\,,
\label{smalllambda}
\end{equation}
then $\mu$ takes two (conjugate) complex values, and
\begin{equation}
\sqrt{\mu}=\sqrt{\frac{e^2}{|\kappa|}+
\frac{\lambda}{2}\,}
\;\mp\; i\,\sqrt{\frac{e^2}{|\kappa|}-\frac{\lambda}{2}\,}\ .
\label{complexmass}
\end{equation}
The (positive) real part yields  the physical mass,
\begin{equation}
m=\sqrt{\frac{e^2}{|\kappa|}+\frac{\lambda}{2}}\,.
\end{equation}

For complex values of its argument, the Bessel function oscillates, with the imaginary part of $\sqrt{\mu}$
as frequency. Taking a real combination,
\begin{equation}
f(r)\sim \frac{c}{\sqrt{r}}e^{-m r}
\cos \left[\big(\frac{e^2}{|\kappa|}-\frac{\lambda}{2}\big)^{1/2}r+\hbox{(shift)}\right].
\label{oscif}
\end{equation}
\item
At the critical value
\begin{equation}
\lambda=\lambda_{SD}=\frac{2e^2}{|\kappa|}\ ,
\label{critlambda}
\end{equation}
all masses coincide,
\begin{equation}
m^2=m_{SD}^2=\frac{2e^2}{|\kappa|}\ .
\label{communZHKmass}
\end{equation}
This is the point
when real mass splitting becomes oscillation.
\end{enumerate}

Numerical calculations \cite{Tafel} (see also \cite{EHI91/2,EHI91/1}) confirm these
investigations -- up to mass splitting, which
seems to be absent. Are both masses physical when
$\lambda\neq\lambda_{SD}$~? The situation is not entirely clear, but it may well happen that,
like in the Paul-Khare case, only
one of the masses can arise for finite-energy,
regular solutions.

\begin{figure}
\begin{center}
\includegraphics[scale=1.1]{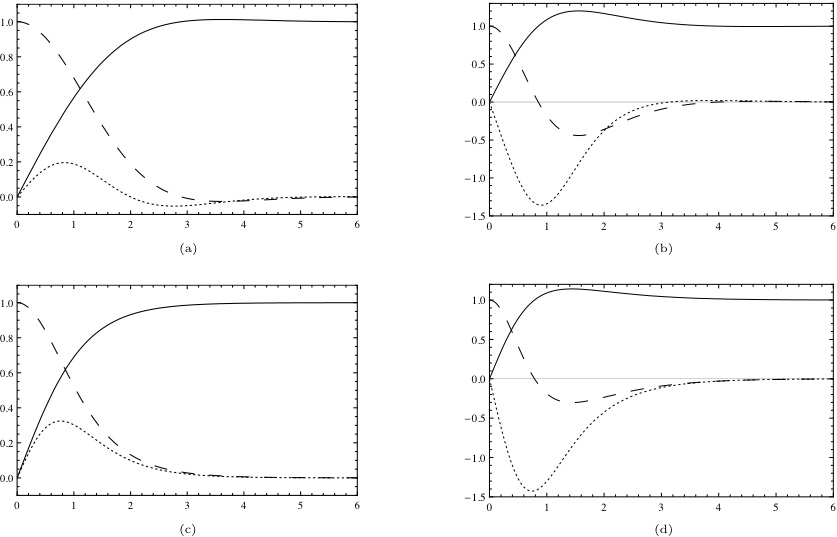}
\caption{{\it ZHK vortices.
(a) $\lambda=0$ and n=1;
(b) $\lambda=0$ and n=-1;
(c) $\lambda=\lambda_{SD}$ and n=1
(SD case)
(d) $\lambda=\lambda_{SD}$ and n=-1.
The heavy/dashed/solid lines refer to the scalar/magnetic/electric fields.}}
\label{ZHKvfig}
\end{center}
\end{figure}

The oscillatory behavior is confirmed numerically, see Fig.
\ref{ZHKvfig}.

In the range (\ref{smalllambda}) of the coupling constant, the solution indeed oscillates.
The relation \cite{Fund3},
\begin{equation}
B\leq1-|\Psi|^2,
\end{equation}
 which implies $|\Psi|\leq1$ and hence the Meissner effect,
is not more valid here, and we
 may have, in particular, a ``shoulder'' when
 $\lambda$ is below the critical value,
 see Fig. \ref{ZHKvfig}.

Another surprise is the {\it asymmetry} in the
winding number~: solutions with $n$ and $-n$
may behave substantially differently.
Donatis and Iengo \cite{DoIe} attribute this
to the breaking of chiral symmetry, P (T), by the
Chern-Simons term. Changing $n$ into $(-n)$ amounts
to performing a parity transformation in the plane,
\begin{equation}
\theta\to\pi-\theta,
\label{parity}
\end{equation}
which doe {\it not} leave the Chern-Simons term
invariant.
The asymmetry in the winding number is  manifest when the coupling
constant takes
the particular value $\lambda=\lambda_{SD}$. Then, choosing the
right sign of $n$, we have the {\it self-dual} solution
constructed below and shown on  Fig. \ref{ZHKvfig} (c). Numerical
integration confirms that its energy is $\cE_c\approx \pi$,
consistently with analytical proof presented below.
Choosing the ``wrong'' sign for $n$,
however, we get some other, {\it non-self dual} solution,
Fig. \ref{ZHKvfig} (d), whose energy, obtained again by numerical integration, is
{\it substantially higher}, $\cE_d\approx 8.670$ \footnote{For completeness, we give
the energies of the other numerical solutions in Fig. \ref{ZHKvfig}~:
$\cE_a\approx1.556$ and $\cE_b\approx 7.611$.}.

\goodbreak
\kikezd{Self-dual solutions}

The critical value (\ref{critlambda}) which separates the simple-exponential-decay regime from the oscillating one, has particular interest. To see this,
let us consider the energy, (\ref{ZHKen}).
Remember the identity (\ref{identity1}), i.e.,
$$
\vert\vD\psi\vert^2=
\vert(D_1\pm iD_2)\psi\vert^2
\pm eB\,\varrho\pm \vnabla\times\vj.
$$
Dropping the surface term and using the Gauss law,
the energy density can be written, therefore, as
\begin{equation}
\2\vert(D_1\pm iD_2)\psi\vert^2
+\frac{\kappa^2}{4e^2}\Big(\lambda
\pm \frac{2e^2}{\kappa}\big)B^2\pm\2eB.
\label{ZHKendec}
\end{equation}
The integral of the last term is
proportional to the \textit{total flux},
$$
\2e\!\int\!B\,d^2\vx=\2e\,\Phi^{(total)}\ ,
$$
assumed to be finite. By the Gauss law, this is also
 proportional to the \textit{vortex number}
 (\ref{vortexnumber}),
$$
{\cal N}=\int(\rho-1)\,d^2\vx=\frac{\kappa}{e}\,
\Phi^{(total)}.
$$
Note that the sign of ${\cal N}$ is the sign of the flux, multiplied by the sign of $\kappa$.

For the specific value
$$
\lambda=\mp\frac{2e^2}{\kappa},
$$
the $B^2$ term drops out, leaving us with
\begin{equation}
\cE=\underbrace{\int\2\vert(D_1\pm iD_2)\psi\vert^2d^2\vx}_{\geq0}
\;\pm\;
\2e\,\Phi^{(total)}.
\label{ZHKbogdec}
\end{equation}

Thus, we get two lower bounds: one of them positive, the other
negative. Since the energy is non-negative,  only the positive one
is relevant here; the other one can obviously not be saturated.
Which is the ``good'' choice of sign~? The one we get with the upper
sign for positive flux, and the one with the lower sign when the
flux is negative. But by (\ref{ZHKflux}) this depends on the winding
number, $n$.

Hence, while
numerical evidence (Fig. \ref{ZHKvfig} (d)) suggests that, for given
$\kappa$, vortex solutions {\it do} exist for
all winding numbers, only {\it one} of them
supports a SD (or ASD) solution, which minimizes
the energy \cite{DoIe}.
Vortices and antivortices behave, hence, differently
 \cite{DoIe}.

The coefficient $\lambda$ of the symmetry-breaking potential
should be non-negative, whereas the Chern-Simons coupling constant
can be negative or positive. The correct choice to suppress
the $B^2$ term in (\ref{ZHKendec}) is, therefore,
\begin{equation}
\lambda=\lambda_{SD}=\frac{2e^2}{|\kappa|}=
\left\{\begin{array}{ccc}
\displaystyle\frac{2e^2}{\kappa}\qquad&\hbox{\small for}
\qquad&\kappa<0,
\\[12pt]
-\displaystyle\frac{2e^2}{\kappa}\qquad
&\hbox{\small for}\qquad&\kappa>0.
\end{array}\right.
\label{ZHKSDlambda}
\end{equation}
which is, once again, the critical value (\ref{critlambda}).
With such a choice,
the energy has a positive Bogomol'ny bound,
\begin{equation}
\cE\;\geq\; \cE_{Bog}=
\2e|\Phi|=\pi e|n|,
\end{equation}
which is attained for SD/ASD fields,
 \begin{equation}
\left\{\begin{array}{cccc}
D_+\psi=0, \qquad&n>0&\hbox{\small for} \qquad&\kappa<0,
\\[12pt]
D_-\psi=0, \qquad&n<0&\hbox{\small for} \qquad&\kappa>0.
\end{array}\right.
\label{ZHKSD/ASD}
\end{equation}

So far, we minimized the energy. But do we get solutions~?
To see this, we have to check the static field equations,
(\ref{ZHKstat1})-(\ref{ZHKstat2})-(\ref{ZHKstat3}).

The last of these equations is precisely the
Gauss law,  which has to be required as a constraint.

The first one is written, in the SD case, using the identity
(\ref{identity2}), i.e.,
$$
{\vD}^2=D_-D_+-eB=D_+D_-+eB,
$$
as,
$$
\left[eA_t+\2\big(
\mp eB+\frac{2e^2}{|\kappa|}(1-\varrho)\big)\right]\psi=0;
$$
with the upper/lower sign for $\kappa<0$ and $\kappa>0$, respectively.
Using once again the Gauss law, this fixes the time component of the total vector potential
as \footnote{
Note that (\ref{ZHKA0}) is consistent with the Ansatz (\ref{ZHKa0Ansatz}).} ,
\begin{equation}
A_t=\frac{e}{2|\kappa|}(\varrho-1).
\label{ZHKA0}
\end{equation}
But now the current is, by (\ref{SDrho}),
$$
\vj=\pm\frac{e}{2}\,\vnabla\times\varrho,
$$
so that (\ref{ZHKstat2}) is also satisfied.

In conclusion, self-dual vortices are solutions of
\begin{equation}\fbox{$
\left\{\begin{array}{lll}
D_1\psi\pm iD_2\psi&=&0,
\\[10pt]
B\pm\displaystyle\frac{e}{|\kappa|}\Big(|\psi|^2-1\Big)&=&0,
\end{array}\right.
\label{ZHKBogeq}
$}
\end{equation}
with the upper or lower sign chosen to be the opposite
of that of $\kappa$.

These equations are, up to a numerical factor,
precisely the SD ``Bogomol'ny'' equations in the Abelian Higgs model \cite{Bogo}, and can be treated in exactly
the same way. The density, $\varrho=|\psi|^2$, satisfies, in particular, the Liouville-type equation
\begin{equation}\fbox{$
\bigtriangleup\log\varrho
=
\displaystyle\frac{2e^2}{|\kappa]}\,\big(\varrho-1\big).
$}\label{ZHKLiouvilletype}
\end{equation}
Note that the coefficient in front of the r.h.s. is precisely the critical value $\lambda_0$ in (\ref{critlambda}),
which separates the oscillatory
regime from the (real)-mass-splitting regime.

For $|\kappa|=2$ the SD solutions of the ZHK model
are, in fact, imbedded solutions of the  Abelian
Higgs model.
Weinberg's Theorem \cite{Weinberg} is still
valid,
so that for topological charge $n$ (of the correct sign), the system admits a $2|n|$-parameter family of (A)SD solutions.

\subsection{Non-relativistic Maxwell-Chern-Simons Vortices}\label{Manton}

Generalizing previous work \cite{LozMM,BaHa1,BaHa2},
 Manton \cite{Mantonmodel1,Mantonmodel2} proposed a modified version
of the Landau-Ginzburg model of Type II superconductivity.
His Lagrange  density is a subtle mixture, blended from
 the usual Landau-Ginzburg expression and augmented with
the Chern-Simons term \footnote{Redefining the
coordinates as
$t/\gamma \to t^{\prime },\;
\gamma A_t\to A_t^{\prime},$
the Lagrangian (\ref{ManLag}) can be presented as
\begin{eqnarray*}
\frac 12B^2-\frac i2(\Psi ^{*}D_t\Psi -\Psi (D_t\Psi
)^{*})+\frac 12|\vec{D}\Psi |^2+\frac \lambda 4
\big(1-|\Psi|^2\big)^2
+\mu\varepsilon^{\mu\nu\lambda }A_\mu\partial_\nu A_\lambda +A^\mu J_\mu ^T,
\end{eqnarray*}
where
$
J_\mu ^T=(1,\vec{J}^T).
$
This form shows that the Manton model
 is the nonrelativistic version of that
Paul and Khare (Section \ref{PaulKhare}), except for the source term.
Below, we use the
notations of Manton in Ref. \cite{Mantonmodel1}, however.},
\begin{eqnarray}
{\cal L}=
&\frac{1}{2}B^2-
\gamma\displaystyle\frac{i}{2}\big(\Psi^*D_t\Psi-\Psi(D_t\Psi)^*\big)
+\frac{1}{2}\big|\vD\Psi\big|^2
+\frac{\lambda}{4}\big(1-|\Psi|^2\big)^2\nonumber
\\[8pt]
&-\mu\big(BA_t+E_2A_1-E_1A_2\big)
+\gamma A_t+{\vA}\cdot\vec{J}^{T},
\label{ManLag}
\end{eqnarray}
where $\mu$, $\gamma>0$, $\lambda>0$ are constants, the
$D_\alpha\Psi=(\partial_\alpha-iA_\alpha)\Psi$  are the covariant derivatives,
$B=\partial_1A_2-\partial_2A_1$ is the magnetic field
and ${\vE}=\vnabla A_t-\partial_t{\vA}$ the electric field.

This Lagrangian  has the usual symmetry-breaking quartic
potential, but differs from the Zhang et al. expression in that
it also contains a magnetic Maxwell term, $B^2/2$.
The Maxwellian electric term, $-{\vE}^2/2$ is missing, however.
This stems from the requirement of Galilean invariance.
It also involves the  terms $\gamma A_{t}$ and
${\vA}\cdot\vec{J}^{T}$, where $\vec{J}^{T}$
is the (constant) transport current.
The term $\gamma A_{t}$ represents the coupling to a
 uniform background electric charge density $\gamma$.
 The term ${\vA}\cdot\vec{J}^{T}$ then
restores the Galilean invariance, assuming that
 the transport current transforms, under a Galilei boost,
as  \cite{Mantonmodel1,HHY},
 \begin{equation}
\vec{J}^{T}\to\vec{J}^{T}+\gamma\vec{v}.
\label{extcurboost}
\end{equation}

The field equations, derived from (\ref{ManLag}), read
\begin{eqnarray}
&i\gamma D_t\Psi=
-\displaystyle\frac{1}{2}\vD^2\Psi
-\displaystyle\frac{\lambda}{2}\big(1-|\Psi|^2\big)\Psi,
\label{ManNLS}
\\[8pt]
&\epsilon_{ij}\partial_{j}B=
J_{i}-J^{T}_{i}+2\mu\,\epsilon_{ij}\,E_j,
\label{ManFCI}
\\[8pt]
&2\mu B=\gamma\big(1-|\Psi|^2\big),
\label{ManGauss}
\end{eqnarray}
where the (super)current is
$
\vec{J}=({1/2i})\big(\Psi^*\vD\Psi-\Psi(\vD\Psi)^*\big).
$
It follows that,
\vskip2mm
$\bullet$  The matter field satisfies a
gauged, planar non-linear Schr\"odinger equation.

$\bullet$ The second equation is Amp\`ere's
law without the displacement current, as usual in
the ``magnetic-type'' Galilean electricity \cite{LBLL}.

$\bullet$ The last equation  is the Gauss law,
  modified due to the inclusion of
the background charge density term $\gamma A_t$.

Under appropriate conditions, the static, second-order field equations of the
Abelian Higgs model can be solved by solving instead first-order ``Bogomol'ny'' equations. Below we show that this is also true
in the Manton case.

Let us first discuss  the finite-energy conditions.
In the frame where $\vec{J}^{T}=0$,
which can always be arranged by a suitable boost, (\ref{extcurboost}), the energy associated
with the Lagrangian (\ref{ManLag}) is~\cite{HHY}
\begin{equation}
\cE=\int\Big\{
\2B^{2}+
\2\big\vert\vD\Psi\big\vert^{2}+U(\Psi)\Big\}
\,d^{2}{\vx},
\qquad
U(\Psi)=\frac{\lambda}{4}\big(1-\vert\Psi\vert^{2}\big)^{2}.
\label{Manenergy1}
\end{equation}
Eliminating the magnetic term $B^2/2$,
using the Gauss law (\ref{ManGauss}),
results in a mere shift of the coefficient of the non-linear term, cf. (\ref{lambdashift}) \footnote{The discussion below could be simplified by
observing that when shifting the coefficient of the
potential as
\begin{equation}
\lambda\to\Lambda=\lambda+\frac{\gamma^2}{2\mu^2},
\label{lambdashift}
\end{equation}
the energy becomes essentially
the one in the ZHK model, (\ref{ZHKen}). To keep this section self-contained, we prefer not to follow this way.},
\begin{equation}
\cE=\int\Big\{
\2\big\vert\vD\Psi\big\vert^{2}
+
\smallover{\Lambda}/{4}\big(1-\vert\Psi\vert^{2}\big)^{2}\Big\}
\,d^{2}{\vx},
\label{Manenergy}
\qquad
\Lambda=\lambda+\frac{\gamma^{2}}{2\mu^{2}}\ .
\end{equation}
Finite energy ``requires'', just like in the
Landau-Ginzburg case,
\begin{equation}
\vD\Psi\to0
\qquad\hbox{\small and}\qquad
\vert\Psi\vert^2\to1.
\label{topMCS}
\end{equation}
Hence,  we get \textit{topological vortices}~:
 the asymptotic values of the scalar field provide us, once again, with a mapping from the circle at infinity ${S}$
into the vacuum manifold $|\Psi|^2=1$. i.e.,
$
\Psi\Big|_{\infty}: S\to S^1.
$

The first of the  equations in (\ref{topMCS})
implies that the angular component
of the vector potential behaves asymptotically as $n/r$.
The integer $n$ here is also the
winding number of the mapping defined by the asymptotic values
of $\Psi$ into the unit circle,
\begin{equation}
n=\frac{1}{2\pi}\oint_{S}{\vA}\cdot d\vec{\ell}
=\frac{1}{2\pi}\int B\,d^2{\vx}.
\end{equation}

The magnetic flux is, therefore, quantized and is related to the (conserved) particle or vortex number
(See the previous Section),
\begin{equation}
{\cal N}\equiv\int\big(\vert\Psi\vert^2-1\big)\,d^2{\vx}
=-\frac{2\mu}{\gamma}\int B\,d^2{\vx}
=-4\pi\big(\frac{\mu}{\gamma}\big)\,n,
\end{equation}
according to (\ref{ManGauss}).

\goodbreak
\kikezd{Self-dual Maxwell-Chern-Simons vortices}

Conventional Landau-Ginzburg theory admits finite-energy, static, purely magnetic vortex solutions.
For a specific value of the coupling constant, one can find purely magnetic vortex solutions
by solving instead the first-order ``Bogomol'ny'' equations,
\begin{equation}
\begin{array}{clc}
(D_1\pm iD_2)\Psi&=&0,
\\[6pt]
2B&=&1-\vert\Psi\vert^2.
\end{array} \label{ManManSD}
\end{equation}

In the frame where $\vec{J}^{T}=0$ the static Manton equations read
\begin{equation}\begin{array}{ll}
&\gamma A_t\Psi=
-\displaystyle\frac{1}{2}\vD^2\Psi
-\displaystyle\frac{\lambda}{2}\big(1-|\Psi|^2\big)\Psi,
\\[12pt]
&\vnabla\times B=\vec{J}+2\mu\vnabla\times A_t,
\\[12pt]
&2\mu B=\gamma\big(1-|\Psi|^2\big).
\end{array}
\label{staticMant}
\end{equation}

The Gauss law here is only slightly different from that in
(\ref{ManManSD}). Manton \cite{Mantonmodel1}-\cite{Mantonmodel2}
observes that, for some specific choice of the parameters,
solutions of the Abelian-Higgs SD equations, (\ref{relSD2}),
 also yield solutions to
the equations, (\ref{staticMant}), of the modified model.
Let us attempt, therefore,  to solve these equations again  by the first-order Ansatz
\begin{equation}\begin{array}{l}
(D_{1}\pm iD_2)\Psi=0,
\\[8pt]
2\mu B=\gamma\big(1-|\Psi|^2\big).
\end{array}\label{ManSD}
\end{equation}
From the first of these relations we infer that
$$
\vD^2=\mp i\big[D_1, D_2\big]=\mp B
\qquad\hbox{\small and}\qquad
\vec{J}=\mp\2\vnabla\times\varrho,
$$
where $\varrho=|\Psi|^2$.
 Inserting the latter into the non-linear Schr\"odinger equation,
we find that it is identically satisfied when
$$
A_t=(\pm \displaystyle\frac{1}{4\mu}-\displaystyle\frac{\lambda}{2\gamma})(1-\varrho).
$$
Then, from Amp\`ere's law, we get that $\lambda$ has to be
given by,
\begin{equation}
\lambda=\lambda_0=
\pm\frac{\gamma}{\mu}-\frac{\gamma^2}{2\mu^2}.
\label{specManlambda}
\end{equation}
Thus, the scalar potential is,
\begin{equation}
A_t=
\smallover1/{4\mu}\big(\mp1+\smallover{\gamma}/{\mu}\big)\,
\big(1-\varrho\big).
\label{Manat}
\end{equation}

The vector potential is expressed, using the ``self-dual''
(SD) Ansatz (\ref{ManSD}), as
\begin{equation}
{\vA}=\pm\2\vnabla\times\log\varrho+\vnabla\omega,
\label{Manvecpot}
\end{equation}
where $\omega$ is an arbitrary real function,
chosen so that ${\vA}$ is regular.
Inserting this into the Gauss law,  we end up with
the ``Liouville-type'' equation
\begin{equation}
\bigtriangleup\log\varrho=\pm\alpha\big(\varrho-1\big),
\qquad
\alpha=\frac{\gamma}{\mu}.
\label{pmLiou}
\end{equation}

Now, if we want a ``confining'' (stable) and bounded-from-below scalar potential,
$\lambda$ must be positive.
Then we see from equation (\ref{specManlambda}) that this means
\begin{equation}
\left\{\begin{array}{c}
\ 0<\alpha<2
\\[8pt]
-2<\alpha<0
\end{array}\right.
\qquad\hbox{\small for the}\qquad
\left\{\begin{array}{l}
\hbox{\small upper sign}
\\[8pt]
\hbox{\small lower sign}
\end{array}\right.\, .
\label{alphacond}
\end{equation}
In both cases, the
coefficient of $(\varrho-1)$ on the r.h.s. of
(\ref{pmLiou}) is  positive
 \footnote{The coefficient in front of $(\varrho-1)$
is
 $\Lambda_0=\lambda_0+\gamma^2/2\mu^2=|\alpha|$,
 consistently
with the one derived for the ZHK model in Section \ref{ZHKvort}.
}, so that (\ref{pmLiou}) reads,
\begin{equation}
\bigtriangleup\log\varrho=\vert\alpha\vert\big(\varrho-1\big).
\label{ManLioutype}
\end{equation}
Using this notation, the critical coupling reads
\begin{equation}
\lambda_{SD}=|\alpha|\big(1-\frac{|\alpha|}{2}\big).
\label{Manlambdabis}
\end{equation}

The solutions look exactly as in the Abelian Higgs
case. They should
be completed with the scalar potential, (\ref{Manat}).

The magnetic field is derived from
(\ref{Manvecpot}).
Note that the electric field, ${\vE}=-{\vnabla}A_{t}$,
only vanishes for $\mu=\pm\gamma$, i.e., when
$\lambda=1/2$, which is Manton's case.

The self-duality equations (\ref{ManSD}) can also be obtained by studying the energy, (\ref{Manenergy}).
Using, once again, the identity
$$
\big|\vD\Psi\big|^2=
\big|(D_1\pm iD_2)\Psi\big|^2
\pm B|\Psi|^{2}
\pm{\vnabla}\times\vec{J},
$$
and assuming that the fields vanish at infinity, the integral of
the current-term can be dropped, so that $\cH$ becomes
\begin{equation}
\int\bigg\{
\frac{1}{2}\Big|(D_1\pm iD_2)\Psi\Big|^2
+\Big[
\big(\mp\frac{\gamma}{4\mu}+
\frac{\Lambda}{4}\big)(1-|\Psi|^2)^2\Big]
\bigg\}d^2{\vx}
\pm\underbrace{\2\int B\, d^2{\vx}}_{\pi n},
\label{ManEnBogodecomp}
\end{equation}
thereby showing that the energy is positive definite  when the
square bracket vanishes, (i.e., for the special value
(\ref{specManlambda}) of $\lambda$). In this case, the energy admits
a lower ``Bogomol'ny'' bound, $\cE\geq\pi|n|$, with equality only
attained when the SD equations hold.

Equation (\ref{ManLioutype}) is essentially that of Bogomol'ny
in the Landau-Ginzburg theory \cite{Bogo}, to which it reduces
when $\vert\alpha\vert=1$.
The proofs of  Weinberg \cite{Weinberg}, and of Taubes \cite{Taubes1,Fund3},
carry over literally, showing that, for each $n$,
the \textit{existence of a $2|n|$-parameter family of solutions}.

Radial solutions can be studied numerically
\cite{BaHa1,BaHa2}, along the same lines as
in the Abelian Higgs and  the ZHK models, respectively.

Just like the ZHK model, that of Manton can also
be viewed as one in an external field. The clue is
 to observe that setting,
\begin{equation}
\cB\equiv
-\frac{\gamma}{2\mu},
\qquad
\cE_{k}=-\frac{\epsilon_{kl}{J^{T}}_{l}}{2\mu},
\end{equation}
transforms the Manton equations of motion
(\ref{ManNLS}-\ref{ManFCI}-\ref{ManGauss}) into
\begin{equation}\begin{array}{llll}
&i\gamma {\cal D}_t\Psi&=
&-\frac{1}{2}\overrightarrow{\cal D}^2\Psi
-\displaystyle\frac{\lambda}{4}\big(1-|\Psi|^2\big)\Psi,
\\[12pt]
&\epsilon_{ij}\partial_{j}\,b&=
&J_{i}+2\mu\,\epsilon_{ij}\,e_j,
\\[12pt]
&2\mu b&=
&-\gamma|\Psi|^2,
\end{array}
\label{extMan}
\end{equation}
\vspace{-3mm}where \vspace{-3mm}
\begin{eqnarray}
b=B+\cB,
\qquad
e_{i}=E_{i}+\cE_{i},
\qquad
{\cal D}_{\alpha}=\partial_{\alpha}-i(a_{\alpha}-\cA_{\alpha}).
\end{eqnarray}
These equations describe
a non-relativistic scalar field
with Maxwell-Chern-Simons dynamics and a symmetry-breaking quartic potential, placed into a constant, external
electromagnetic field.
For details, and for a discussion
of the Manton model and its relation
to other similar situations,
 \cite{BaHa1,BaHa2},  the reader is referred to \cite{HHY}.

 It is worth mentioning that the Manton model can also be obtained as the non-relativistic limit of a suitable relativistic Maxwell-Chern-Simons model \cite{HHY}. The latter must, however, contain an auxiliary neutral field, as found previously for other relativistic models \cite{LLM,DuTru,Bum}.

Radial vortices in the Manton model
for the critical coupling $\lambda=\lambda_{SD}$
in (\ref{Manlambdabis}) can be easily analyzed.
Linearizing the Liouville-type equation (\ref{ManLioutype}),
we get for  the  deviation from the vacuum value, $\varphi=1-f$, Bessel's equation of order zero,
\begin{equation}
\varphi''+\frac{1}{r}\varphi'-\vert\alpha\vert\varphi=0,
\end{equation}
The solution, and its asymptotic behavior, are, therefore
\begin{equation}
1-\varphi(r)\;\sim\;
\begin{array}{llll}
1-K_{0}(mr)
&\sim&
1-\displaystyle\frac{C}{\sqrt{r}}e^{-mr},
\qquad
&m=\sqrt{\vert\alpha\vert}.
\label{Manlarger}
\end{array}
\end{equation}

The Higgs and the magnetic field behave
as in the Abelian Higgs model;
 the (radial) electric field is
\begin{equation}
E_r=\frac{\alpha\mp1}{4\mu}
\frac{\ d}{dr}(1-\varphi)^2.
\end{equation}

It is, however, more convenient to study the first-order equations.
For the radial Ansatz $a_r=0,\
a_\theta=a(r),$
the self-duality equations read,
\begin{equation}
f'=\pm\frac{n+a}{r}\,f,
\qquad
\frac{a'}{r}=\pm2(f^2-1).
\end{equation}
For small $r$, we get
\begin{equation}
f(r)\sim \beta r^{|n|},
\qquad
a\sim\mp r^2,
\end{equation}
where $\beta$ is some real parameter.
The large-$r$ behavior (\ref{Manlarger}) of the scalar field
is confirmed and for the magnetic and electric fields we get,
\begin{eqnarray}
B&=&\frac{\alpha}{2}(1-f^2)\sim
\frac{\alpha}{\sqrt{r}}
e^{-mr},
\\[6pt]
{\vE}&=&\smallover1/{4\mu}\big(\alpha\mp1\big)\,
\vnabla f^2\sim
\frac{1}{\sqrt{r}}e^{-mr}\,\frac{\vx}{r}.
\end{eqnarray}

Off the SD case, an analysis similar to that
in Section \ref{ZHKvort} would suggest to
find both oscillatory
as well as purely exponential solutions,
depending on the value of $\lambda$.

To conclude this Section, we mention that further
aspects of Maxwell-Chern-Simons vortices are
discussed in Refs. \cite{HH,LLM,DuTru,Bum}.

\subsection{Time-dependent Jackiw-Pi Vortices
in External Field}\label{t-dVort}

The  static, non-relativistic non-topological ``Jackiw-Pi'' solitons,
studied in Section \ref{JPmodel} above, can be
``exported'' to construct \textit{time-dependent vortex solutions
in a constant external magnetic field}
${\cal B}$ \cite{EHIt-d,EHI91/3,JPt-d1,JPt-d2}.
To this end, consider the non-relativistic Zhang et al. Lagrangian (\ref{ZHKLag})
in a background magnetic field $\cB$, but
with the symmetry breaking potential, \break
$\frac{\lambda}{4}(1-|\Psi|^2)^2$,
replaced by
the symmetric one, $-\frac{\lambda}{2}|\Psi|^4$, of the Jackiw-Pi model (\ref{NRCSlag}),
\begin{equation}
\frac{1}{i}\Psi^*\big(\p_t-i(a_t-\cA_t)\big)\Psi
+\2|\big(\vnabla-ie(\va-\vcA)\big)\Psi|^{2}
-\frac{\lambda}{2}|\Psi|^4
-\frac{\kappa}{2}
\epsilon^{\mu\nu\sigma}a_{\mu}\p_{\nu}a_{\sigma},
\label{BJPLag}
\end{equation}
where $a_\alpha$ and $\cA_\alpha$ are  the ``statistical'' and background vector
potential, respectively
\footnote{We use units where $e=m=1$.}.

Let us  assume that the background
 magnetic field $\cB$ is constant, so that we may choose
$\cA_0=0$ and $\cA_i=-\2\epsilon_{ij}x^j\cB$.
The equations of motion are
\begin{equation}
i\big(\p_t-i(a_t-\cA_t)\big)\Psi=\left[
-\frac{1}{2}\big(\vnabla-i(\va-\vcA)\big)^2
-\lambda\,|\Psi|^2
\right]\Psi,
\label{efNLS}
\end{equation}
 supplemented with the field-current identities,
\begin{eqnarray}
\kappa \,b 
=\varrho,
\qquad
\kappa \,e_i
=\epsilon_{ij}j_j,
\label{FCIt-d}
\end{eqnarray}
where density and current are the familiar expressions
$\varrho=\Psi^*\Psi,
$ and
$
\vec{\jmath}=
\frac{1}{2i}[\Psi^*\vD
\Psi-\Psi(\vD\Psi)^*],
$
respectively, and the covariant derivative
involves the total gauge field,
$
\vD=\vnabla-i(\va-\vcA),
$
cf. (\ref{Bcovder}). The density and current always
satisfy the continuity equation,
$
\p_t\varrho+\vnabla\cdot\vJ=0,
$
so that the mass (\ref{nontopmass}),
$$
{\cal M}=\int\varrho\, d^2\vx,
$$
is conserved (provided the integral converges).
By (\ref{FCIt-d}), this is equivalent to requiring the
\textit{statistical flux},
\begin{equation}
\Phi^{(stat)}=\int b\,d^2\vx=
\frac{1}{\kappa}{\cal M},
\label{statflux}
\end{equation}
to converge \footnote{
Remember  that, in the topological case, it is rather the vortex number,
${\cal N}=\displaystyle\int(\varrho-1)\,d^2\vx$ in  (\ref{vortexnumber}), which is required to be finite.}.

 Unexpectedly, the full time-dependent
equations can also be solved \textit{exactly}, namely by applying an appropriate
coordinate transformation to a ``free'' solution,
$\Psi^0$ and $a_\alpha^0$ [i.e., the one
with $\cB=0$], studied in
Section \ref{LiouVort}, \cite{EHI91/1,EHIt-d,JPt-d1,JPt-d2}.
We have
\footnote{The formulae in \cite{JPt-d1} also contain,
as a result of gauge fixing, an additional factor, which involves the vortex number.},
\begin{eqnarray}
\Psi(t,{\vx})&=&\frac{1}{\cos\2\cB t}\,
\exp\left\{-i\cB\frac{r^2}{4}\tan{\2\cB t}\right\}\,
\Psi^0(\vX,T),
\label{psiexport}
\\[10pt]
a_\alpha&=&\frac{\partial X^\beta}{\partial x^\alpha}\,a_\beta^0,
\label{Aexport}
\end{eqnarray}
with
\begin{equation}
{\vX}=\frac{1}{\cos\2\cB t}\,R(\2\cB t)\,{\vx},
\qquad
T=\frac{2}{\cB}\tan\2\cB t,
\label{TtXx}
\end{equation}
where
\begin{equation}
R(\alpha)=\left(\begin{array}{cc}
\cos\alpha&\sin\alpha
\\[6pt]
-\sin\alpha&\cos\alpha
\end{array}\right)
\end{equation}
 is the matrix of a planar rotation by angle
$\alpha$.

The solution admits a nice physical interpretation.
Let $\Psi^0(\vx),\, a^0_\alpha(\vx)$ be, for example, the radially symmetric static Jackiw-Pi solution
  (\ref{radrho}). Then
the ``exported'' time-dependent solution (\ref{psiexport}) is ``breathing'',
 i.e. has a periodically oscillating radius,
 \begin{equation}
 |\Psi(\vx,t)|^2=\frac{1}{\cos^2(\2\cB t)}\ .
\end{equation}

More generally, we can consider the shifted JP solution,
$$
\Psi^0(\vX-2\vR_0),\qquad
a^0_\alpha(\vX-2\vR_0),
$$
with center at $2\vR_0$, and ``export'' it using
(\ref{psiexport}). The new soliton will be centered at
\begin{equation}
2\vr(t)=2\cos(\2\cB t)\left(\begin{array}{cc}
\cos(\2\cB t)&-\sin(\2\cB t)
\\[8pt]
\sin(\2\cB t)&\cos(\2\cB t)
\end{array}\right)\vR_0=
\vR_0-R(-\cB t)\vR_0,
\end{equation}
which effectuates, therefore,  a \textit{cyclotronic} motion with radius $R_0$
around $\vR_0$ and frequency $\omega=\cB$. It will be also
 ``breathing'' according to (\ref{psiexport}).

This construction can be understood by considering the following
sequence of transformations.

\begin{itemize}
\item
Let us start with the problem in a background magnetic field $\cB$,
(\ref{efNLS}) with the background potential,
\begin{equation}
\vcA=-\frac{\cB}{2}\epsilon_{ij}r^j,\qquad\cA_0=0.
\end{equation}
Consider the time-dependent rotation with angle $\2\cB t$,
\begin{equation}
\vx\to\vX'=
\left(\begin{array}{cc}
\cos(\2\cB t)&\sin(\2\cB t)
\\[8pt]
-\sin(\2\cB t)&\cos(\2\cB t)
\end{array}\right),
\qquad
t\to T'=t,
\label{tdrotat}
\end{equation}
implemented as,
\begin{eqnarray}
\Psi(\vx,t)=\Psi'(\vX',T'),
\qquad
a_\mu=a'_\nu\displaystyle\frac{\p {x'}^\nu}{\p X^\mu},
\label{tdrotimp}
\end{eqnarray}
cf. (\ref{Aexport}).
This transformation carries a solution of the problem in the background magnetic field $\cB$ into one in an oscillator background with frequency $\omega=\cB/2$.

\item
Consider the problem in an oscillator background with constant frequency $\omega$ for the fields
$(\Psi',a_\alpha')$, with Lagrangian (\ref{efNLS}), where
\begin{equation}
\vcA'=0,\qquad\cA_0'=\2\omega^2r^2.
\end{equation}
Then the time-dependent dilatation,
\begin{eqnarray}
\vx\to\vX=\frac{1}{\cos\omega t}\,\vx{},
\qquad
t\to T=\frac{\tan\omega t}{\omega}\ ,
\label{osctr}
\end{eqnarray}
implemented as
\begin{eqnarray}
\Psi(\vx,t)&=&\frac{1}{\cos\omega t}e^{-i\frac{\omega}{2}r^2\tan\omega t}
\Psi'(\vX,T),
\label{osctrimp}
\\[8pt]
a_\mu&=&a'_\nu\frac{\p x^ \nu}{\p{x'}^\mu},
\end{eqnarray}
(first considered by Niederer \cite{NiedererOsc}),
transforms the harmonic background problem into a ``free'' Jackiw-Pi problem (\ref{BJPLag}).
\end{itemize}

Composing these two transformations yields
(\ref{psiexport}-\ref{Aexport}-\ref{TtXx}), which removes
the background gauge field altogether, leaving us with
the ``free'' JP problem.

Conversely, the first step followed backwards yields in a harmonic,
and in a constant, electric background
\cite{JPt-d2}.
Composing the two transformations
backwards provides us with (\ref{psiexport}),
which thus carries
 an \textit{arbitrary} solution $(\Psi^0,a_\alpha^0)$ of the Jackiw-Pi model into a solution in a constant background
 magnetic field.

Note that both transformations
can be generalized to
the time-dependent cases $\omega=\omega(t)$ and $\cB=\cB(t)$, respectively \cite{BDPOsc,Hotta}.

The strange form of these transformations, which
``export'' the solution with $\cB=0$ to an external-field,  will
be explained in Section \ref{Bargmann}.

 It is worth mentioning that, for the critical value
$$
\lambda=\lambda_{SD}=\mp\frac{2}{\kappa},
$$
of the coupling, the system also admits static,
self-dual solutions such that the mass is finite
\cite{EHI91/1}. They can be studied along the same
lines as in Section \ref{ZHKvort} and are omitted
here.


\section{SPINOR VORTICES}\label{spinorvort}

\subsection{Relativistic Spinor Vortices}\label{RSpinor}

In Ref. \cite{ChoRelSpin} Cho et al. obtain,
by dimensional reduction from Minkowski
space,
a $(2+1)$-dimensional system. After notational changes and simplification, their Lagrangian and equations,  respectively,  read
\begin{equation}
{\cal L}_{Cho}=\frac\kappa4\varepsilon^{\alpha\beta \gamma }A_\alpha F_{\beta \gamma
}+\bar \psi_+i\gamma_+^\alpha D_\alpha\psi+\bar\psi_-i \gamma_-^\alpha D_\alpha\psi_-
-m(\bar\psi_+\psi_++\bar\psi_- \psi_-),
\label{ChoLag}
\end{equation}
and
\begin{eqnarray}
\2\kappa\epsilon^{\alpha\beta\gamma}F_{\beta\gamma}
&=&
-e\big(
\bar\psi_+\gamma_+^\alpha\psi_++\bar\psi_-\gamma_-^\alpha\psi_-
\big),
\label{ChoFCI}
\\[8pt]
\big(i\gamma^\alpha_+D_\alpha-m\big)\psi_+
&=&0,\label{Cho+}
\\[8pt]
\big(i\gamma^\alpha_-D_\alpha-m\big)\psi_-
&=&0,
\label{Cho-}
\end{eqnarray}
where the two sets of $2\times2$ Dirac matrices are,
\begin{equation}
(\gamma^\alpha_\pm)
=
(\pm\sigma_3, i\sigma_2,-i\sigma_1),
\label{gammamat}
\end{equation}
($c=1$)
\footnote{We use the convention
$$
\sigma_1=\left(\begin{array}{cc}
0&1\\ 1&0\end{array}\right),
\qquad
\sigma_2=\left(\begin{array}{cc}
0&-i\\ i&0\end{array}\right),
\qquad
\sigma_3=\left(\begin{array}{cc}
1&0\\0&-1\end{array}\right).
$$
The $\psi_\pm$ correspond, ``before'' dimensional reduction, to the
chiral components of $4$-component spinors - see the next Section.
}. These equations describe two $2$-spinor scalar fields, $\psi_+$
and $\psi_-$,
 interacting through the Chern-Simons FCI.
Although the Dirac equations satisfied by the
$\psi_\pm$ are  decoupled, the latter
still coupled through the Chern-Simons
equation, which involves the sum of the currents.

It is particularly interesting to search for {\it stationary} solutions.
For
\begin{equation}
A_0=0,\qquad
\partial_tA_i=0,\qquad
\psi_{\pm}=e^{-imt}\left(\begin{array}{c}
F_\pm\\ G_\pm
\end{array}\right),
\end{equation}
where the
$F_\pm$ and $G_\pm$ are time-independent,
the {\it relativistic} system
(\ref{ChoFCI})-(\ref{Cho+})-(\ref{Cho-}) becomes, \begin{eqnarray}
\kappa\epsilon^{ij}\partial_iA_j&=&e\big(
|F_+|^2-|G_+|^2-|F_-|^2+|G_-|^2\big),
\label{ChoFCIbis}
\\[12pt]
\big(D_1+iD_2\big)G_+&=&0,
\label{ChoSD+1}
\\[12pt]
\big(D_1-iD_2\big)F_+&=&2mG_+,
\label{ChoSD+2}
\\[12pt]
\big(D_1-iD_2\big)F_-&=&0,
\label{ChoSD-1}
\\[12pt]
\big(D_1+iD_2\big)G_-&=&2mF_-.
\label{ChoSD-2}
\end{eqnarray}
These equations admit vortex solutions
representing imbedded Jackiw-Pi vortices.
Choosing purely chiral spinors with just one component,
\begin{equation}
\psi=\psi_+=e^{-imt}\left(\begin{array}{c}
F_{+}\\0
\end{array}\right)
\qquad\hbox{\small or}\qquad
\psi=\psi_-=e^{-imt}\left(\begin{array}{c}
0\\ G_{-}
\end{array}\right),
\end{equation}
the  system (\ref{ChoFCIbis})-(\ref{ChoSD+1})
-(\ref{ChoSD+2})-(\ref{ChoSD-1})-(\ref{ChoSD-2})
 becomes,
\begin{eqnarray}
\kappa\epsilon^{ij}\partial_iA_j=e|F_+|^2,
\qquad
&\big(D_1-iD_2\big)F_+=0,
\label{ChoSD+spec}
\\[8pt]
\kappa\epsilon^{ij}\partial_iA_j=e|G_-|^2,\qquad
&
\big(D_1+iD_2\big)G_-=0.
\label{ChoSD-spec}
\end{eqnarray}
These equations are exactly to those,
(\ref{NRCSSD}),
which describe  non-relativistic self-dual, scalar vortices
in the Jackiw - Pi model \cite{JaPi1,JaPi2}.

For further details, the reader is referred to the literature
\cite{ChoRelSpin,LiBha}.

\subsection
{Non-relativistic Spinor Vortices}\label{NRSpinor}

Non-relativistic spinor vortices
can also be constructed along the same lines \cite{DHPSpinor1,DHPSpinor2}.
Following L\'evy-Leblond \cite{LL},
a  non-relativistic spin $\2$ field has
$4$-components,
$\psi=\left(\begin{array}{c}
\Phi\\ \chi\end{array}\right)$,
where  $\Phi$ and $\chi$ are two-component ``Pauli'' spinors, and satisfies
 the   equations, 
\begin{equation}\left\{
\begin{array}{lllll}
(\vec{\sigma}\cdot\vD)\,\Phi
&+&2m\,\chi&=&0,
\\[8pt]
D_t\,\Phi&+&i(\vec{\sigma}\cdot\vD)\,\chi&=&0.
\end{array}\right.
\label{LLequations}
\end{equation}
We consider a $2+1$  dimensional theory, and
couple our
spinors are to the Chern-Simons gauge field through
the density $\varrho=|\Phi|^2$~\footnote{The $\chi$-component does {\it not} contribute to the density \cite{LL}.},
  as well as through the spatial components of the current,
\begin{equation}
\vec{J}=i\big(\Phi^\dagger\vec{\sigma}\,\chi
-\chi^\dagger\vec{\sigma}\,\Phi\big),
\label{spinorcurrent}
\end{equation}
according to the Chern-Simons equations,
$
\kappa \epsilon_{ik}E_k=ej^i,
\
\kappa B=e\rho\ .
$
The chiral components of a $4$-component Dirac
spinor, $\psi$,
\begin{equation}
\psi=\psi_++\psi_-,
\qquad
\psi_\pm=\2(1\pm i\Gamma)\psi_\pm,
\label{chiralcomp}
\end{equation}
are eigenvectors of the chirality operator,
\begin{equation}
i\Gamma\psi_\pm=\pm\psi_\pm,
\quad\hbox{\small where}\quad
\Gamma=\left(\begin{array}{cc}
-i\sigma_3&0
\\
0&i\sigma_3
\end{array}\right).
\label{ChiralOp}
\end{equation}

 Note that
$\Phi$ and $\chi$ in equation (\ref{LLequations}) are {\it not} the chiral components
of $\psi$; the latter are
\begin{equation}
\psi_+=\left(\begin{array}{c}
\Phi_1\\0\\0\\\chi_2
\end{array}\right)
\quad
\psi_-=\left(\begin{array}{c}
0\\\Phi_2\\\chi_1\\0
\end{array}\right)
\quad\hbox{\small for}\quad
\psi=\left(\begin{array}{c}
\Phi\\
\chi
\end{array}\right)
=
\left(\begin{array}{c}
\begin{array}{c}
\Phi_1\\\Phi_2\end{array}
\\
\begin{array}{c}
\chi_1\\
\chi_2\end{array}
\end{array}\right)\ .
\end{equation}

It is easy to see that the equations (\ref{LLequations})
split into chiral components.
Choosing either
$\psi_+$ or $\psi_-$
the equation (\ref{LLequations})
provides us with two uncoupled systems,
each of which describes (in
general different) physical phenomena in $2+1$ dimensions.
For the ease of presentation,
we nevertheless  keep all four components of $\psi$.

Now the ``lower'' $\chi$ can be eliminated using the upper equation
in (\ref{LLequations}),
\begin{equation}
\chi=-\frac{1}{2m}(\vec{\sigma}\cdot\vD)\Phi,
\label{chiequation}
\end{equation}
and everything can be expressed in terms of the ``upper''component, $\Phi$, alone.
For example, the current can be written as,
\begin{equation}
\vec{J}=\frac{1}{2im}\Big(
\Phi^\dagger\vD\Phi-(\vD\Phi)^\dagger\Phi\Big)
+{\vnabla}\times\Big(\frac{1}{2m}\,\Phi^\dagger\sigma_3\Phi\Big).
\label{spinorcurrentbis}
\end{equation}
Using the identity
$$
(\vD\cdot\vec{\sigma})^2=
\vD^2+eB\sigma_3,
$$
we find that the component-spinors satisfy
\begin{equation}\left\{
\begin{array}{lll}
iD_t\Phi&=&-\displaystyle\frac{1}{2m}\Big[\vD^2+eB\sigma_3\Big]\Phi,
\\[12pt]
iD_t\chi&=&-\displaystyle\frac{1}{2m}\Big[\vD^2+eB\sigma_3\Big]\chi
-\frac{e}{2m}\,(\vec{\sigma}\cdot{\vE})\,\Phi.
\label{LLeqbis}
\end{array}\right.
\end{equation}

Thus, $\Phi$ solves a ``Pauli equation'', while
$\chi$ couples through the term
$\vec{\sigma}\cdot{\vE}$.
Expressing ${\vE}$ and $B$ through the
Chern-Simons equations (\ref{Gauss}-\ref{FCI}),
and inserting them into our equations,
we get, finally,
\begin{equation}\left\{
\begin{array}{ll}
iD_t\Phi=
&\Big[-\displaystyle\frac{1}{2m}\,\vD^2
+\displaystyle\frac{e^2}{2m\kappa}\,|\Phi|^2\,\sigma_3
\Big]\Phi,
\\[14pt]
iD_t\chi=
&\Big[-\displaystyle\frac{1}{2m}\,\vD^2
+\displaystyle\frac{e^2}{2m\kappa}\,|\Phi|^2\,\sigma_3
\Big]\chi
-\displaystyle\frac{e^2}{2m\kappa}\,\big(
\vec{\sigma}\times\vec{J}\big)\Phi.
\end{array}\right.
\label{LLeqtris}
\end{equation}

We emphasize that the ``lower'' component, $\chi$, is
not and independent physical field~: owing to
(\ref{chiequation}), it is a mere auxiliary field, determined
by the ``upper'' component $\Phi$.

If the chirality of $\psi$ is restricted to $+1$ (or $-1$),
this system describes
non-relativistic spin $s=+\2$ (resp. $s=-\2$) fields,
 interacting with a
Chern-Simons gauge field.
Leaving the chirality of $\psi$ unspecified, it
describes {\it two} spinor fields of spin $\pm\,\2$,
interacting with each other
and the Chern-Simons gauge field.

By (\ref{chiequation}), it is enough to solve the
$\Phi$-equation.
For
\begin{equation}
\Phi_+=\left(\begin{array}{c}
\ \Psi_+\\
0\; \end{array}\right)
\qquad\hbox{\small and}\qquad
\Phi_-=\left(\begin{array}{c}
0\; \\
\ \Psi_-\end{array}\right),
\end{equation}
respectively, [which amounts to working with the $\pm$ chirality
components], the Pauli equation in (\ref{LLeqtris})
reduces to
\begin{equation}
iD_t\Psi_\pm=
\Big[-\frac{\vD^2}{2m}
\pm\lambda\,(\Psi_\pm^\dagger\Psi_\pm)\Big]\Psi_\pm,
\qquad
\lambda\equiv\frac{e^2}{2m\kappa}.
\label{NLSvort}
\end{equation}

This is again (\ref{gNLS}), but with non-linearity
$\pm\lambda$, {\it half} of the
special value $\lambda$ in (\ref{NRBog}), used by Jackiw and Pi.
For this reason, our solutions (presented below)
will be {\it purely magnetic}, ($A_t\equiv0$), as in the Abelian Higgs model and unlike as in the case
of Jackiw and Pi.

In detail, let us consider the static system,
\begin{equation}\left\{\begin{array}{l}
\Big[-\displaystyle\frac{1}{2m}(\vD^2+eB\sigma_3)-eA_t\Big]\Phi=0,
\\[12pt]
\vec{J}=-\displaystyle\frac{\kappa}{e}\vnabla\times A_t,
\\[12pt]
\kappa B=-e\varrho,
\end{array}\right.
\label{staticvortSD}
\end{equation}
and try the first-order Ansatz
\begin{equation}
\big(D_1\pm iD_2\big)\Phi=0,
\label{vortSDAns}
\end{equation}
which allows us to replace $\vD^2=D_1^2+D_2^2$
by $\mp eB$, so that the first
equation in (\ref{staticvortSD}) can be written as
\begin{equation}
\Big[\frac{1}{2m}eB(\mp 1+\sigma_3)+eA_t\Big]\Phi=0,
\label{vortA_t}
\end{equation}
while the current is
\begin{equation}
\vec{J}
=
\frac{1}{2m}\vnabla\times\Big[\Phi^\dagger(\mp1+\sigma_3)\Phi\Big].
\label{spinSDcur}
\end{equation}

 It is readily seen from equation (\ref{vortA_t}) that any solution has a definite chirality, and
choosing
$\Phi\equiv\Phi_+$
($\Phi\equiv\Phi_-$) for the upper (lower) cases, respectively,
makes $(-1+\sigma_3)\Phi$ (resp. $(1+\sigma_3)\Phi$),
and thus also the current,  $\vec{J}$, vanish.
The two upper equations in (\ref{staticvortSD}) are,
therefore, satisfied with $A_t=0$.

The remaining task is to solve the first-order conditions,
\begin{equation}
(D_1+iD_2)\Psi_+=0,\qquad{\rm or}\qquad(D_1-iD_2)\Psi_-=0,
\label{spinorSD}
\end{equation}
which is done in the same way as before~:
\begin{equation}{\vA}=
\pm\frac{1}{2e}{\vnabla}\times\log\varrho+{\vnabla}\omega,
\qquad
\bigtriangleup\log\varrho=\pm\frac{2e^2}{\kappa}\varrho.
\label{spinorA}
\end{equation}

A normalizable solution is obtained for
$\Psi_+$ when $\kappa<0$, and for $\Psi_-$ when $\kappa>0$.
(These correspond to attractive non-linearity
in equation (\ref{NLSvort})).
The lower components  vanish in both cases, as seen from
the $\chi$-equation (\ref{chiequation}).
Both solutions only involve
{\it one} of the $2+1$ dimensional spinor fields $\psi_\pm$, depending on the sign of $\kappa$.

As a consequence of self-duality,
our solutions have {\it vanishing energy}, just like their scalar counterparts in Ref. \cite{JaPi2}.

The physical properties such as symmetries and conserved quantities
can be conveniently studied by noting that our equations are, in
fact, obtained by variation of a $2+1$-dimensional action
\cite{DHPSpinor1,DHPSpinor2}, $\displaystyle\int\!d^3x\cL$, with,
\begin{eqnarray}\label{NRspinorLag}
&{\cal L}&= \displaystyle\frac{\kappa}{4}\epsilon^{\mu\nu\rho}A_\mu F_{\nu\rho}+
\\[8pt]
&&\Im\left\{\bigl[
\psi_+^\dagger
\big(\Sigma^t_+D_t+\Sigma^i_+D_i-2im\Sigma^m_+\big)\psi_+
\bigr]
+\bigl[
\psi_-^\dagger
\big(\Sigma^t_-D_t+\Sigma^i_-D_i-2im\Sigma^m_-\big)\psi_-
\bigr]\right\}\nonumber
\end{eqnarray}
  where, with some abuse of notation, we identified the chiral
components $\psi_\pm$ with 2-component spinors and introduced
the $2\times 2$ matrices
\begin{equation}\begin{array}{cccc}
\Sigma^t_+=\Sigma^t_-=
&\2(1+\sigma_3),\qquad
&\Sigma^m_+=\Sigma^m_-=
&\2(1-\sigma_3),
\\[12pt]
\Sigma^1_+=-\sigma_2,
&\Sigma^2_+=\sigma_1,
&\Sigma^1_-=-\sigma_2,
&\Sigma^2_-=-\sigma_1.
\end{array}
\end{equation}

Note that the matter action in Eq. (\ref{NRspinorLag}) decouples into chiral components,
consistently with the decoupling of
the L\'evy-Leblond equation (\ref{LLequations}).

A conserved energy-momentum tensor can be constructed, and used to
derive conserved quantities associated with the generator of the
Schr\"odinger group \cite{DHPSpinor2},
\begin{equation}
\left\{
\begin{array}{lll}
\cM=&
m\displaystyle\int|\Phi|^2\,d^2\!\vx
\hfill&\hbox{\small mass,}
\\[16pt]
\vec{\cP}=&\displaystyle\int\underbrace{\left\{\frac{1}{2i}
\left(\Phi^\dagger\vD\Phi-(\vD\Phi)^\dagger\Phi\right)
\right\}}_{\vec{p}}\,d^2\!\vx
&\hbox{\small linear momentum,}
\\[16pt]
\cJ=&\underbrace{\displaystyle\int
\vec{r}\times\vec{p}\,d^2\!\vx
}_{{\small orbital}}
\;+\;
\underbrace{
\2\displaystyle\int\Phi^\dagger\sigma_3\Phi\,d^2\!\vx
}_{{\small spin}}
\hfill
&\hbox{\small angular momentum,}\hfill
\\[16pt]
\vec{\cG}=&t\vec{\cP}
-m\displaystyle\int|\Phi|^2\,\vx\,d^2\!\vx
\hfill
&\hbox{\small boost,}
\\[16pt]
\cE
=&
\displaystyle\int\left\{
\frac{1}{2m}|\vD\Phi|^2
+\lambda
|\Phi|^2\Phi^\dagger\sigma_3\Phi
\right\}\,d^2\!\vx\;
\hfill
\hfill&\hbox{\small energy,}
\\[16pt]
\cD=&2t\,\cE-\displaystyle\int\vec{p}\cdot\vx\,d^2\!\vx
&\hbox{\small dilation,
}\hfill
\\[16pt]
\cK=&-t^2\cE+t\cD+\displaystyle\frac{m}{2}\displaystyle\int
|\Phi|^2\,\vx{\,}^2\,d^2\!\vx
&\hbox{\small expansion,
}\hfill
\end{array}\right.
\label{SpinorSchgen}
\end{equation}
cf. the scalar case, (\ref{GalGen})-(\ref{NRconfGen}) in Section \ref{JPmodel}.

For a static solution, one finds that the {\it mass}, ${\cal M}$,
determines the actual values of all the conserved charges. For the
radially symmetric solution, e.g., the magnetic flux $-e{\cal
M}/\kappa m$ is the same as for the scalar JP soliton in Section
\ref{JPmodel}. The orbital angular momentum vanishes and total
angular momentum reduces to the spin part,
\begin{equation}
\cJ=\mp\frac{\cal M}{2m}\ ,
\label{SPangmom}
\end{equation}
which {\it half} of the corresponding value, (\ref{SDcQ}),
of the
scalar soliton, consistently with the spinorial character of our
solution.

Our non-relativistic spinor
model here can  be derived from the relativistic theory
of Cho et al \cite{ChoRelSpin}.
Setting
\begin{equation}
\psi_+=e^{-imc^2t}\left(\begin{array}{c}
\Psi_+\\\widetilde{\chi}_+
\end{array}\right)
\qquad\hbox{\small and}\qquad
\psi_-=e^{-imc^2t}\left(\begin{array}{c}
\widetilde{\chi}_-\\\Psi_-
\end{array}\right),
\end{equation}
the Cho et al. equations (\ref{ChoFCI})-(\ref{Cho+})-
(\ref{Cho-}) become
\begin{equation}\left\{\begin{array}{l}
iD_t\Phi-c\vec{\sigma}\cdot\vD\widetilde{\chi}=0,
\\[8pt]
iD_t\widetilde{\chi}+c\vec{\sigma}\cdot\vD\Phi
+2mc^2\widetilde{\chi}
=0,
\end{array}\right.
\end{equation}
where
$
\Phi=\left(\begin{array}{c}
\Psi_+\\\Psi_-
\end{array}\right)
$
and
$\widetilde{\chi}=\left(\begin{array}{c}
\widetilde{\chi}_-
\\\widetilde{\chi}_+
\end{array}\right).
$
In the non-relativistic limit,
$$
mc^2\widetilde{\chi}\gg iD_t\widetilde{\chi},
$$
so that this latter term can be dropped from
the second equation. Redefining $\widetilde{\chi}$ as
$\chi=c\widetilde{\chi}$ yields precisely our
equation (\ref{LLequations}).

This also explains why one gets the same (namely the Liouville) equation both in
the relativistic and the non-relativistic cases:
for static and purely magnetic fields, the terms containing $D_t$
are automatically zero.

\kikezd{Spinor vortices with Maxwell-Chern-Simons dynamics}

 The construction above can be extended to
spinor vortices in nonrelativistic Maxwell-Chern-Simons theory \cite{HHY}.
Let $\Phi$ denote a 2-component Pauli spinor.
We postulate the following equations of motion,
\begin{equation}\left\{
\begin{array}{ll}
i\gamma D_t\Phi=-\frac{1}{2}\big[\vD^2+B\sigma_3\big]\Phi
&\hbox{\small Pauli Equation},
\\[14pt]
\epsilon_{ij}\partial_{j}B
=
J_{i}-J^{T}_{i}+2\mu\,\epsilon_{ij}\,E_j\qquad\quad
&\hbox{\small Amp\`ere's Equation},
\\[14pt]
2\mu B=\gamma\big(1-|\Phi|^2\big)
&\hbox{\small Gauss' Law},
\end{array}\right.
\label{spinorequations}
\end{equation}
where the current is now
\begin{equation}
\vec{J}=\frac{1}{2i}\Big(
\Phi^\dagger\vD\Phi-(\vD\Phi)^\dagger\Phi\Big)
+{\vnabla}\times
\Big(\frac{1}{2}\,\Phi^\dagger\sigma_3\Phi\Big).
\end{equation}

The model  admits self-dual vortex solutions,
as we show now.
The transport current can again be eliminated by a
Galilean boost. For fields which are static in
the frame where $\vec{J}^T=0$, the equations of motion become
\begin{equation}\left\{\begin{array}{ll}
&\big[{\2}(\vD^2+B\sigma_3)+\gamma A_t\big]\Phi=0,
\\[10pt]
&\vnabla\times B=
\vec{J}+2\mu\,\vnabla\times A_t,
\\[10pt]
&2\,(\frac{\mu}{\gamma})B=1-\Phi^\dagger\Phi.
\end{array}\right.
\label{staticSpinor}
\end{equation}
Now we attempt to solve these equations
by the first-order Ansatz
\begin{equation}
\big(D_1\pm iD_2\big)\Phi=0.
\label{SDtris}
\end{equation}
Equation (\ref{SDtris}) implies that
\begin{equation}
\vD^2=\mp B
\qquad\hbox{\small and}\qquad
\vec{J}
=
{\2}\vnabla\times\Big[\Phi^\dagger(\mp1+\sigma_3)\Phi\Big],
\end{equation}
so that the Pauli equation in (\ref{staticSpinor}) requires
\begin{equation}
\Big[(\mp 1+\sigma_3)B+2\gamma A_t\Big]\Phi=0,
\qquad
\Phi=
\left(\begin{array}{c}
\Psi_+\\ \Psi_-\end{array}\right)
\label{statPauli}
\end{equation}
Written in component form this requires, for the upper sign, e.g.,
\begin{equation}\left\{
\begin{array}{c}
2\gamma A_t\Psi_+=0
\\[6pt]
(-2B+2\gamma A_t)\Psi_-=0
\end{array}\right.
\label{upcond}
\end{equation}
and if $\Psi_+\neq0$, necessarily we have $\Psi_-=0$.
Things work similarly for the lower sign. Hence,
Equation (\ref{statPauli}) requires again that $\Phi$ has a definite chirality,
\begin{equation}
\Phi=\Phi_+=\left(\begin{array}{c}
\ \Psi_+\\ 0
\end{array}\right),
\quad\hbox{\small or}\quad
\Phi=\Phi_-=\left(\begin{array}{c}
0 \\ \ \Psi_-\ \end{array}\right).
\end{equation}

Let us assume that $\Phi=\Phi_+$.
Then, for the upper sign, (\ref{upcond}) implies
that $A_t=0$, just like in the non-topological case considered before.  The current also vanishes,
$
\vJ=0.
$
 Amp\`ere's equation in (\ref{staticSpinor})
implies therefore that $B=0$, and the solution is trivial.

For the lower sign, however, the Pauli equation it requires instead
\begin{equation}\left\{
\begin{array}{c}
(2B+2\gamma A_t)\Psi_+=0,
\\[6pt]
2\gamma A_t\Psi_-=0,
\end{array}\right.
\label{lowcond}
\end{equation}
so that $A_t=-\frac{1}{\gamma}B$. Then $\vJ=\vnabla\times|\Psi_+|^2$
so that Amp\`ere's equation  amounts to
\begin{equation}
{\vnabla}\times\Big(\big[1-\smallover{2\mu}/{\gamma}\big]B
-\big|\Psi_{+}\big|^2\Big)=0.
\nonumber
\end{equation}
The same argument applied to  $\Phi=\Phi_-$
allows us to conclude, instead,
\begin{equation}
A_t=\mp\smallover1/\gamma\,B
\qquad\hbox{\small and}\qquad
\left\{\begin{array}{ll}
\Phi\equiv\Phi_+\quad
&\hbox{\small for the lower sign}
\\[12pt]
\Phi\equiv\Phi_-\quad
&\hbox{\small
for the upper sign}
\end{array}\right.\ .
\end{equation}
Note that the sign choice is the opposite of the one which works in
the non-topological case. Then
$
\vec{J}=\pm\vnabla\times\big|\Psi_{\pm}\big|^2,
$
so that Amp\`ere's law requires,
\begin{equation}
{\vnabla}\times\Big(\big[1\mp\smallover{2\mu}/{\gamma}\big]B
\mp\big|\Psi_{\pm}\big|^2\Big)=0.
\label{Ampbis}
\end{equation}
But
\begin{equation}
\big|\Phi_{\pm}\big|^2=\big|\Psi_\pm\big|^2=
1-\frac{2\mu}{\gamma}B
\end{equation}
by the Gauss law, so that (\ref{Ampbis}) holds, when
\begin{equation}
\alpha\equiv\pm\frac{\gamma}{\mu}=4.
\label{alfa}
\end{equation}
In conclusion, for the particular value
(\ref{alfa}), the second-order field equations
can be solved by solving one or the
other of the first-order equations in (\ref{staticSpinor}).
These latter conditions fix the gauge potential as,
\begin{equation}
{\vA}=\pm\2{\vnabla}\times\log\varrho+\vnabla\omega,
\qquad
\varrho\equiv\big|\Phi\big|^2
=\big|\Phi_{\pm}\big|^2,
\end{equation}
and  then the Gauss law yields,
\begin{equation}
\bigtriangleup\log\varrho=4(\varrho-1),
\end{equation}
which is again the ``Liouville-type'' equation (\ref{ManLioutype}),
studied previously. Note that the sign, the same for both choices,
is automatically positive, since $\alpha=4$.

The equations of motion (\ref{spinorequations}) can be derived from the Lagrangian
\begin{equation}
\begin{array}{ll}
{\cal L}=
&\frac{1}{2}B^2-\displaystyle\frac{i\gamma}{2}
\big[\Phi^\dagger(D_t\Phi)-(D_t\Phi)^\dagger\Phi\big]
+\displaystyle\frac{1}{2}(\vD\Phi)^\dagger(\vD\Phi)
\\[12pt]
&-\displaystyle\frac{B}{2}\Phi^\dagger\sigma_3\Phi
-\mu\big(BA_t+E_2A_1-E_1A_2\big)
+\gamma a_t+{\vA}\cdot\vec{J}^{T}.
\label{NRMCSSpinorL}
\end{array}
\end{equation}
Then, in the frame where $\vec{J}^T=0$,
the energy is
\begin{equation}
\cE=\frac{1}{2}\int\left\{B^2+\big|\vD\Phi\big|^2
-B\,\Phi^\dagger\sigma_3\Phi
\right\}\,d^2{\vx}.
\end{equation}
Using the identity
\begin{equation}
\big|\vD\Phi\big|^2=\big|(D_{1}\pm iD_{2})\Phi\big|^2
\pm B\,\Phi^\dagger\Phi,
\end{equation}
(valid up to surface terms), the energy is rewritten as
$$
\cE=
\2\,\int\left\{B^2
+\big|(D_{1}\pm iD_{2})\Phi\big|^2
-B\Big[\Phi^\dagger(\mp1+\sigma_3)\Phi\Big]
\right\}\,d^2{\vx}.
$$
Eliminating $B$ using the Gauss law, we finally get,
for purely chiral fields, $\Phi=\Phi_{\pm}$,
\begin{equation}
\cE=
\2\,\int\left\{
\big|(D_{1}\pm iD_{2})\Phi_{\pm}\big|^2
+\frac{\gamma}{4\mu}\big[\mp4+
\frac{\gamma}{\mu}\big]
\big(1-\vert\Phi_{\pm}\vert^2\big)^{2}
\right\}\,d^2{\vx}
\pm\,\int B\,d^2{\vx}.
\end{equation}
The last integral here yields the topological charge
$\pm2\pi n$. Also, the
integral is positive definite when $\pm\gamma/\mu\geq4$,
depending on the chosen sign, yielding the Bogomol'ny bound,
\begin{equation}
\cE\geq2\pi\vert n\vert.
\end{equation}
Hence, the Pauli term  {\it doubles}
the Bogomol'ny bound with respect to the scalar case.
The bound can be saturated when $\pm\gamma/\mu=4$
and the self-dual equations (\ref{SDtris}) hold.

The action (\ref{NRMCSSpinorL}) can be used to
show that the coupled L\'evy-Leblond --- Chern-Simons system is,
just like its scalar counterpart, Schr\"odinger symmetric,
proving that our theory is indeed non-relativistic.
A conserved energy-momentum tensor can be constructed and used to derive conserved quantities.


\section{NON-RELATIVISTIC KALUZA-KLEIN FRAMEWORK
}\label{Bargmann}

\subsection{Relativistic Framework for Non-relativistic Physics}

How do the extra symmetries of JP vortices come about~?
Can one derive the
energy-momentum tensor (\ref{Tmunu}),
 together with its strange
properties (\ref{NRsymEM}), in a systematic way, just like in a relativistic theory~?

The answer is affirmative; it is enough to use a
``relativistic'' framework to describe non-relativistic physics.
Below we present such a
 ``non-relativistic Kaluza-Klein'' approach
 \cite{DBKP,DGH,Gomis}.

Our starting point is to rewrite the free Schr\"odinger
equation in $D$ space dimensions,
\begin{equation}
\left[2mi\p_t+\bigtriangleup\right]\Psi=0,
\label{fS}
\end{equation}
in a ``relativistic'' form in a $D+2$ dimensional
space. To this end, let us introduce a new, ``vertical''
variable $s$, and consider,
\begin{equation}
\psi(\vx,t,s)=e^{ims}\Psi(\vx,t).
\label{liftwf}
\end{equation}
The new, ``lifted'' wave function $\psi$, defined on extended space, satisfies,
by construction, the ``equivariance'' condition
\begin{equation}
\p_s\psi=im\psi.
\label{sequivar}
\end{equation}
It follows that
$\psi$ in (\ref{liftwf}) satisfies,
\begin{equation}
\left[\bigtriangleup+2\p_t\p_s\right]\psi=0.
\label{liftfS}
\end{equation}
$\vx,t,s$ can now be viewed as coordinates on a
$(D+2)$ dimensional extended space  $M$
made into Minkowski space, $\IR^{D+1,1}$, by setting
\begin{equation}
ds^2=d\vx^2+2dtds.
\label{Minkowski}
\end{equation}
Note that $\vx$ is space-like, but $s$ and $t$ are
 light-cone
coordinates.
Comparing  with the light-cone expression of the
Laplace-Beltrami operator on Minkowski space,
we recognize that (\ref{liftfS}) is precisely the
\textit{free, massless Klein-Gordon equation} in
$(D+1,1)$ dimensions,
\begin{equation}
\dAlembert\psi=0,
\label{m0KlGg}
\end{equation}
for the equivariant function $\psi$.
Conversely, let $\psi$ be equivariant.
Then
\begin{equation}
\Psi(\vx,t)=e^{-ims}\psi(x)
\end{equation}
[where  $x=(\vx,t,s)$] is a function of $\vx$ and $t$ alone, since
$\p_s\Psi=0.$
therefore,  the massless
Klein-Gordon equation for $\psi$ on Minkowski space reduces
 to the free Schr\"odin\-ger equation for $\Psi$.

Intuitively, the lifted wave function, $\psi$,
incorporates the gauge degrees of freedom. Let
$G(\vx,t)$ denote an arbitrary real function.
 Then
\begin{equation}
g(x)=(\vx,t,s+G(\vx,t))
\end{equation}
is a ``vertical'' transformation of extended space
$M$. Thus,
$
\psi(g(x))=e^{imG(\vx,t)}\psi(x)
$
owing to equivariance,
so that it carries the ``ordinary'' wave function $\Psi$ into its gauge transform,
$$
\Psi(\vx,t)\to e^{imG(\vx,t)}\Psi(\vx,t).
$$
\goodbreak

\kikezd{Galilean symmetry in the Kaluza-Klein-type
framework}.

Let us observe that the ``vertical vector''
\begin{equation}
\xi=\p_s
\label{vvec}
\end{equation}
is lightlike as well as covariantly constant, $\nabla\xi=0$ w.r.t. the Minkowski metric.

Next, let us remind the Reader that the isometries
of Minkowski space $M$ form the $(D+1,1)$ dimensional
Poincar\'e
group, and that all conformal transformations
of  $M$, i.e. such that,
\begin{equation}
f^*ds^2=\Omega^2(x)ds^2,
\label{metconftransf}
\end{equation}
for some real function $\Omega$, form the
\textit{conformal group} O$(D+2,2)$.
The infinitesimal conformal transformations form
the $\o(4,2)$ Lie algebra, which acts on $M$ according to,
\begin{equation}
\left\{
\begin{array}{llll}
P_\mu&=&-i\partial_\mu
&\hbox{\small translations,}
\\[12pt]
M_{\mu\nu}&=&-i\left(
x_\mu\partial_\nu-x_\nu\partial_\mu\right)
&\hbox{\small Lorentz transformations},
\\[12pt]
d&=&-ix^\mu\partial_\mu-\smallover3i/2
&\hbox{\small dilations},
\\[12pt]
K_\mu&=&-i\left(
x_\nu x^\nu\partial_\mu
-x_\mu(3+2x^\nu\partial_\nu)\right)
&\hbox{\small conformal transformations}.
\end{array}
\right.
\label{confalg}
\end{equation}
\goodbreak
One can prove, furthermore, the following~:

\kikezd{Theorem} \cite{DBKP,DGH}~:
{\it The subgroup of the $(D+1,1)$ dimensional Poincar\'e
group that leaves the ``vertical vector'' $(\xi^\mu)=\p_s$
invariant is isomorphic to the $1$-parameter central extension of the Galilei group in $D+1$ dimensions.}

{\it The stability subgroup of the ``vertical vector''
$(\xi^\mu)=\p_s$ in
the $(D+1,1)$ dimensional conformal group  O$(D+2,2)$  is isomorphic to the {\it Schr\"odinger group}, which is in fact the
$1$-parameter central extension of the Galilei group in $D+1$ dimensions, augmented with dilatations and expansions. The conformal factor (\ref{metconftransf}) is $s$-independent,
$\Omega=\Omega(\vx,t)$. The Schr\"odinger algebra is the subalgebra of (\ref{confalg}) which commutes with $\xi$.
}
\vskip4mm

$\bullet$ For $D=3$ the (centrally extended)
Schr\"odinger group has 13 generators.
 It acts on the extended spacetime according to
\begin{equation}\left\{
\begin{array}{ll}
\vx' &=\displaystyle\frac{R\vx-{\vb}t+\vc}{ft+g}
\\[12pt]
t' &=\displaystyle\frac{dt+e}{ft+g}
\\[12pt]
s' &=s+\displaystyle
\frac{f}{2}\displaystyle\frac{(R\vx-{\vb}t+{\vc})^2}
{ft+g}+{\vb}\cdot R\vx
-\displaystyle\frac{t}{2}\,{\vb}^2+h,
\end{array}\right.
\label{SchronB}
\end{equation}
where $R\in SO(3); {\vb},\ {\vc} \in \IR^3; d,\ e,\ f,\ g,\ h \in \IR$ and s.t. $dg-ef=1$.

The corresponding conformal factor is
\begin{equation}
\Omega =\frac{1}{ft+g}\ .
\end{equation}

The 11 dimensional subgroup defined by
$d=g=1, f=0$ is the centrally extended Galilei group.
The infinitesimal action of the extended Schr\"odinger group,
also obtained as those vector fields in (\ref{confalg}) which commute with vertical translations generated by $\xi$,  reads
\begin{equation}\left\{
\begin{array}{ll}
\vX\, &= \vec{\Omega}\times\vx +
\displaystyle\frac{\delta}{2}\,\vx+\kappa t\,\vx
+{\vbeta}t +\vgamma,
\\[12pt]
X^t &=\kappa t^2+\delta t + \epsilon,
\\[12pt]
X^s &=-\kappa\displaystyle\frac{r^2}{2}+\vbeta\cdot\vx
+\eta
\end{array}\right.
\end{equation}
($\vec{\Omega},\ \vbeta,\ \vgamma \in \IR^3Ê;\  \kappa,\ \delta,\ \epsilon,
\eta \in \IR$).

Let $f~: M\to M$ denote a conformal transformation of
$(D+1,1)$ dimensional extended space.  It is naturally implemented on a lifted wave function, $\psi$, according to,
\begin{equation}
\big(\hat{f}\psi\big)\,(\vx,t,s)=\Omega^{D/2}(\vx,t)\psi(\vx',t',s')
\end{equation}
where the conformal factor is included to enforce normalization~: it takes care of
the change of measure.
Expressed using $\Psi(\vx,t)=e^{ims}\psi(\vx,t,s)$,
we find,
\begin{equation}
\big(\hat{f}\Psi\big)\,(\vx,t)=\Omega^{D/2}(\vx,t)
e^{im(s'-s)}\Psi(\vx',t').
\label{Bimp}
\end{equation}
In particular, setting in (\ref{SchronB}),
yields,
\begin{equation}
\begin{array}{lll}
R=Id,\ \vc=0,\ &e=f=0,\, g=d=1
&\hbox{\small boost},
\\[8pt]
R=Id,\ \vb=\vc=0,\ &f=e=0, \ d=\lambda,
\ g=\lambda^{-1}\qquad\qquad
&\hbox{\small dilatation},
\\[8pt]
R=Id,\ \vb=\vc=0, &e=0,\ f=-\kappa, d=1,\ g=1
&\hbox{\small expansion}.
\end{array}
\end{equation}
The implementations,
 deduced from (\ref{Bimp}), read,
\begin{equation}
\begin{array}{ll}
\Psi'(\vx,t)=e^{im(\vb\cdot\vx-\vb^2t)}
\Psi(\vx',t')
&\hbox{\small boost,}
\\[12pt]
\Psi'(\vx,t)=\lambda^{D/2}\Psi(\vx',t')
&\hbox{\small dilatation,}
\\[12pt]
\Psi'(\vx,t)=\big(\displaystyle\frac{1}
{1-\kappa t}\big)^{-D/2}
e^{-i\frac{m\kappa}{2(1-\kappa t)}}\Psi(\vx',t')\qquad
&\hbox{\small expansion,}
\end{array}
\end{equation}
which are precisely the implementations
 found before \cite{JaPi1}.

More generally, $D+1$ dimensional non-relativistic spacetime can be
obtained from a \hfill\break
$(D+1,1)$ dimensional relativistic spacetime, $M$,
endowed with a Lorentz-signature metric $g_{\mu\nu}$
and a covariantly constant, lightlike vector $\xi^\mu$,
called ``Bargmann space'' \cite{DBKP,DGH}
\footnote{Requiring that the metric satisfies the
Einstein equations provides us with a gravitational pp wave \cite{DGH}.
These metrics provide us with exact
string vacua \cite{DGH,Bstring}.}.
Then non-relativistic spacetime is the factor space of $M$,
obtained by factoring out the integral curves of the
``vertical vector'' $(\xi^\mu)$ \footnote{Choosing the ``vertical'' vector
$\xi^\mu$ to be spacelike would provide us with a relativistic
theory ``downstairs''.}.

The most general metric with a covariantly constant
lightlike vector is
\begin{equation}
ds^2=g_{ij}dx^idx^j+2dt(ds+{\cal A}_i(\vx,t)dx^i)-
2{\cal U}(\vx,t)dt^2,
\label{genBmetric}
\end{equation}
where $g_{ij}$ is some $D$-dimensional spatial metric
and ${\cal A}_i$ and ${\cal U}$ are a vector and a scalar potential, respectively.
The coordinate $\vx$ can be viewed as position, $t$ as
non-relativistic time, and $s$ as an ``internal,
Kaluza-Klein-type coordinate'', directed
along the ``vertical'' vector
$\xi^\mu=\p_s$. Quotienting $M$ by the integral curves
of $\xi^\mu$ amounts, intuitively, to ``forgetting''
the vertical coordinate, $s$.

One can show \cite{DBKP,DGH} that the projection of
the null-geodesics of $M$, endowed with the metric
(\ref{genBmetric}), satisfy
the usual equations of motion of a non-relativistic
particle in a (static) ``electromagnetic'' field
\begin{equation}
\overrightarrow{\cal B}=\rot\overrightarrow{\cal A},
\qquad
\overrightarrow{\cal E}=-\grad{\cal U}.
\label{Bemfield}
\end{equation}
There is, however, one strange detail~: the coupling constant is
not the electric charge, $e$, but the mass, $m$ \footnote{
In this Section the mass will be denoted by $m$.}.

For $\overrightarrow{\cal A}=0$, in particular, we recover Newton's
equations, as observed by Eisenhart in 1929 \cite{Eisen}.

Null geodesics are conformally invariant and, hence, their
projections are invariant w.r.t. $\xi$-pre\-serving
conformal transformations which are, therefore,
symmetries of the projected system.

All
conformal transformations of $(M, g_{\mu\nu})$ span, in particular, the conformal algebra; those which preserve the lightlike vector
$(\xi^\mu)=\p_s$
define the generalized
 Schr\"odinger group,
centrally extended with the mass.

Turning to quantum mechanics [field theory], the
equivariance condition (\ref{sequivar}) is generalized to
\begin{equation}
\nabla_\xi\psi=im\psi,
\label{gequivar}
\end{equation}
where $\nabla_\mu$ is the metric-covariant derivative.

Finally, the massless wave equation (\ref{m0KlGg}) is generalized to
\begin{equation}
\nabla_\mu\nabla^\mu\psi=0.
\label{covwaveeq}
\end{equation}
Note that while $\nabla_\mu\psi=\p_\mu\psi$
for a \textit{function}  $\psi$, the second-order operator
in (\ref{covwaveeq}) is not trivial. In fact,
\begin{equation}
\nabla_\mu\nabla^\mu\psi=
g^{\mu\nu}\nabla_\mu\nabla_\nu\,\psi
=g^{\mu\nu}\Big(\p_\mu\p_\nu-\Gamma_{\mu\nu}^\rho\p_\rho\Big)\psi,
\label{Bwaveop}
\end{equation}
where the
$\Gamma_{\mu\nu}^\rho$ are the Christoffel symbols of the
metric.
\goodbreak

\subsection{Vortices in the K-K-type Framework}\label{NTvortKK}

The ``non-relativistic Kaluza-Klein'' framework,
useful for studying the Schr\"odinger symmetry of
classical systems, can also be adapted to Chern-Simons field theory \cite{DHP1}. We first examine the case of Jackiw-Pi vortices.
Begin by choosing, on 4-dimensional Minkowski space
$M$, a four-vector potential
$a_\mu$   with field strength $f_{\mu\nu}$ and let
$j_\mu$ be a four-current.

\vskip2mm
$\bullet$  Let us posit the relation
\begin{equation}
\kappa f_{\mu\nu}=e\sqrt{-g}\epsilon_{\mu\nu\rho\sigma}\,
\xi^\rho j^\sigma.
\label{4dFCI}
\end{equation}
Then $f_{\mu\nu}$ is the lift
from space-time with coordinates $\vx$ and $t$ of a
closed two-form $F_{\mu\nu}$.  The potential
$a_\mu$ can be chosen, therefore, as the pull-back of a 3-potential
$A_\alpha=(A_t,\vA)$. The four-current $j^\mu$
projects, in turn, onto a 3-current $J^\alpha=(\varrho,\vJ)$.
Then (\ref{4dFCI}) is readily seen to project precisely to
the Chern-Simons equations (\ref{Gauss})-(\ref{FCI}).

 $\bullet$  Similarly, let $\psi$ denote a scalar field on $M$ and let
us suppose that $\psi$ satisfies the (massless) non-linear Klein-Gordon wave equation,
\begin{equation}
\left[D_\mu D^\mu-\frac{R}{6}+\lambda(\psi^*\psi)\right]\psi=0,
\label{m0NLKG}
\end{equation}
where $D_\mu=\nabla_\mu-iea_\mu$ is the metric- and
gauge-covariant derivative on $M$ and we have also added,
for the sake of generality, a term which involves the
scalar curvature, $R$, of the Riemann manifold $(M,g_{\mu\nu})$.
The scalar field $\psi$ is required to be equivariant,
\begin{equation}
\xi^\mu D_\mu\psi=im\psi.
\label{equivar}
\end{equation}
Then
$
\Psi=e^{-ims}\psi
$
is a function of $\vx$ and $t$, and
(\ref{m0NLKG}) becomes, for the Minkowski
metric (\ref{Minkowski}), the gauged non-linear
Schr\"odinger equation (\ref{gNLS}).

$\bullet$  The equations (\ref{4dFCI})
and (\ref{m0NLKG}) are coupled through
\begin{equation}
j^\mu=\frac{1}{2mi}\left[\psi^*(D^\mu\psi)-\psi(D^\mu\psi)^*\right],
\label{4dcur}
\end{equation}
which projects to the correct NR current.

Equations (\ref{4dFCI})-(\ref{m0NLKG})-(\ref{equivar})-(\ref{4dcur}) form a self-consistent
system, allowing us to lift our non-relativistic
coupled scalar field-Chern-Simons system to  relativistic
extended spacetime, $M$.
It can now be shown \cite{DHP1}
that the latter is invariant
w.r.t. any conformal transformation of the metric of $M$
that preserves the ``vertical'' vector, $(\xi^\mu)$.
Thus, we have established the generalized Schr\"odinger
invariance of the non-relativistic Chern-Simons + scalar field system.

The theory on $M$ is relativistic and admits, therefore,
a conserved, traceless and symmetric energy-momentum tensor
$\theta_{\mu\nu}$. In the present case, a lengthy calculation \cite{DHP1} shows that the canonical procedure
yields
\begin{equation}
\begin{array}{lll}
3m\theta_{\mu\nu}&=&
(D_\mu\psi)^*D_\nu\psi+D_\mu\psi(D_\nu\psi)^*
\\[12pt]
&&-\displaystyle\frac{1}{2}\left(
\psi^*D_\mu D_\nu\psi+\psi(D_\mu D_\nu)^*\right)
\\[12pt]
&&+\,\displaystyle\frac{1}{2}|\psi|^2\left(R_{\mu\nu}-\frac{R}{6}g_{\mu\nu}\right)
-\frac{1}{2}g_{\mu\nu}(D^\rho\psi)^*D_\rho\psi
-\displaystyle\frac{\lambda}{4}g_{\mu\nu}|\psi|^4.
\end{array}
\end{equation}

In Ref \cite{DHP1} a version of
Noether's theorem was proved. It says that, for any $\xi$-preserving
conformal vector field $(X^\mu)$ on Bargmann space, the quantity
\begin{eqnarray}
Q_X=\int_{\Sigma_t}
\vartheta_{\mu\nu}X^\mu\xi^\nu\,\sqrt{\gamma}\,d^2{\vx},
\label{consQ}
\end{eqnarray}
is a constant of the motion. (Here
$\gamma$ is the determinant
of the metric $(g_{ij})$ induced by $(g_{\mu\nu})$
on  ``transverse space'' $t={\rm const.}$).
The charge (\ref{consQ}) is conveniently calculated using
\begin{eqnarray}
\vartheta_{\mu\nu}\xi^\nu
=\frac{1}{2i}\left[
\psi^*\,(D_\mu\psi)-\psi\,(D_\mu\psi)^*\right]
-\frac{1}{6}\,\xi_\mu\left(
\frac{R}{6}|\psi|^2
+(D^\nu\psi)^*\,D_\nu\psi
+\frac{\lambda}{2}\,|\psi|^4
\right).
\end{eqnarray}

\vskip2mm
$\bullet$ Our first illustration is when $M$ is ordinary,
4-D Minkowski space. The general theory yields the
Schr\"odinger symmetry of the coupled Chern-Simons +
matter system.
The energy-momentum tensor $\theta_{\mu\nu}$
constructed above is related to the
$T^{\alpha\beta}$ found by Jackiw and Pi \cite{JaPi1}
according to
\begin{equation}
\begin{array}{ll}
T^{00}=-\theta^0_{\ 0},\qquad\qquad
&
T^{i0}=-\theta^i_{\ 0}-\displaystyle\frac{1}{6m}\p_i\p_t\varrho,
\\[12pt]
T^{0j}=\theta^0_{\ j},
&
T^{ij}=\theta^i_{\ j}+\displaystyle\frac{1}{3m}
\left(\delta^i_j\Delta-\p^i\p_j\right)\varrho,
\end{array}
\end{equation}
where $\Delta$ is the spatial Laplace operator.
These formulae allow us
to rederive all  characteristic properties of $T^{\alpha\beta}$ listed  in Section \ref{JPmodel}, see Ref. \cite{DHP1} for detail.

Interestingly,  our proof used the {\it field equations}.
Is it possible to  use instead a {\it variational principle}~?
On $M$, we could attempt to use  the $4$d
``Chern-Simons type'' expression \cite{Carroll},
\begin{equation}
\cL=\frac{\kappa}{2}\epsilon^{\mu\nu\rho\sigma}
\xi_\mu a_\nu f_{\rho\sigma}+\hbox{matter}.
\label{4dCS}
\end{equation}
Curiously, this correctly reproduces the
{\it relativistic} Chern-Simons equations (\ref{relCSEL2}) if $\xi^\mu$
is {\it spacelike}, but {\it fails} when it is {\it lightlike},
$\xi_\mu\xi^\mu=0$  \cite{HH} --- which is
precisely the {\it non-relativistic} case we study here.
 Varying the gauge field in (\ref{4dCS}) yields in fact the field equations
\begin{equation}
\kappa\,\sqrt{-g}\,\epsilon^{\mu\rho\sigma\nu}\xi_\mu f_{\rho\sigma}=-j^\nu,
\label{wrongfe}
\end{equation}
supplemented with the matter equation.
Assuming that $\psi$ is equivariant
and that $f_{\mu\nu}$ is the lift from $Q$
of a two-form  $F_{\alpha\beta}$
\footnote{In the approach presented above, this follows
automatically from the field equation.}.
Hence
$
f_{\mu\nu}\xi^\mu=0.
$
\goodbreak

The equations (\ref{wrongfe})  are {\it similar} to those in (\ref{4dFCI}), 
 except for the
``wrong'' position of the $\epsilon_{\mu\nu\rho\sigma}$ tensor.
Transferring the
$\epsilon_{\mu\nu\rho\sigma}$ to the other side of
(\ref{wrongfe})
and contracting with $\xi^\rho$ we get, using
$
f_{\mu\nu}\xi^\mu=0,
$
\begin{equation}
\kappa\,(\xi_\rho\xi^\rho)f_{\mu\nu}
=\sqrt{-g}\,\epsilon_{\mu\nu\rho\sigma}\,\xi^\rho j^\sigma.
\label{wrongfebis}
\end{equation}

$\bullet$ If $\xi$ is {\it spacelike} or {\it timelike},
 it can be normalized as
$\xi_\mu\xi^\mu=\pm1$. Then Eq. (\ref{wrongfebis}) is (possibly up to a sign) our posited equation (\ref{4dFCI}). 
  In the spacelike case, for example,
we get a well-behaved {\it relativistic}
model~: the quotient
is a Lorentz   manifold.
Let $M$ be, for example,  Minkowski space with metric
$dx^2+dy^2-dz^2+dw^2$. The vector $\xi=\partial_w$ is space-like
and covariantly constant.
The quotient is $(2+1)$-dimensional Minkowski space
 with metric $dx^2+dy^2-dz^2$ and the equations reduce to,
\begin{eqnarray}
\kappa\,\epsilon^{\alpha\beta\gamma}F_{\alpha\beta}
=J^\gamma,
\end{eqnarray}
[where $J^\gamma$ is a suitably defined $3$-current],
plus the matter equation. 
 These are indeed the correct 
 Chern-Simons and Klein-Gordon equations
for a  relativistic model in (2+1)-dimensional flat space.

$\bullet$  If, however, $\xi$ is {\it lightlike},
\begin{equation}
\xi_\mu\xi^\mu=0
\end{equation}
then the $f_{\mu\nu}$ has a {\it vanishing} coefficient,
so the term behind $\kappa$ drops out,
and Eqn. (\ref{wrongfebis}) does not reproduce
the Chern-Simons equation (\ref{4dFCI}).

In conclusion, (\ref{4dCS}) is
a correct Lagrangian in the {\it relativistic} case, but
{\it fails} to work in the lightlike, {\it nonrelativistic} case~:
the $(2+1)$D Chern-Simons Lagrangian does
{\it not} come by lightlike reduction from  an action on Minkowski space.

\goodbreak
\kikezd{Time-dependent vortices in the K-K framework}

Now we explain the results found previously for
time dependent vortices in external fields in our
``Kaluza-Klein-type'' framework.
Consider the coupled system
(\ref{4dFCI})-(\ref{m0NLKG})-(\ref{equivar})-(\ref{4dcur})
on a  ``Bargmann'' metric (\ref{genBmetric})
with $g_{ij}=\delta_{ij}$,
$$
ds^2=g_{ij}dx^idx^j+2dt(ds+{\cal A}_i(\vx,t)dx^i)-
2{\cal U}(\vx,t)dt^2.
$$
A calculation analogous to that of the preceding section shows that,
after reduction, the covariant derivative
\begin{equation}
D_\alpha=\nabla_\alpha-iea_\alpha,
\label{gmcovder}
\end{equation}
(where $\nabla_\alpha$ is the metric-covariant
derivative) becomes precisely the background-field expression
in (\ref{Bcovder}),
$$
D_\alpha=\p_\alpha-ieA_\alpha,\qquad
A_\alpha=a_\alpha-\cA_\alpha.
$$
The equation of motion is, therefore, a
 generalization of (\ref{efNLS}).

\vskip2mm
$\bullet$ Choose, for example, the ``oscillator''
metric
\begin{equation}
d{\vx}_{osc}^2+2dt_{\rm osc}ds_{\rm osc}-\omega^2r_{\rm osc}^2dt_{\rm osc}^2,
\label{oscMet}
\end{equation}
where ${\vx}_{\rm osc}\in\IR^2$, $r_{\rm osc}=|{\vx}_{\rm osc}|$
and $\omega$ is a constant. The null geodesics  in this case
correspond  to a non-relativistic,
spinless particle in an oscillator background \cite{DBKP,DGH}.
Requiring equivariance, (\ref{equivar}), the wave equation
(\ref{m0NLKG}) reduces to
\begin{equation}
i\partial_{t_{\rm osc}}\Psi_{\rm osc}=\left\{
-\frac{\vD^2}{2}+\frac{\omega^2}{2}{r_{\rm osc}}^2
-\Lambda\,\Psi_{\rm osc}\Psi_{\rm osc}^*\right\}\Psi_{\rm osc},
\end{equation}
($\vD=\vec{\partial}-i{\vA}$, $\Lambda=\lambda/2$),
which describes Chern-Simons vortices in a harmonic-force background,  Ref. \cite{JPt-d2}.

\vskip2mm
$\bullet$ Let us consider instead the ``magnetic'' metric
\begin{equation}
d{\vx} {}^2+2dt\Big[ds+
\2\epsilon_{ij}{\cal B}{x}^jd{x}^i\Big],
\label{BMet}
\end{equation}
where ${\vx}\in\IR^2$ and ${\cal B}$
is a constant, whose null geodesics
describe a charged particle in a uniform magnetic
field in the plane \cite{DGH}.
Imposing equivariance, equation~(\ref{m0NLKG})
reduces
to equation~(\ref{efNLS}) with $\Lambda=\lambda/2$ and the
 total covariant derivative.

Returning to the general theory, let $\varphi$ denote a conformal
Bargmann diffeomorphism between \textit{two}
Bargmann spaces, i.e. let $\varphi~:~(M,g,\xi)\to(M',g',\xi')$ be
such that
\begin{equation}
\varphi^\star g'=\Omega^2g
\qquad
\xi'=\varphi_\star\xi.
\label{Bconftransf}
\end{equation}
Such a mapping projects to a diffeomorphism of the quotients,
$Q$ and $Q'$, which we will denote by $\Phi$.
Then the same proof as in Ref. \cite{DHP1} allows one to
show  that, if
$(a'_\mu,\psi')$ is a solution of the field equations on
$M'$, then
\begin{equation}
a_\mu=(\varphi^\star a')_\mu,
\qquad
\psi=\Omega\,\varphi^\star\psi'
\label{apsiexp}
\end{equation}
is a solution of the analogous equations on $M$ \cite{DHP2}.

 Locally,
$$
\varphi(t,{\vx},s)=(t',{\vx}',s')\qquad\hbox{\small with}\qquad
(t',{\vx}')=\Phi(t,{\vx}),
\quad
s'=s+\Sigma(t,{\vx}),
$$
so that
$\psi=\Omega\,\varphi^\star\psi'$ reduces to
\begin{equation}
\Psi(t,{\vx})=\Omega(t)\,e^{i\Sigma(t,{\vx})}\Psi'(t',{\vx}'),
\qquad
A_\alpha=\Phi^\star A'_\alpha
\end{equation}
($\alpha=0,1,2$).
Note that $\varphi$ takes a $\xi$-preserving conformal
transformation of $(M,g,\xi)$ into a $\xi'$-preserving conformal
transformation of $(M',g',\xi')$.
Conformally related Bargmann spaces
have, therefore, isomorphic symmetry groups.

The conserved quantities can be related by comparing
the expressions in (\ref{consQ}).
Using the transformation properties of the scalar curvature,
a short calculation shows that the conserved quantities
associated with $X=(X^\mu)$ on $(M,g,\xi)$ and with
$X'=\varphi_\star X$ on
$(M',g',\xi')$ coincide,
\begin{equation}
Q_X=\varphi^\star Q'_{X'}.
\label{Qexport}
\end{equation}
The labels of the generators
are, however, different (see the examples below).

\vskip2mm
$\bullet$ As a first application, we note the
 lift to Bargmann space of Niederer's mapping
\cite{NiedererOsc, BDPOsc}
\begin{eqnarray}
&\varphi(t_{\rm osc},{\vx}_{\rm osc},s_{\rm osc})
=(T,\vX,S),&\nonumber
\\[10pt]
&T=\displaystyle\frac{\tan\omega\,t_{\rm osc}}{\omega},
\qquad
\vX= \displaystyle\frac{{\vx}_{\rm osc}}{\cos\omega t_{\rm osc}},
\qquad
&S=s_{\rm osc}-\displaystyle\frac{\omega r_{\rm osc}^2}{2}\tan\omega t_{\rm osc}
\label{oscfree}
\end{eqnarray}
carries the oscillator metric (\ref{oscMet})
Bargmann-conformally
($\varphi_\star\partial_{s_{\rm osc}}=\partial_S$)
into the free form (\ref{Minkowski}),
with conformal factor
$\Omega(t_{\rm osc})=|\cos\omega t_{\rm osc}|^{-1}$.
 A solution in the harmonic background can be obtained by
using equation~(\ref{apsiexp}).

A subtlety arises, however. The mapping (\ref{oscfree}) is many-to-one~:  it maps each ``open strip''
\begin{equation}
I_j=\Big\{
({\vx}_{\rm osc},t_{\rm osc},s_{\rm osc})\;\big|\;
(j-\2)\pi<\omega t_{\rm osc}<(j+\2)\pi
\Big\},
\qquad
\,j=0,\pm1,\ldots,
\end{equation}
corresponding to a \textit{half oscillator-period}, onto
full Minkowski space.
Application of (\ref{apsiexp}) with $\Psi$ an ``empty-space'' solution
yields, in each $I_j$, a solution, $\Psi^{(j)}_{\rm osc}$. However,
at the contact points
$t_j\equiv(j+1/2)(\pi/\omega)$, these fields may not match.
For example, for the ``empty-space'' solution obtained by an expansion,
equation~(\ref{expansImp}) with ${\vB}=0,\,k\neq0$,
\begin{equation}
\lim_{t_{\rm osc}\to t_j-0}\Psi^{(j)}_{\rm osc}=
(-1)^{j+1}\frac{\omega}{ k}
e^{-i\frac{\omega^2}{2k}r_{\rm osc}^2}
\Psi_0(-\frac{\omega}{k}\vX)=
-\lim_{t_{\rm osc}\to t_j+0}\Psi^{(j+1)}_{\rm osc}.
\end{equation}
Hence, the left and right limits differ by a sign.
The continuity of the wave functions is restored
by including the ``Maslov'' phase correction \cite{Maslov}~:
\begin{equation}
\begin{array}{lll}
\Psi_{\rm osc}(t_{\rm osc},{\vx}_{\rm osc})&=&
(-1)^{j}\,\displaystyle\frac{1}{\cos\omega t_{\rm osc}}\,
\exp\left\{-\frac{i\omega}{2}r_{\rm osc}^2\tan{\omega t_{\rm osc}}\right\}\,
\Psi(T,\vX),
\\[18pt]
(A_{\rm osc})_0(t_{\rm osc},{\vx}_{\rm osc})
&=& \displaystyle\frac{1}{\cos^2\omega t_{\rm osc}}
\big[
A_0(T,{\vx})-\omega\sin\omega t_{\rm osc}\;
{\vx}_{\rm osc}\cdot{\vA}(T,\vX)
\big],
\\[18pt]
{\vA}_{\rm osc}(t_{\rm osc},{\vx}_{\rm osc})
&=& \displaystyle\frac{1}{\cos\omega t_{\rm osc}}\,
{\vA}(T,{\vx}),
\end{array}
\label{MaslovExport}
\end{equation}

 Equation~(\ref{MaslovExport})
extends the results in \cite{JPt-d2}, valid for
$|t_{\rm osc}|<\pi/2\omega$,
to any $t_{\rm osc}$ (\footnote{For the static solution in \cite{JaPi1}
or for that obtained from it by
a boost, $\lim_{t_{\rm osc}\to t_j}\Psi^{(j)}_{\rm osc}=0$,
and the inclusion of the correction factor is not mandatory.
}).

Since the oscillator metric (\ref{oscMet}) is
Bargmann-conformally related to
Minkowski space, Chern-Simons theory in the oscillator background again
has a Schr\"odinger symmetry, but with ``distorted''
generators. The latter are,
\begin{eqnarray}
J_{\rm osc}=\cJ,
\qquad
\cE_{\rm osc}={\cH}+\omega^2\cK,
\qquad
N_{\rm osc}=\cN,
\end{eqnarray}
completed by
\begin{eqnarray}
(C_{\rm osc})_\pm=
\cH-\omega^2\cK\pm 2i\omega\,\cD,
\qquad
(\vec{P}_{\rm osc})_\pm=
\vec{\cP}\pm i\omega\,\vec{\cG}.
\end{eqnarray}
In particular,
the oscillator-Hamiltonian, $H_{\rm osc}$,
is  a combination of the ``empty-space'' ($\omega=0$)
Hamiltonian ${\cH}$  and the expansion, $\cK$, etc.

\vskip2mm
$\bullet$ Turning to the magnetic case,
observe that the ``magnetic'' metric (\ref{BMet}) is
readily transformed into an oscillator metric (\ref{oscMet})
by the mapping
$
\varphi(t,{\vx},s)=(t_{\rm osc},{\vx}_{\rm osc},s_{\rm osc}),
$
\begin{equation}
t_{\rm osc}=t,
\qquad
x_{\rm osc}^i=x^i\cos\omega t+\epsilon^i_jx^j\sin\omega t,
\qquad
s_{\rm osc}=s,
\label{tdrot}
\end{equation}
which amounts to switching
to a rotating frame with angular velocity $\omega={\cal B}/2$.
The vertical vectors
$\partial_{s_{\rm osc}}$ and $\partial_s$ are permuted.

Composing the two steps, we see that the time-dependent rotation (\ref{tdrot}),
followed by the transformation (\ref{oscfree}),
[which projects to the coordinate transformation (\ref{TtXx})],
conformally carries the
constant-${\cal B}$ metric (\ref{BMet}) into the free
($\omega=0$) metric.
It carries, therefore,
the ``empty'' space so\-lution $e^{is}\Psi$, with $\Psi$ as
in (\ref{expansImp}), into a solution in
a uni\-form mag\-netic field back\-ground
according to equation~(\ref{apsiexp}). Taking into account the equivariance,
we recover the formulas of \cite{EHIt-d},
duly corrected by the Maslov factor $(-1)^j$.

\vskip2mm
The setup  allows us to
``export'' the Schr\"odinger symmetry to
non-relativistic Chern-Simons theory in
the constant magnetic field background. The (rather
complicated) generators \cite{Hotta} can be
 obtained using equation~(\ref{Qexport}).

$\bullet$ For example, time-translation $t\to t+\tau$ in
the ${\cal B}$-background amounts to a time translation for the
oscillator with parameter $\tau$, followed by a rotation with angle $\omega\tau$. Hence,
\begin{equation}
\cE_{\cal B}=\cE_{\rm osc}-\omega\cJ=\cH+\omega^2\cK-\omega\cJ.
\end{equation}

$\bullet$ Similarly, a space translation for ${\cal B}$ amounts, in ``empty'' space,
to a space translation and a rotated boost~:
\begin{equation}
P_B^i=\cP^i+\omega\,\epsilon^{ij}\cG^j,
\end{equation}
etc.

All our preceding results apply to any Bargmann space which can
be Bargmann-conformally mapped into Minkowski space.
All these
``Bargmann-conformally flat'' spaces can be determined \cite{DHP2}.
In $D=n+2>3$ dimensions, conformal
flatness is guaranteed by the
vanishing of the conformal Weyl tensor $C^{\mu\nu}_{\ \ \rho\sigma}$.
A tedious calculation \cite{DHP2} yields
\begin{equation}
C^{\mu\nu}_{\ \ \rho\sigma}=
R^{\mu\nu}_{\ \ \rho\sigma}-\frac{4}{D-2}\,
\varrho\,\delta^{[\mu}_{\ [\rho}\,\xi^{\nu]}\xi^{ }_{\sigma]}.
\end{equation}
 Skipping technical details, here we only
state that Schr\"odinger-conformal flatness requires \cite{DHP2},
\begin{eqnarray}
&{\cal A}_i=\2\epsilon_{ij}{\cal B}(t)x^j+a_i,
\qquad
{\vnabla}\times{\va}=0,
\qquad\partial_t{\va}=0,
\\[12pt]
&U(t,{\vx})=\2 C(t)r^2+\vec{F}(t)\cdot{\vx}+K(t).
\label{confflat}
\end{eqnarray}

The  metric (\ref{genBmetric})-(\ref{confflat}) describes a
uniform magnetic field ${\cal  B}(t)$,
an attractive or  repulsive,  $C(t)=\omega^2(t)$ or
$C(t)=-\omega^2(t)$,
isotropic oscillator and a uniform force field $\vec{F}(t)$ in the plane, all of
which may depend on time. It also includes a curl-free
vector potential~${\va}({\vx})$ that can be gauged away if the
transverse space is simply connected: $a_i=\partial_if$ and the coordinate transformation $(t,{\vx},s)\to(t,{\vx},s+f)$
results in the ``gauge'' transformation
\begin{equation}
{\cal{A}}_i\to{\cal{A}}_i-\partial_if=
-\2{\cal B}\,\epsilon_{ij}x^j.
\end{equation}
 However, if the space is not simply connected, we can
also include an external Aharonov-Bohm-type vector potential,
that can not be gauged away by an everywhere-defined
transformation.

Being conformally related, all these metrics share the symmetries of flat
Bargmann space. For example, if the transverse space is $\IR^2$, we get the
full Schr\"odinger symmetry; for $\IR^2\setminus\{0\}$ the symmetry is reduced instead to
\begin{equation}
{\rm o}(2)\times{\rm o}(2,1)\times\IR,
\end{equation}
as found for a magnetic vortex \cite{MagVort}.

\vskip2mm
$\bullet$
The case of a constant ``electric'' field
is quite amusing. Its metric,
\begin{equation}
d{\vx}^2+2dtds-2\vec{F}\cdot{\vx}dt^2,
\end{equation}
can be brought to the free
form by switching to an accelerated coordinate system,
\begin{equation}
{\vX}={\vx}+\2\vec{F}\,t^2,
\qquad
T=t,
\quad
S=s-\vec{F}\cdot{\vx}\,t-\smallover1/6\vec{F}^2t^3.
\label{cEfield}
\end{equation}
This example
also shows that the action of the Schr\"odinger group --- e.g. a rotation
--- looks quite different in the inertial and in the moving frame.

In conclusion, our ``non-relativistic Kaluza-Klein'' approach provides
a unified view on the various known constructions
and explains the common origin of their symmetries.

\goodbreak
\subsection{Spinors on Bargmann space}

The (2+1) dimensional
spinor model presented in Section \ref{spinorvort} can also be obtained
in the Kaluza-Klein-type framework. For simplicity, we limit ourself to
 Minkowski space; the general case is discussed in Ref. \cite{DHPSpinor2}.
Expressing the metric  in
light-cone coordinates, $d\vx^2+2dtds$, we have the
Dirac matrices,
\begin{equation}
\gamma^t=\left(\begin{array}{cc}0&0\\1&0\end{array}\right),
\qquad
\vec{\gamma}=\left(\begin{array}{cc}
-i\vec{\sigma} &0\\\,0&i\vec{\sigma}\end{array}\right),
\qquad
\gamma^s=\left(\begin{array}{cc}0&-2\\0&\ 0\end{array}\right).
\end{equation}
These matrices satisfy $\{\gamma^\mu,\gamma^\nu\}=-2g^{\mu\nu}$ as they should
and are, just like the 4d chirality operator,
\begin{equation}
\Gamma\equiv \gamma^5=-\frac{\sqrt{-g}}{4!}\epsilon_{\mu\nu\rho\sigma}
\gamma^\mu\gamma^\nu\gamma^\rho\gamma^\sigma
=\left(
\begin{array}{cc}-i\sigma_3&0\\ \,0&i\sigma_3\end{array}\right),
\label{chirality}
\end{equation}
hermitian,
$
\overline{\gamma^\mu}=\gamma^\mu,
$
with respect to the hermitian structure defined by the Gram matrix
$
G=\left(\begin{array}{ll}
0&1\\1&0
\end{array}\right).
$

\vskip2mm
Let us now posit~:

$\bullet$ the gauged {\it massless Dirac equation} on $M$ \cite{DuvClaus},
\begin{equation}
\D\psi=0,
\qquad
\D=\gamma^\mu D_\mu\ ;
\label{m0Dirac}
\end{equation}

$\bullet$ the Bianchi identity
\begin{equation}
\epsilon_{\mu\nu\rho}\p_\mu f_{\nu\rho}=0;
\label{SPBianchi}
\end{equation}

$\bullet$ the lifted Chern-Simons equation (\ref{4dFCI}), i.e.,
 \begin{equation}
\kappa f_{\mu\nu}=e\sqrt{-g}\epsilon_{\mu\nu\rho\sigma}\,
\xi^\rho j^\sigma,
\label{SP4dFCI}
\end{equation}
where the current is
\begin{equation}
j^\sigma=\bar{\psi}\,\gamma^\sigma\psi.
\label{SP4dcur}
\end{equation}

Then our clue is that  when also

$\bullet$ equivariance, (\ref{equivar}),  i.e.,
\begin{equation}
D_\xi\psi=m\psi,
\label{SPequiv}
\end{equation}
is imposed, a self-consistent system, namely
the one considered in Section (\ref{NRSpinor}) is obtained.
In particular, the massless Dirac equation, (\ref{m0Dirac}), reduces to the  first-order L\'evy-Leblond system (\ref{LLequations}).

This framework also explains the simultaneous presence of the two fields
$\psi_\pm$ in Section \ref{spinorvort}~: they are the chiral components,
\begin{equation}
i\Gamma\psi_\pm=\pm\psi_\pm,
\label{4dchir}
\end{equation}
cf. (\ref{ChiralOp}),
 of the $4$-component Dirac spinors ``before'' dimensional reduction.
In $4d$, Dirac spinors have in fact $4$ components, while in $(2+1)d$
they only have $2$ components. Dimensional reduction yields, hence,
{\it two} planar systems, distinguished by their chirality
``before'' reduction.

\kikezd{Conformal symmetry in spinor representation}

The setup can also be used to
rederive the Schr\"odinger symmetry of the spinor system \cite{DHPSpinor2}
along the lines presented, in the scalar case, earlier in this Section.
In detail, we have seen those conformal transformations  of Minkowski space
which preserve the ``vertical vector'' $\xi=\partial_s$  form the $9$-parameter (extended)
Schr\"odinger group as a $9$-dimensional subgroup of the
(relativistic) conformal group
$\O(4,2)$ which acts on $M$ according to
(\ref{SchronB}).

Its induced action on spacetime
is given by $(t,\vx)\to(t^*,\vx{\,}^*)$, i.e. by ``forgetting''
about the $s$ coordinate in (\ref{SchronB}).
Here, $R\in{\rm SO}(2)$ is a rotation in the plane, and $\vb,\vc\in\IR^2$.  Denote by
\begin{equation}
u=(a,\vb,\vc,d,e,f,g,h)
\label{SpinBarg}
\quad\hbox{\small where}\quad
a=\exp(\smallover i/2 \theta\sigma_3) \quad\hbox{\small is
s.t.}\quad~~ a(\vsigma\cdot\vx)a^{-1}=\vsigma\cdot(R\vx)
\end{equation}
a typical element of the  ``spin-Schr\"odinger'' group,


We first consider dilations and expansions.

$\bullet$ For expansions one gets from Eq.~(\ref{SchronB})
$t^*=t/(ft+1)$, $\vx{\,}^*=\vx/(ft+1)$
and $s^*=s+\frac{1}{2}f\vx{\,}^2/(ft+1)$.
A tedious calculation then
yields
\begin{equation}
\psi_f(t,\vx,s)
=\left(\begin{array}{cc}
(ft+1)^{-1}& 0
\\[14pt]
\displaystyle\frac{f}{2i}\,(\vsigma\cdot\vx)(ft+1)^{-2}
& (ft+1)^{-2}
\end{array}\right)
\psi(t^*,\vx{\,}^*,s^*).
\end{equation}

Putting $\Psi(t,\vx)\equiv\psi(t,\vx,s)e^{-ims}$, one readily gets,
\begin{equation}
\Psi_{\!f}(t,\vx)
=
\frac{1}{(ft+1)^2}
\left(\begin{array}{cc}
ft+1 & 0
\\[14pt]
\displaystyle\frac{f}{2i}\,(\vsigma\cdot\vx) &1
\end{array}\right)
\Psi\left(\frac{t}{ft+1},\frac{\vx}{ft+1}\right)
\exp\left[-\frac{im f\vx{\,}^2}{2(ft+1)}\right],
\end{equation}
and one verifies that we have, indeed, obtained a representation
$(\Psi_{\!f})_{f'}=\Psi_{\!f+f'}$ of the (additive) group of pure
expansions on the solutions of the LL equation.

$\bullet$ As for  dilations, a straightforward calculation yields
\begin{equation}
\Psi_{\!d}(t,\vx)=
\left(\begin{array}{cc}
d\hfill &  0
\\[12pt]
0& d^2
\end{array}\right)
\Psi(d^2t,d\,\vx),
\end{equation}
which, again, turns out to be a genuine representation.

On the other hand, a simple calculation gives the following
(anti-)representation of the
 Bargmann group
(a mere group of isometries), originally due to L\'evy-Leblond~
 \cite{LL}
\begin{equation}
\psi_u(t,\vx,s)
=
\left(\begin{array}{cc}
a^{-1}\hfill &  0
\\[12pt]
{\displaystyle\frac{i}{2}\,a^{-1}\vsigma\cdot\vb}\hfill &a^{-1}
\end{array}\right)
\psi(t^*,\vx{\,}^*,s^*),
\end{equation}
where $(t^*,\vx{\,}^*,s^*)=u\cdot(t,\vx,s)$ with
$u=(a,\vb,\vc,1,e,0,1,h)$ in the spin-Bargmann group,
(\ref{SpinBarg}).
Combining the two results, one ends up with
the following (anti-)representation of the full spin-Schr\"odinger group,
\begin{equation}
\psi_u(t,\vx,s)
=
\frac{1}{(ft+g)^2}
\left(\begin{array}{ll}
a^{-1}(ft+g) & 0
\\[12pt]
\displaystyle\frac{1}{2i}\,
a^{-1}\vsigma\cdot(fR\vx-g\vb+f\vc)\qquad &a^{-1}
\end{array}\right)
\psi(t^*,\vx{\,}^*,s^*),
\end{equation}
where $(t^*,\vx{\,}^*,s^*)=u\cdot(t,\vx,s)$ with
$u=(a,\vb,\vc,d,e,f,g,h)$.

By equivariance, one finally obtains the `natural'
(anti-)representation of the full spin-Schr\"odinger group on
the space of solutions  of the free L\'evy-Leblond equation, that is,
\begin{equation}
\begin{array}{lll}
\Psi_u(t,\vx)
&=
& \displaystyle\frac{1}{(ft+g)^2}\left(
\begin{array}{cc}
a^{-1}(ft+g)\hfill &0
\\[12pt]
\displaystyle\frac{1}{2i}\,
a^{-1}\vsigma\cdot(fR\vx-g\vb+f\vc)  &a^{-1}
\end{array}\right)
\Psi\left(\frac{dt+e}{ft+g},\frac{R\vx+\vb t+\vc}{ft+g}\right)
\\[36pt]
&&\times
\exp\left(-im\left[\displaystyle\frac{f(R\vx+\vb t+\vc)^2}{2(ft+g)}-
\vb\cdot{R\vx}-\displaystyle\frac{t}{2}\vb{\,}^2+h\right]\right).
\end{array}
\label{SchrOnSpinors}
\end{equation}
The implementation (\ref{SchrOnSpinors}) is a {\it spinor
representation} of the Schr\"odinger group for spin $\2$ \footnote{An infinite-component, Majorana-type representation of the two-fold
centrally extended Galilei group is presented in \cite{HPMaj}.}.

The action of the Schr\"odinger group commutes with
 $\Gamma$, and descends therefore to the chiral components $\psi_\pm$.
This chiral representation is also of the `lower triangular' form:
though the $\Phi_\pm$ components are mapped into themselves,
the $\chi_+$ (resp.~$\chi_-$) components transform into combinations
of $\Phi_+$ and $\chi_+$  (resp.~$\Phi_-$ and $\chi_-$) under boosts
and expansion. It is, therefore, not possible to further reduce the
representation provided by the $\psi_\pm$.

The infinitesimal action, i.e., the action of the Schr\"odinger Lie algebra, can also be obtained from the spinor representation,
\begin{equation}
\left\{
\begin{array}{llll}
P_\mu&=&-i\partial_\mu
&\hbox{\small translations,}
\\[12pt]
M_{\mu\nu}&=&-i\left(
x_\mu\partial_\nu-x_\nu\partial_\mu\right)
+\smallover{i}/4[\gamma_\mu,\gamma_\nu]\quad
&\hbox{\small Lorentz transformations},
\\[12pt]
d&=&-ix^\mu\partial_\mu-\smallover3i/2
&\hbox{\small dilations},
\\[12pt]
K_\mu&=&-i\left(
x_\nu x^\nu\partial_\mu
-x_\mu(3+2x^\nu\partial_\nu)\right)
-\smallover{i}/2[\gamma_\mu,\gamma_\nu]x^\nu\qquad
&\hbox{\small conformal transformations},
\end{array}
\right.
\label{spinconfalg}
\end{equation}
of the conformal algebra $\o(4,2)$ in (\ref{confalg}).

Remember now that the massless Dirac equation can be
obtained by varying the {\sl matter action}
\begin{equation}
\cS
=
\int_M\Im\big\{\overline{\psi}\,\D\psi\big\}\sqrt{-g}\,d^4\!x.
\label{DiracAction}
\end{equation}
The variational derivative with respect to the vector
potential, $\delta\cS/\delta a_\mu$, yields the current $j^\mu$ in
Eq.~(\ref{SP4dcur}), whose conservation, $\nabla_\mu j^\mu=0$, follows
from the invariance of $\cS$ with respect to gauge transformations.

Denote $\cL$ the {\sl Lagrangian density} in  (\ref{DiracAction}). Then a short calculation  \cite{DHPSpinor2} shows that
$
\delta_X\cL=0
$
modulo a surface term. Thus, for any conformal vector field $X$, we have
\begin{equation}
\delta_X\cS=0
\end{equation}
at the critical points of the matter action, $\cS$, which again proves
the conformal symmetry of our system.
The energy-momentum tensor
$\vartheta_{\mu\nu}=-2\,\delta\cS/\delta g^{\mu\nu}$, {\it viz.}
\begin{equation}
\vartheta_{\mu\nu}
=-2\,\frac{\delta\cS\;\;}{\delta g^{\mu\nu}}
=\2\,\Im\Big\{
\overline{\psi}\big(\gamma_\mu D_\nu+\gamma_\nu D_\mu\big)\psi\Big\},
\end{equation}
is therefore traceless as well as conserved and is automatically symmetric.
Hence, a conserved quantity, namely
\begin{equation}
{\cal Q}_X=
\int\vartheta_{\mu\nu}X^\mu\xi^\nu\sqrt{\gamma}\,d^2\!\vx,
\end{equation}
where $\gamma=\det(g_{ij})$,
 is associated
to each $\xi$-preserving conformal vector field.
In detail, we get exactly those quantities listed above in
(\ref{SpinorSchgen}).

\vskip10mm\goodbreak
\kikezd{Acknowledgements}.
P.A.H. is indebted to the {\it Institute of Modern Physics
of the Chinese Academy of Sciences at Lanzhou} (China)
and to the  {\it Chern Institute of Mathematics of Nankai University} (China).
P-M.Z. is supported by the National Natural Science Foundation of China
(Grant No. 10604024),  and he is also
 indebted to the {\it International
Center for Theoretical Physics} (ICTP) of Trieste (Italy) for
hospitality. We would like to thank Noel Gorman, Kimyeong Lee and
Erick Weinberg for correspondence and Zhao Zhenhua for his help to
produce a computer plot.

\newpage

\section{REFERENCES}


\end{document}